%% file: ai-adoption-paper.tex
\documentclass[12pt]{article}
\usepackage[a4paper,pdftex]{geometry}
\usepackage[T1]{fontenc}
\usepackage[english]{babel}
\usepackage[authoryear]{natbib}
\usepackage[small]{titlesec}
\usepackage[justification=centering]{caption}
\usepackage[final]{pdfpages}

\usepackage{amsmath,amsthm,amssymb,graphicx,enumerate,booktabs,bigstrut,rotating,multirow,float,etoolbox,lmodern,comment,listings,qtree,a4wide,moresize,bbm,subcaption,tikz,pdfescape,mathtools,flafter,enumitem,setspace,ragged2e,colortbl,lineno,lscape,pdflscape,stackengine}

\usepackage{datatool}
\DTLsetseparator{,}
\DTLloaddb[keys={key}]{est}{results/kv_store.csv}
\newcommand{\getval}[1]{\DTLfetch{est}{key}{#1}{value}}

\usetikzlibrary{shapes.geometric, arrows, positioning, decorations.pathreplacing}

\newcolumntype{L}[1]{>{\raggedright\let\newline\\\arraybackslash\hspace{0pt}}m{#1}}
\newcolumntype{C}[1]{>{\centering\let\newline\\\arraybackslash\hspace{0pt}}m{#1}}
\newcolumntype{R}[1]{>{\raggedleft\let\newline\\\arraybackslash\hspace{0pt}}m{#1}}

\usepackage{hyperref}
\hypersetup{colorlinks,linkcolor={red},citecolor={blue},urlcolor={blue}}

\makeatletter
\renewcommand\subsubsection{\@startsection{subsubsection}{3}{\z@}%
	{-3.25ex\@plus -1ex \@minus -.2ex}%
	{-1.5ex \@plus -.2ex}%
	{\normalfont\normalsize\bfseries}
}
\def\@biblabel#1{\hspace*{-\labelsep}}

\makeatother

\def\sym#1{\ifmmode^{#1}\else\(^{#1}\)\fi}

\pdfpageheight\paperheight
\pdfpagewidth\paperwidth

\definecolor{redcomment}{RGB}{213,94,0}
\definecolor{yellowcomment}{RGB}{240,228,66}
\definecolor{greencomment}{RGB}{0,158,115}
\definecolor{bluecomment}{RGB}{0,114,178}

\begin{document}
\begin{filecontents*}{results/kv_store.csv}
key,value,timestamp,label
survey_start,December 2024,,Survey start month (sec 2.1)
survey_end,February 2025,,Survey end month (sec 2.1)
incentive_low_usd,50,,Amazon gift card lottery low end (sec 2.1)
incentive_high_usd,500,,Amazon gift card lottery high end (sec 2.1)
n_enrolled_students,2760,,"Total Middlebury enrollment (Fall 2024, admin records)"
n_midd_students_approx,"2,800",,Approximate enrollment cited in intro (sec 1)
n_majors,49,,Number of majors at Middlebury (admin)
chatgpt_launch_date,November 2022,,ChatGPT public launch (sec 3.3)
semester_current,Fall 2024,,Data-collection reference semester (sec 3.3)
usd_chatgpt_monthly,20,,ChatGPT Plus monthly price at survey time (sec 3.4)
usd_subscription_high,40,,Highest subscription bracket threshold ($40/mo) (sec 3.4)
n_tasks,10,,Number of academic tasks in AI-usage module (sec 4.1)
n_scale_points,five-point,,Likert scale length for AI-purpose items (sec 4.1)
n_prolific_students,56,,Prolific student classification sample (Appendix C)
n_prolific_instructors,51,,Prolific instructor classification sample (Appendix C)
usd_prolific_pay,5,,Prolific compensation per respondent (Appendix C)
min_prolific_duration,5,,Prolific survey duration in minutes (Appendix C)
prolific_date,February 2026,,Prolific data-collection date (Appendix C)
n_classification_dims,4,,Dimensions respondents rated per task (Appendix C)
scale_importance_low,0,,Importance-slider minimum (Appendix C)
scale_importance_high,100,,Importance-slider maximum (Appendix C)
rank_low,1,,Rank 1 = most 'works with' (Appendix C)
rank_high,10,,Rank 10 = most 'does for' (Appendix C)
n_tasks_agreed,8,,Tasks with student-instructor classification agreement (Appendix C)
n_aug_tasks,4,,Tasks classified as augmentation
n_auto_tasks,6,,Tasks classified as automation
n_weight_cells_gc,8,,Gender-cohort weighting cells (Appendix B)
n_weight_cells_race,4,,Race weighting cells (Appendix B)
n_cohorts,4,,Number of cohorts
pct_cds_female,53.3,,Middlebury CDS female share (%) (Appendix B)
pct_adoption_us_workers,40,,Bick et al. 2026 — U.S. workers AI-at-work (%)
pct_adoption_us_adults,23,,McClain 2024 — U.S. adults used ChatGPT (%)
zeitgeist_version,7.0,,Zeitgeist survey version (§3.1)
zeitgeist_period,Spring 2025,,Zeitgeist survey period (§3.1)
n_zeitgeist,1026,,Zeitgeist respondents (§3.1)
pct_zeitgeist_rate,37,,Zeitgeist response rate (%) (§3.1)
pct_zeitgeist_adoption,85.2,,Zeitgeist AI-use (%) (§3.1)
pct_harvard_adoption,87.5,,Hirabayashi 2024 — Harvard adoption (%)
pct_ihe_adoption,85,,Flaherty 2025 Inside Higher Ed (%)
pct_uk_adoption,88,,Freeman 2025 UK undergraduates (%)
pct_bestcolleges_adoption,56,,Nam 2023 BestColleges Fall 2023 (%)
pct_ravselj_adoption,71.4,,Ravselj et al. 2025 multi-country (%)
pct_sweden_adoption,63,,Stöhr et al. 2024 Sweden (%)
pct_norway_adoption,68.9,,Carvajal 2024 Norway (%)
pct_pew_adults_2025,34,,McClain 2025 Pew U.S. adults (%)
pct_gallup_workers,45,,Gallup 2025 U.S. workers (%)
pct_gallup_daily,10,,Gallup 2025 U.S. workers daily use (%)
pct_chatgpt_world,10,,Chatterji et al. 2025 ChatGPT world adult share (%)
pct_bick_low,36,,Bick/Hartley 2026 low end workforce (%)
pct_bick_high,45,,Bick/Hartley 2026 high end workforce (%)
pct_bick_tech_low,56,,Bick/Hartley 2026 info/tech industry low (%)
pct_bick_tech_high,76,,Bick/Hartley 2026 info/tech industry high (%)
pct_humlum_adoption,41,,Humlum & Vestergaard 2025 AI-exposed workers (%)
pct_pew_under30,58,,McClain 2025 Pew under-30 adults (%)
pct_pew_overall,34,,McClain 2025 Pew overall (%)
age_range_young,18--29,,Bick et al. young-worker age range
age_range_old,50--64,,Bick et al. older-worker age range
n_otis_studies,18,,Otis 2024 meta-analysis # studies
pp_otis_low,10,,Otis 2024 low-end gender gap (pp)
pp_otis_high,20,,Otis 2024 high-end gender gap (pp)
pct_nam_business,62,,Nam 2023 business majors (%)
pct_nam_stem,59,,Nam 2023 STEM majors (%)
pct_nam_humanities,52,,Nam 2023 humanities majors (%)
pct_bick_stem,46.0,,Bick et al. 2026 STEM graduates (%)
pct_bick_business,40.0,,Bick et al. 2026 Business/Econ graduates (%)
pct_bick_liberal,22.4,,Bick et al. 2026 Liberal Arts graduates (%)
n_years_computers,20,,Years for computers to reach 80% (Bick et al. 2026)
pct_computer_benchmark,80,,80% benchmark adoption (comparison)
n_years_internet,15,,Years for internet to reach 80% (Bick et al. 2026)
pct_pew_feb2024,23,,Pew Feb 2024 ChatGPT adult use (%)
pew_feb_date,February 2024,,Pew survey reference date
pct_pew_jul2023,18,,Pew July 2023 ChatGPT adult use (%)
pew_jul_date,July 2023,,Pew earlier survey reference date
pct_harvard_chatgpt,95,,Hirabayashi 2024 Harvard ChatGPT share (%)
pct_bick_chatgpt,28.5,,Bick et al. 2026 ChatGPT U.S. adults (%)
pct_bick_gemini,16.3,,Bick et al. 2026 Gemini U.S. adults (%)
pct_ravselj_pay,8,,Ravselj et al. 2025 paid AI share (%)
pct_harvard_pay,29.7,,Hirabayashi 2024 Harvard paid AI (%)
pct_harvard_officehours,25,,Hirabayashi 2024 Harvard substitute for office hours (%)
pct_handa_technical,33.5,,Handa et al. 2025 technical explanations (%)
pct_handa_design,39.3,,Handa et al. 2025 design practice questions (%)
n_ammari_logs,10536,,Ammari et al. 2025 ChatGPT logs analyzed
pct_handa_aug,57,,Handa et al. 2025 workplace augmentation (%)
pct_handa_auto,43,,Handa et al. 2025 workplace automation (%)
pct_ravselj_aug,64.7,,Ravselj et al. 2025 augmentation (%)
pct_ravselj_auto,63.2,,Ravselj et al. 2025 automation (%)
pct_ravselj_bottom_aug,74,,Ravselj bottom quintile augmentation (%)
pct_ravselj_bottom_auto,65,,Ravselj bottom quintile automation (%)
pp_carvajal_female,37.2,,"Carvajal 2024 ban reduces usage, females (pp)"
pp_carvajal_male,20.7,,"Carvajal 2024 ban reduces usage, males (pp)"
pp_carvajal_overall,28.9,,"Carvajal 2024 ban reduces usage, overall (pp)"
pct_stohr_policy,19.1,,Stöhr et al. 2024 Swedish students with policy guidance (%)
pct_ravselj_general_know,68.8,,Ravselj general-knowledge belief (%)
pct_ravselj_specific_know,62.7,,Ravselj specific-knowledge belief (%)
pct_ravselj_deadlines,57.4,,Ravselj meet-deadlines belief (%)
pct_stohr_effective,47.7,,Stöhr 2024 effective-learner belief (%)
pct_stohr_grades,17.3,,Stöhr 2024 grades-improve belief (%)
pct_coding_benchmark,37.0,,"Coding-task usage, major poststrat weights (%) (App B)"
pct_coding_gcweight,45.3,,"Coding-task usage, gender x cohort weights (%) (App B)"
pct_arts_low,57.6,,Arts usage min across weighting schemes (%) (App B)
pct_arts_high,73.8,,Arts usage max across weighting schemes (%) (App B)
n_analysis_sample,634,,Analysis sample size (sec 2.2)
n_survey_started,739,,Students who began the survey (sec 2.2)
n_respondents,739,,"Total survey respondents, alias for n_survey_started (sec 1)"
n_excluded,105,,Dropped before AI-usage module (sec 2.2)
pct_response_rate,26.8,,Response rate: started/enrolled (%) (sec 2.2)
pct_analysis_sample,23.0,,Analysis sample as % of enrollment (sec 2.2)
pct_sample_male_unw,44.6,,Unweighted % male (sec 2.2)
pct_sample_female_unw,50.8,,Unweighted % female (sec 2.2)
pct_sample_white_unw,61.8,,Unweighted % white (sec 2.2)
pct_sample_asian_unw,15.5,,Unweighted % Asian (sec 2.2)
pct_sample_hispanic_unw,9.9,,Unweighted % Hispanic (sec 2.2)
pct_sample_black_unw,3.6,,Unweighted % Black (sec 2.2)
pct_sample_public_hs,54.3,,Unweighted % public high school (sec 2.2)
pct_sample_private_hs,42.0,,Unweighted % private high school (sec 2.2)
n_majors_represented,43,,# majors represented in sample (sec 2.2)
n_fields,seven,,Number of fields of study (sec 2.2)
pct_undeclared,3.5,,% not yet declared (sec 2.2)
pct_white_sample_cmp,61.8,,Sample % white (cmp) 
pct_asian_sample_cmp,15.5,,Sample % Asian (cmp)
pct_black_sample_cmp,3.6,,Sample % Black (cmp)
pct_hispanic_sample_cmp,9.9,,Sample % Hispanic (cmp)
pct_white_admin,53.8,,Admin % white (sec 2.2)
pct_asian_admin,7.3,,Admin % Asian (sec 2.2)
pct_black_admin,5.2,,Admin % Black (sec 2.2)
pct_hispanic_admin,12.4,,Admin % Hispanic (sec 2.2)
pct_freshman_admin,25.5,,Admin % freshman (sec 2.2)
pct_senior_admin,28.7,,Admin % senior (sec 2.2)
pct_freshman_sample,31.1,,Sample % freshman (sec 2.2)
pct_senior_sample,21.3,,Sample % senior (sec 2.2)
pct_adoption_any,82.5,,Overall AI adoption (%) (sec 3.1)
pct_adoption_any_cmp,82.5,,"Our rate (repeat, cmp) (sec 3.1)"
pct_adoption_any_cmp2,82.5,,"Our rate (repeat 2, cmp) (sec 3.1)"
pct_adoption_round,80,,Rounded adoption rate (sec 3.1)
pct_adoption_round2,80,,Rounded adoption rate (sec 3.3)
pct_adoption_2yr_mark,80,,Adoption after 2 years (§1 headline)
pct_adoption_abstract,80,,Headline adoption used in abstract
pct_adoption_intro,80,,Headline adoption used in intro
pct_freq_rarely,23.5,,% rarely use (sec 3.1)
pct_freq_occasionally,22.2,,% occasionally (sec 3.1)
pct_freq_frequently,26.2,,% frequently (sec 3.1)
pct_freq_veryfrequently,10.6,,% very frequently (sec 3.1)
pct_midd_fall2023_adoption,39,,% Middlebury adopters by Fall 2023 (sec 3.1)
pct_adoption_male,88.7,,Adoption rate among male (%) (sec 3.2)
pct_adoption_female,78.4,,Adoption rate among female (%) (sec 3.2)
pct_adoption_white,80.2,,Adoption rate among white (%) (sec 3.2)
pct_adoption_black,92.3,,Adoption rate among black (%) (sec 3.2)
pct_adoption_asian,91.3,,Adoption rate among asian (%) (sec 3.2)
pct_adoption_latino,77.9,,Adoption rate among latino (%) (sec 3.2)
pct_adoption_pub_hs,80.4,,Adoption rate among pub_hs (%) (sec 3.2)
pct_adoption_priv_hs,84.1,,Adoption rate among priv_hs (%) (sec 3.2)
pct_adoption_arts,73.8,,Adoption rate among arts (%) (sec 3.2)
pct_adoption_hum,76.5,,Adoption rate among hum (%) (sec 3.2)
pct_adoption_lang,59.5,,Adoption rate among lang (%) (sec 3.2)
pct_adoption_lit,53.8,,Adoption rate among lit (%) (sec 3.2)
pct_adoption_natsci,87.4,,"Natural Sciences adoption (sec 1,3.2)"
pct_adoption_socsci,84.5,,Social Sciences adoption (sec 3.2)
pct_adoption_male_sec2,88.7,,Male adoption (sec 3.2 repeat)
pct_adoption_female_sec2,78.4,,Female adoption (sec 3.2 repeat)
pct_adoption_private_hs,84.1,,Private HS adoption (sec 3.2)
pct_adoption_public_hs,80.4,,Public HS adoption (sec 3.2)
pct_adoption_natsci_sec2,87.4,,Natural Sciences adoption (sec 3.2 repeat)
pct_adoption_languages,59.5,,"Languages adoption (sec 1,3.2)"
pct_adoption_languages_sec2,59.5,,Languages (sec 3.2 repeat)
pct_adoption_literature,53.8,,"Literature adoption (sec 1,3.2)"
pct_adoption_literature_sec2,53.8,,Literature (sec 3.2 repeat)
pct_adoption_lowgpa,87.1,,"Below-median GPA adoption (sec 1,3.2)"
pct_adoption_highgpa,80.3,,"Above-median GPA adoption (sec 1,3.2)"
pct_adoption_lowgpa_sec2,87.1,,Below-median GPA (sec 3.2 repeat)
pct_adoption_highgpa_sec2,80.3,,Above-median GPA (sec 3.2 repeat)
pp_gender_gap_col1,10.3,,Male coef (pp) Col 1 (sec 3.2)
pp_gender_gap_cmp,10.3,,"Male coef (repeat, sec 3.2)"
pp_black_col1,13.6,,Black coef (pp) Col 1 (sec 3.2)
pp_asian_col1,11.4,,Asian coef (pp) Col 1 (sec 3.2)
pp_public_hs_col1,4.1,,abs(Public-HS coef) (pp) Col 1 — paper uses 'less likely' phrasing (sec 3.2)
pct_cumadopt_pre,8.9,,Cumulative adoption before Spring 2023 (%)
pct_adoption_pre_spring_2023,8.9,,"Same, alias used in sec 1"
pct_cumadopt_fall2024,79.9,,Cumulative adoption by Fall 2024 (%)
pct_adopted_fall2024,31.2,,Share of current users adopting in Fall 2024
pct_adopted_spring2024,24.1,,Share of current users adopting in Spring 2024
pct_adopted_fall2023,19.8,,Share of current users adopting in Fall 2023
pp_gender_earlyadopt,8.7,,Male coef early-adoption (pp) Col 1 (sec 3.3)
pct_chatgpt_free,88.0,,% AI users on free ChatGPT (sec 3.4)
pct_gemini,13.5,,% AI users on Gemini (sec 3.4)
pct_copilot_use,7.7,,% AI users on Microsoft Copilot (sec 3.4)
pct_other_models_ceiling,5,,Other platforms each <5% (sec 3.4)
pct_pay_ai,11.4,,% AI users paying any (sec 3.4)
pct_pay_ai_cmp,11.4,,Repeat (footnote)
pct_pay_chatgpt,12.8,,% AI users on paid ChatGPT (sec 3.4 footnote)
pct_task_explain,80.3,,% AI users explain concepts (sec 4.1)
pct_task_explain_cmp,80.3,,Repeat (sec 4.1)
pct_task_summarize,74.0,,% AI users summarize texts (sec 4.1)
pct_task_summarize_cmp,74.0,,Repeat (sec 4.1)
pct_task_findinfo,63.1,,% AI users find information (sec 4.1)
pct_task_genideas,61.9,,% AI users generate ideas (sec 4.1)
pct_task_proofread,54.1,,% AI users proofread (sec 4.1)
pct_task_editessays,47.3,,% AI users edit essays (sec 4.1)
pct_task_editessays_cmp,47.3,,Repeat (sec 4.1)
pct_task_coding,34.4,,% AI users coding (sec 4.1)
pct_task_writeessays,23.5,,% AI users write essays (sec 4.1)
pct_task_images,20.4,,% AI users images (sec 4.1)
pct_avg_augmentation,56.0,,Avg augmentation usage (%) (sec 4.2)
pct_avg_automation,45.4,,Avg automation usage (%) (sec 4.2)
pp_aug_auto_gap,10.6,,Augmentation-automation gap (pp) (sec 4.2)
pct_users_augmentation_intro,56.0,,Aug rate (intro)
pct_users_automation_intro,45.4,,Auto rate (intro)
pp_male_aug,8.0,,Male coef augment_avg (pp) Table 4 col 2 (sec 4.3)
beta_black_aug,0.446,,Black coef augment_intense Table 4 col 3 (sec 4.3)
pp_male_auto,9.5,,Male coef automate_avg (pp) Table 4 col 5 (sec 4.3)
beta_black_auto,0.477,,Black coef automate_intense Table 4 col 6 (sec 4.3)
pct_openend_efficiency,33.9,,% open-ended responses mentioning efficiency (sec 4.2)
pct_openend_explain,30,,% classified motivation responses tagged 'explainer' (App D)
pct_openend_timesaving,25,,% classified motivation responses tagged 'time-saver' (App D)
pct_openend_tutor,12,,% classified motivation responses tagged 'tutor' (App D)
pct_policy_unrestricted,52.5,,% likely to use if unrestricted (sec 5.1)
pct_policy_cite,40.9,,% likely to use if citation required (sec 5.1)
pct_policy_noexplicit,42.2,,% likely to use if no explicit policy (sec 5.1)
pct_policy_prohibited,13.4,,% likely to use if prohibited (sec 5.1)
pct_policy_prohibited_unlikely,72.9,,% unlikely to use if prohibited (sec 5.1)
pp_ban_overall,39.1,,Ban effect overall (pp) (sec 5.1)
pp_prohibition_effect_intro,39,,"Ban effect (rounded, intro)"
pp_ban_female,38.4,,Ban effect among females (pp) (sec 5.1)
pp_ban_male,40.1,,Ban effect among males (pp) (sec 5.1)
pct_understand_policy,79.1,,% understand AI class policy (sec 5.2)
pct_policy_unclear,19.2,,"% policy unclear (sec 1, 5.2)"
pct_policy_unclear_body,19.2,,Repeat (sec 5.2)
pct_copilot_aware,10.0,,"% know Copilot free access (sec 1, 5.2)"
pct_copilot_aware_body,10.0,,Repeat (sec 5.2)
pct_cite_aware,31.3,,"% know how to cite AI (sec 1, 5.2)"
pct_cite_aware_body,31.3,,Repeat (sec 5.2)
pct_understand_female,81.7,,Female policy understanding (sec 5.2)
pct_understand_male,75.8,,Male policy understanding (sec 5.2)
pct_belief_understand,68.5,,% believe AI improves understanding (sec 6.1)
pct_belief_understand_cmp,68.5,,Repeat (sec 6.1)
pct_belief_improves_understand,68.5,,"Same stat, intro alias"
pct_belief_learn,57.6,,% believe AI improves learning (sec 6.1)
pct_belief_learn_cmp,57.6,,Repeat (sec 6.1)
pct_belief_improves_learn,57.6,,"Same stat, intro alias"
pct_belief_ontime,56.4,,% believe AI improves on-time completion (sec 6.1)
pct_belief_ontime_cmp,56.4,,Repeat (sec 6.1)
pct_belief_grades,37.6,,% believe AI improves grades (sec 6.1)
pct_belief_improves_grades,37.6,,"Same stat, intro alias"
pct_belief_grades_noeffect,59.2,,% report no effect on grades (sec 6.1)
pct_belief_grades_negative,3.2,,% report negative effect on grades (sec 6.1)
pp_slope_belief_learn,4.9,,Belief-learn adoption slope (pp per 10pp) (sec 6.1)
pp_slope_belief_understand,5.0,,Belief-understand adoption slope (pp per 10pp) (sec 6.1)
pp_slope_belief_grades,6.2,,Belief-grades adoption slope (pp per 10pp) (sec 6.1)
pp_slope_belief_ontime,5.4,,Belief-ontime adoption slope (pp per 10pp) (sec 6.1)
pp_slope_belief_learn_sec4,4.9,,Alias of pp_slope_belief_learn (sec 6.1)
pp_belief_change_unit,10-pp,,Unit used in belief-slope interpretations
pp_black_belief_learn,36.1,,Black coef belief-learn (pp) (footnote §6.1)
pp_lit_belief_grades,31.4,,abs(Literature coef belief-grades) (pp) — paper uses 'less likely' (footnote §6.1)
pp_male_belief_range_low,2.7,,Min male coef across belief outcomes (footnote §6.1)
pp_male_belief_range_high,18.1,,Max male coef across belief outcomes (footnote §6.1)
pct_actual_peer_schoolwork,61.6,,Actual peer schoolwork use (sec 6.2)
pp_schoolwork_overestimate,4,,"Schoolwork belief minus actual (pp, rounded)"
pct_belief_peer_schoolwork,65.2,,% peers using AI for schoolwork (belief) (sec 6.2)
pct_belief_nopolicy,62.7,,"Belief: peer use, no policy (sec 6.2)"
pct_belief_nopolicy_fn,62.7,,Repeat (footnote)
pct_belief_permissive,72.2,,"Belief: peer use, permissive (sec 6.2)"
pct_belief_prohibit,43.5,,"Belief: peer use, prohibit (sec 6.2)"
pct_belief_prohibit_cmp,43.5,,Repeat (sec 6.2)
pct_belief_prohibit_fn,43.5,,Repeat (footnote)
pp_belief_ban_effect,19.2,,Belief-based ban effect (pp) (footnote §6.2)
pct_self_prohibit,13.1,,Self: likely use if prohibited (sec 6.2)
pct_self_prohibit_fn,13.1,,Repeat (footnote)
pct_self_nopolicy,42.6,,Self: likely use if no policy (footnote)
pp_self_ban_effect,29.5,,Self-reported ban effect (pp) (footnote §6.2)
pp_nopolicy_overestimate,20,,"Belief minus self, no policy (pp, rounded) (sec 6.2)"
pp_prohibit_overestimate,30,,"Belief minus self, prohibit (pp, rounded) (sec 6.2)"
pct_actual_range_low,53,,"Actual peer-use range, low (sec 6.2)"
pct_actual_range_high,81,,"Actual peer-use range, high (sec 6.2)"
pct_male_peer_actual,74.9,,Actual peer-use among male (at least occasionally) (sec 6.2)
pct_female_peer_actual,52.8,,Actual peer-use among female (at least occasionally) (sec 6.2)
pct_white_peer_actual,56.3,,Actual peer-use among white (at least occasionally) (sec 6.2)
pct_black_peer_actual,81.0,,Actual peer-use among black (at least occasionally) (sec 6.2)
pct_latino_peer_actual,61.2,,Actual peer-use among latino (at least occasionally) (sec 6.2)
pct_belief_cluster_low,64,,"Belief range, low (sec 6.2)"
pct_belief_cluster_high,69,,"Belief range, high (sec 6.2)"
pct_male_peer_belief,66.6,,Believed peer-use among male (sec 6.2)
pct_female_peer_belief,65.1,,Believed peer-use among female (sec 6.2)
pct_white_peer_belief,64.0,,Believed peer-use among white (sec 6.2)
pct_black_peer_belief,69.1,,Believed peer-use among black (sec 6.2)
pct_latino_peer_belief,65.4,,Believed peer-use among latino (sec 6.2)
pp_male_gap,8.3,,|Actual-Belief| gap among male (sec 6.2)
pp_female_gap,12.2,,|Actual-Belief| gap among female (sec 6.2)
pp_white_gap,7.7,,|Actual-Belief| among white (sec 6.2)
pp_black_gap,11.9,,|Actual-Belief| gap among black (sec 6.2)
pp_latino_gap,4.1,,|Actual-Belief| gap among latino (sec 6.2)
pp_hispanic_gap,4.1,,|Actual-Belief| among Hispanic (sec 6.2)
pp_male_proofread,4.2,,Male coef task=proofread (pp) (App B)
pp_black_proofread,14.8,,Black coef task=proofread (pp) (App B)
pp_latino_proofread,7.2,,Latino coef task=proofread (pp) (App B)
pp_asian_proofread,8.0,,Asian coef task=proofread (pp) (App B)
pp_socsci_proofread,8.7,,Social-Sci coef task=proofread (pp) (App B)
pp_socsci_proofread_abs,8.7,,abs(Social-Sci coef) task=proofread (pp) (App B)
pp_public_proofread,9.0,,abs(Public-HS coef) task=proofread (pp) (App B)
pp_arts_proofread,18.8,,abs(Arts coef) task=proofread (pp) (App B)
pp_hum_proofread,4.4,,abs(Humanities coef) task=proofread (pp) (App B)
pp_lang_proofread,7.2,,abs(Languages coef) task=proofread (pp) (App B)
pp_lit_proofread,2.6,,abs(Literature coef) task=proofread (pp) (App B)
pp_male_ideas,4.9,,Male coef task=ideas (pp) (App B)
pp_black_ideas,-11.3,,Black coef task=ideas (pp) (App B)
pp_latino_ideas,21.4,,Latino coef task=ideas (pp) (App B)
pp_asian_ideas,3.8,,Asian coef task=ideas (pp) (App B)
pp_socsci_ideas,4.8,,Social-Sci coef task=ideas (pp) (App B)
pp_socsci_ideas_abs,4.8,,abs(Social-Sci coef) task=ideas (pp) (App B)
pp_public_ideas,0.1,,abs(Public-HS coef) task=ideas (pp) (App B)
pp_arts_ideas,42.4,,abs(Arts coef) task=ideas (pp) (App B)
pp_hum_ideas,33.3,,abs(Humanities coef) task=ideas (pp) (App B)
pp_lang_ideas,42.5,,abs(Languages coef) task=ideas (pp) (App B)
pp_lit_ideas,2.9,,abs(Literature coef) task=ideas (pp) (App B)
pp_male_essays,4.4,,Male coef task=essays (pp) (App B)
pp_black_essays,9.3,,Black coef task=essays (pp) (App B)
pp_latino_essays,-1.4,,Latino coef task=essays (pp) (App B)
pp_asian_essays,0.6,,Asian coef task=essays (pp) (App B)
pp_socsci_essays,11.0,,Social-Sci coef task=essays (pp) (App B)
pp_socsci_essays_abs,11.0,,abs(Social-Sci coef) task=essays (pp) (App B)
pp_public_essays,0.0,,abs(Public-HS coef) task=essays (pp) (App B)
pp_arts_essays,4.8,,abs(Arts coef) task=essays (pp) (App B)
pp_hum_essays,6.0,,abs(Humanities coef) task=essays (pp) (App B)
pp_lang_essays,5.5,,abs(Languages coef) task=essays (pp) (App B)
pp_lit_essays,13.3,,abs(Literature coef) task=essays (pp) (App B)
pp_male_editing,9.7,,Male coef task=editing (pp) (App B)
pp_black_editing,30.2,,Black coef task=editing (pp) (App B)
pp_latino_editing,7.9,,Latino coef task=editing (pp) (App B)
pp_asian_editing,6.1,,Asian coef task=editing (pp) (App B)
pp_socsci_editing,5.7,,Social-Sci coef task=editing (pp) (App B)
pp_socsci_editing_abs,5.7,,abs(Social-Sci coef) task=editing (pp) (App B)
pp_public_editing,2.5,,abs(Public-HS coef) task=editing (pp) (App B)
pp_arts_editing,37.0,,abs(Arts coef) task=editing (pp) (App B)
pp_hum_editing,2.5,,abs(Humanities coef) task=editing (pp) (App B)
pp_lang_editing,4.3,,abs(Languages coef) task=editing (pp) (App B)
pp_lit_editing,19.1,,abs(Literature coef) task=editing (pp) (App B)
pp_male_coding,9.7,,Male coef task=coding (pp) (App B)
pp_black_coding,-9.1,,Black coef task=coding (pp) (App B)
pp_latino_coding,-15.6,,Latino coef task=coding (pp) (App B)
pp_asian_coding,5.4,,Asian coef task=coding (pp) (App B)
pp_socsci_coding,-17.0,,Social-Sci coef task=coding (pp) (App B)
pp_socsci_coding_abs,17.0,,abs(Social-Sci coef) task=coding (pp) (App B)
pp_public_coding,4.8,,abs(Public-HS coef) task=coding (pp) (App B)
pp_arts_coding,59.2,,abs(Arts coef) task=coding (pp) (App B)
pp_hum_coding,31.8,,abs(Humanities coef) task=coding (pp) (App B)
pp_lang_coding,35.6,,abs(Languages coef) task=coding (pp) (App B)
pp_lit_coding,31.9,,abs(Literature coef) task=coding (pp) (App B)
pp_male_images,16.3,,Male coef task=images (pp) (App B)
pp_black_images,-11.7,,Black coef task=images (pp) (App B)
pp_latino_images,12.2,,Latino coef task=images (pp) (App B)
pp_asian_images,-8.1,,Asian coef task=images (pp) (App B)
pp_socsci_images,-3.5,,Social-Sci coef task=images (pp) (App B)
pp_socsci_images_abs,3.5,,abs(Social-Sci coef) task=images (pp) (App B)
pp_public_images,2.1,,abs(Public-HS coef) task=images (pp) (App B)
pp_arts_images,17.6,,abs(Arts coef) task=images (pp) (App B)
pp_hum_images,6.9,,abs(Humanities coef) task=images (pp) (App B)
pp_lang_images,23.9,,abs(Languages coef) task=images (pp) (App B)
pp_lit_images,1.7,,abs(Literature coef) task=images (pp) (App B)
pp_male_explain,7.6,,Male coef task=explain (pp) (App B)
pp_black_explain,11.4,,Black coef task=explain (pp) (App B)
pp_latino_explain,15.8,,Latino coef task=explain (pp) (App B)
pp_asian_explain,11.2,,Asian coef task=explain (pp) (App B)
pp_socsci_explain,0.8,,Social-Sci coef task=explain (pp) (App B)
pp_socsci_explain_abs,0.8,,abs(Social-Sci coef) task=explain (pp) (App B)
pp_public_explain,8.9,,abs(Public-HS coef) task=explain (pp) (App B)
pp_arts_explain,12.2,,abs(Arts coef) task=explain (pp) (App B)
pp_hum_explain,12.0,,abs(Humanities coef) task=explain (pp) (App B)
pp_lang_explain,37.7,,abs(Languages coef) task=explain (pp) (App B)
pp_lit_explain,15.9,,abs(Literature coef) task=explain (pp) (App B)
pp_male_emails,4.1,,Male coef task=emails (pp) (App B)
pp_black_emails,22.2,,Black coef task=emails (pp) (App B)
pp_latino_emails,19.8,,Latino coef task=emails (pp) (App B)
pp_asian_emails,24.7,,Asian coef task=emails (pp) (App B)
pp_socsci_emails,8.9,,Social-Sci coef task=emails (pp) (App B)
pp_socsci_emails_abs,8.9,,abs(Social-Sci coef) task=emails (pp) (App B)
pp_public_emails,7.4,,abs(Public-HS coef) task=emails (pp) (App B)
pp_arts_emails,21.9,,abs(Arts coef) task=emails (pp) (App B)
pp_hum_emails,12.0,,abs(Humanities coef) task=emails (pp) (App B)
pp_lang_emails,3.6,,abs(Languages coef) task=emails (pp) (App B)
pp_lit_emails,6.0,,abs(Literature coef) task=emails (pp) (App B)
pp_male_summarize,12.0,,Male coef task=summarize (pp) (App B)
pp_black_summarize,4.2,,Black coef task=summarize (pp) (App B)
pp_latino_summarize,6.4,,Latino coef task=summarize (pp) (App B)
pp_asian_summarize,3.4,,Asian coef task=summarize (pp) (App B)
pp_socsci_summarize,9.9,,Social-Sci coef task=summarize (pp) (App B)
pp_socsci_summarize_abs,9.9,,abs(Social-Sci coef) task=summarize (pp) (App B)
pp_public_summarize,5.7,,abs(Public-HS coef) task=summarize (pp) (App B)
pp_arts_summarize,4.3,,abs(Arts coef) task=summarize (pp) (App B)
pp_hum_summarize,14.3,,abs(Humanities coef) task=summarize (pp) (App B)
pp_lang_summarize,49.7,,abs(Languages coef) task=summarize (pp) (App B)
pp_lit_summarize,2.9,,abs(Literature coef) task=summarize (pp) (App B)
pp_male_findinfo,16.2,,Male coef task=findinfo (pp) (App B)
pp_black_findinfo,26.5,,Black coef task=findinfo (pp) (App B)
pp_latino_findinfo,8.4,,Latino coef task=findinfo (pp) (App B)
pp_asian_findinfo,1.4,,Asian coef task=findinfo (pp) (App B)
pp_socsci_findinfo,1.7,,Social-Sci coef task=findinfo (pp) (App B)
pp_socsci_findinfo_abs,1.7,,abs(Social-Sci coef) task=findinfo (pp) (App B)
pp_public_findinfo,0.7,,abs(Public-HS coef) task=findinfo (pp) (App B)
pp_arts_findinfo,40.9,,abs(Arts coef) task=findinfo (pp) (App B)
pp_hum_findinfo,13.8,,abs(Humanities coef) task=findinfo (pp) (App B)
pp_lang_findinfo,1.0,,abs(Languages coef) task=findinfo (pp) (App B)
pp_lit_findinfo,16.5,,abs(Literature coef) task=findinfo (pp) (App B)
n_openend_motiv,309,,N motivation open-end responses (App D)
pct_openend_response_motiv,48.7,,% responded to motivation open-end (App D)
n_openend_policy,194,,N policy open-end responses (App E)
beta_vader_acad,0.169,,VADER academic-use coefficient (App D)
pct_vader_neg_ever_low,83,,"VADER negative-sentiment, ever use, low end (App D)"
pct_vader_neg_ever_high,92,,"VADER negative-sentiment, ever use, high end (App D)"
pct_vader_neg_acad_low,69,,"VADER negative-sentiment, academic use, low (App D)"
pct_vader_neg_acad_high,85,,"VADER negative-sentiment, academic use, high (App D)"
pct_vader_pos_ever,100,,"VADER positive-sentiment, ever use, ceiling (App D)"
pct_vader_pos_acad_low,87,,"VADER positive-sentiment, academic use, low (App D)"
pct_vader_pos_acad_high,100,,"VADER positive-sentiment, academic use, high (App D)"
\end{filecontents*}


	\title{Generative AI in Higher Education: \\ Evidence from an Elite College\thanks{\noindent Contractor: Middlebury College (\href{mailto:zcontractor@middlebury.edu}{zcontractor@middlebury.edu}). Reyes (corresponding author): Middlebury College and IZA (\href{mailto:greyes@middlebury.edu}{greyes@middlebury.edu}). We thank Jeff Carpenter, Amy Collier, Christa Deneault, Michael Linderman, Peter Matthews, Caitlin Myers, Andrea Robbett, and participants at the Cornell Behavioral Economics Group and LACBEE for helpful discussions and comments. Coco Kitai and Gavin Randolph provided excellent research assistance. Financial support from Middlebury College's Office of the Provost is gratefully acknowledged. This study was approved by the Middlebury College Institutional Review Board.}}

	\author{Zara Contractor \and Germán Reyes} 
	
	\renewcommand{\today}{\ifcase \month \or January\or February\or March\or %
		April\or May \or June\or July\or August\or September\or October\or November\or %
		December\fi \ \number \year}

	\maketitle
	
	\begin{abstract}
		\begin{singlespace}
			\noindent Generative AI is transforming higher education, yet systematic evidence on student adoption, usage patterns, and perceived learning impacts remains scarce. Using survey data from a selective U.S.\ college, we document rapid generative-AI adoption, reaching over \getval{pct_adoption_abstract} percent within two years of ChatGPT's release. Adoption varies sharply across disciplines, demographics, and achievement levels. Students use AI both to augment their learning---by obtaining explanations and feedback---and to automate coursework by generating final outputs, with augmentation more common than automation. Students generally perceive AI as benefiting their learning, and these beliefs are strongly correlated with adoption. Institutional policies shape usage but have uneven effects, in part because awareness and compliance vary across student groups. These findings suggest that effective AI policies must distinguish between uses that enhance learning and those that substitute for it.
		\end{singlespace}		
	\end{abstract}

	
	
	\clearpage

	\section{Introduction} \label{sec:intro}

	ChatGPT and similar generative artificial intelligence (AI) tools can perform tasks central to academic assessment and learning---write essays, solve problems, and explain concepts---instantly and at near-zero marginal cost. Yet systematic evidence on how this new technology is reshaping student learning remains scarce. How widespread is AI adoption among students, and what factors drive it? Do students primarily use AI to augment their learning or to automate coursework, potentially harming human capital development? Could disparities in access to paid AI resources amplify existing educational inequalities? These questions are central to ongoing debates about AI in education, yet the evidence base remains thin.
	
	In this paper, we examine generative AI adoption, usage patterns, beliefs, and policy responses among students at Middlebury College, a selective liberal arts college in Vermont with about \getval{n_midd_students_approx} undergraduates across \getval{n_majors} majors. Our survey ran from \getval{survey_start} to \getval{survey_end} and yielded \getval{n_respondents} responses (a \getval{pct_response_rate} percent response rate). To address sample selection, we construct poststratification weights from the major distribution in administrative records. Main findings are robust to alternative weighting schemes and to unweighted estimates.
	
	We present five main descriptive findings. First, AI adoption is rapid and approaching universality. Over \getval{pct_adoption_intro} percent of students use AI for academic purposes, up from less than 10 percent before Spring 2023. This rate far exceeds the \getval{pct_adoption_us_workers} percent among U.S.\ workers \citep{bick.etal2026} and the \getval{pct_pew_adults_2025} percent among all U.S.\ adults \citep{mcclain2025}. Our rate is in line with the upper end of estimates from other universities \citep{nam2023ai, hirabayashi2024, stohr.etal2024, flaherty2025, freeman2025, ravselj.etal2025}.

	Second, AI adoption is unequal across academic disciplines and demographic groups. Field of study is the strongest predictor of adoption, likely reflecting AI's usefulness for each field's tasks. Adoption ranges from over 90 percent in Natural Sciences majors (including mathematics and computer science) to less than 50 percent in Literature. Adoption also varies across demographic groups. Males adopt AI at higher rates than females (\getval{pct_adoption_male} versus \getval{pct_adoption_female} percent), consistent with gender gaps in AI adoption documented elsewhere \citep{otis2024global}. Lower-achieving students also adopt AI at higher rates than their higher-achieving peers (\getval{pct_adoption_lowgpa} versus \getval{pct_adoption_highgpa} percent); thus, AI could narrow achievement gaps if it enhances learning, or widen them if it undermines skill development.
	
	Third, AI is reshaping students' learning production function through two channels: augmentation, in which AI works \textit{with} the student, and automation, in which AI does the work \textit{for} the student. To classify each task into one of these two categories, we conducted a supplementary Prolific survey in which undergraduates and college instructors classified each task as one where AI primarily ``works with'' the student or ``does the work for'' them. We then apply these classifications to our main sample to measure how much students rely on each. Among AI users, the average usage rate across augmentation tasks is \getval{pct_users_augmentation_intro} percent, compared to \getval{pct_users_automation_intro} percent across automation tasks. Qualitative evidence reinforces these patterns: students describe AI as an ``on-demand tutor'' for augmentation, particularly when traditional resources like office hours are unavailable, and turn to automation to save time during periods of high workload. These patterns align with actual usage data from Claude conversation logs \citep{handa2025education} and ChatGPT interaction logs at another U.S.\ university \citep{ammari2025students}.
	
	Fourth, institutional policies can influence AI adoption, though information gaps undermine their effectiveness. Explicit prohibitions reduce self-reported intended use by \getval{pp_prohibition_effect_intro} percentage points (pp). The size of this effect and intended usage under prohibition vary by student characteristics like gender, race, and field of study, suggesting that uniform policies may have uneven effects. Information gaps compound these challenges: \getval{pct_policy_unclear} percent of students misunderstand their institution's AI rules, only \getval{pct_copilot_aware} percent know about the college's premium AI resources, and just \getval{pct_cite_aware} percent know how to cite AI properly.
	
	Fifth, students hold positive beliefs about AI's learning impact, and these beliefs are strongly correlated with adoption. Most students believe that AI improves their understanding of course materials (\getval{pct_belief_improves_understand} percent) and learning ability (\getval{pct_belief_improves_learn} percent), though fewer believe it improves grades (\getval{pct_belief_improves_grades} percent). Students who believe AI improves academic outcomes are more likely to adopt: a \getval{pp_belief_change_unit} increase in the share of students who believe AI improves learning is associated with a \getval{pp_slope_belief_learn} pp increase in adoption.
	
	Our analysis is descriptive, documenting stylized facts to inform future causal work. Yet several patterns carry direct policy relevance. The information gaps we document---in permitted uses and available resources---suggest straightforward opportunities for intervention. Heterogeneity across disciplines and student groups, however, means uniform policies risk unintended consequences. Blanket prohibitions risk disproportionately harming students who would benefit most from AI as a learning aid---if perceived benefits translate into actual learning gains---while unrestricted use may encourage automation that hinders skill development.

	Although our findings are specific to a liberal arts college, the setting offers insights into AI's broader societal impact. Rapid AI adoption in higher education may reflect a generalizable principle: adoption is fastest when AI consolidates fragmented tools into a unified platform. Students have long had access to various tools for the tasks AI now handles.\footnote{Examples include Chegg and Course Hero for homework help and essay writing; Grammarly for proofreading and grammar checking; Khan Academy, Coursera, and YouTube tutorials for concept explanations; faculty office hours for personalized instruction; SparkNotes and CliffsNotes for text summaries; and Stack Overflow for coding assistance.} AI delivers all these services through a single interface at near-zero marginal cost. This consolidation may partly explain students' positive perception of AI's learning benefits. If the interpretation holds, the principle extends beyond education: industries with fragmented, specialized tools could adopt AI just as quickly, even if AI does not introduce new capabilities.
	
	Our findings contribute to a rapidly growing literature on the adoption and impacts of generative AI. Recent work has examined AI's effects on worker productivity \citep{dellacqua.etal2023, noy.zhang2023, peng.etal2023, cui.etal2024, brynjolfsson.etal2025, cruces.etal2026} and its potential to transform occupations \citep{felten.etal2021, felten.etal2023, eloundou.etal2024}. Several papers document AI adoption in workplace settings \citep{humlum.vestergaard2025, bick.etal2026, hartley.etal2026} and at the firm level \citep{mcelheran.etal2023, bonney.etal2024, kharazian2025rampai}. We examine adoption in higher education---a setting in which future high-skilled workers develop human capital.

	We add to an emerging and concurrent literature on generative AI in education. Some studies examine AI's learning impacts and find mixed results across settings \citep[e.g.,][]{bastani.etal2025, contractor.reyes2025learning, desimone.tiberti2025, kestin.etal2025, kim.etal2025, lehmann.etal2025, lira.etal2025}. A complementary strand documents adoption across diverse educational contexts, including Australian \citep{kelly2023generative}, Ghanaian \citep{bonsu2023consumers}, Norwegian \citep{carvajal2024}, Swedish \citep{stohr.etal2024}, U.S.\ \citep{hirabayashi2024, arum.etal2025, flaherty2025}, U.K.\ \citep{freeman2025}, and multi-country \citep{ravselj.etal2025} universities. A few studies analyze AI interaction logs rather than self-reports \citep{ammari2025students, chatterji.etal2025, handa2025education, handa2025economic}.
	
	Our paper contributes in three ways. First, unlike most studies that rely on convenience samples with unclear selection, we survey a well-defined population and address representativeness through poststratification weighting. Second, we examine adoption, usage, beliefs, and policy responses together within the same sample, whereas existing work typically focuses on one dimension at a time. Third, we distinguish between AI uses that \textit{augment} student effort and those that \textit{automate} academic tasks \citep{brynjolfsson2017machine, acemoglu2019automation}---a framework that, to our knowledge, has not been systematically applied in higher education.
	
	Finally, our findings speak to the literature on technology diffusion. Unlike the S-shaped adoption curves documented for most technologies \citep{griliches1957hybrid, rogers1962diffusion}, we observe \getval{pct_adoption_2yr_mark} percent AI adoption within two years of ChatGPT's release. This pattern contrasts with historical general purpose technologies like electricity, which required decades to reach widespread use \citep{david1990dynamo, bresnahan1995general}. It aligns, though, with evidence that newer technologies diffuse faster than older ones \citep{comin2010international, comin2010technology}. Two factors may explain this speed. First, unlike electricity or steam power, generative AI requires minimal physical infrastructure---students access it through existing devices---and is available at no cost, eliminating the financial barriers that typically slow adoption. Second, AI consolidates multiple specialized tools into a single platform, making its benefits immediately apparent without specialized training or organizational restructuring \citep{brynjolfsson2000beyond}.

	\section{Data: Student Survey} \label{sec:survey}	
	
	\subsection{Recruitment and Structure}
	
	We conducted the survey from \getval{survey_start} to \getval{survey_end}. We contacted all students by email and sent two follow-up reminders. To encourage participation, we entered respondents into a lottery for Amazon gift cards (\$\getval{incentive_low_usd}--\$\getval{incentive_high_usd}).
	
	The survey has three main sections (see Appendix~Figure~\ref{fig:flow} for the flow and Appendix~\ref{app:survey} for the full survey instrument). First, we gather demographic and academic information: gender, race/ethnicity, high school type (private or public), current academic year, and declared or intended major.\footnote{We asked students to report their primary major; some listed two in the open-text box. In those cases, we keep the first. Results are similar if we include each major-student pair in the dataset.} We also collect self-reported weekly study hours and first-year GPA.\footnote{We asked about first-year GPA rather than cumulative GPA to measure academic performance at a comparable point in students' college careers, unaffected by grade trajectories across class years.}
	
	Second, we measure students' experience with generative AI tools. We ask whether students have ever used tools like ChatGPT or Claude and, for users, collect usage frequency, specific models, payment for AI services, and how they use AI across academic tasks (e.g., writing assistance, learning support, coding).
	
	The final part elicits students' beliefs about generative AI's academic impacts and diffusion. We ask students to evaluate AI's effects on their academic experience across four dimensions---learning, grades, time management, and understanding of course material---and how different policy environments influence their likelihood of using AI. We also elicit beliefs about peer AI use, including estimates of the share of Middlebury students who use AI for schoolwork and leisure. The survey concludes with two open-ended questions inviting students to share their experiences with AI and to comment on institutional policies and support services.
	
	\subsection{Sample and Summary Statistics}
	
	Of Middlebury's \getval{n_enrolled_students} enrolled students, \getval{n_survey_started} began the survey (a \getval{pct_response_rate} percent response rate). This response rate is in line with the U.S.\ institutional average \citep{nsse2020response} and exceeds those at institutions like Harvard \citep{hirabayashi2024}. We exclude \getval{n_excluded} respondents who left before reaching the generative AI usage module, leaving an analysis sample of \getval{n_analysis_sample} students. To make the sample more representative, we construct poststratification weights from the distribution of declared majors in administrative records.\footnote{We target representativeness at the field-of-study level given substantial evidence that AI adoption varies systematically across disciplines and occupations, with usage patterns tied to field-specific tasks \citep{stohr.etal2024, humlum.vestergaard2025, ravselj.etal2025, bick.etal2026}.} We weight observations by the ratio of each major's share in the student population to its share in our responses.\footnote{We normalize weights to sum to total enrollment (\getval{n_enrolled_students}) rather than total declared majors, which differs because some students have multiple majors.}
	
	Table~\ref{tab:summ_stats} presents summary statistics. Column 1 reports unweighted survey averages, column 2 shows poststratification-weighted averages, and column 3 provides administrative-record benchmarks where available. Our unweighted sample is \getval{pct_sample_male_unw} percent male and \getval{pct_sample_female_unw} percent female. The racial/ethnic composition is \getval{pct_sample_white_unw} percent white, \getval{pct_sample_asian_unw} percent Asian, \getval{pct_sample_hispanic_unw} percent Hispanic, and \getval{pct_sample_black_unw} percent Black. Most students (\getval{pct_sample_public_hs} percent) attended public school; \getval{pct_sample_private_hs} percent attended private school. The sample spans \getval{n_majors_represented} majors across \getval{n_fields} fields, with \getval{pct_undeclared} percent of students not yet having declared a major. We group undeclared students by their intended field of study as reported in the survey.
	
	Comparing the unweighted sample to administrative records reveals differences. The sample overrepresents white (\getval{pct_white_sample_cmp} versus \getval{pct_white_admin} percent) and Asian (\getval{pct_asian_sample_cmp} versus \getval{pct_asian_admin} percent) students, and underrepresents Black (\getval{pct_black_sample_cmp} versus \getval{pct_black_admin} percent) and Hispanic (\getval{pct_hispanic_sample_cmp} versus \getval{pct_hispanic_admin} percent) students. First-year students are overrepresented (\getval{pct_freshman_sample} versus \getval{pct_freshman_admin} percent) and seniors underrepresented (\getval{pct_senior_sample} versus \getval{pct_senior_admin} percent). Our weighting partially addresses these differences---weighted figures closely match administrative records for academic fields---but some demographic gaps persist. Appendix~\ref{app:weights} shows that our main results hold under alternative weighting schemes (by gender and cohort, or by race) and under unweighted estimates.
	
	\section{Generative AI Usage Patterns Among Students}  \label{sec:usage}
	
	\subsection{Adoption of Generative AI}
	
	Generative AI is widely adopted at Middlebury. Figure~\ref{fig:ai_use} shows the distribution of AI usage frequency during the academic semester across four levels: ``Rarely'' (a few times per semester), ``Occasionally'' (a few times per month), ``Frequently'' (a few times per week), and ``Very Frequently'' (daily or almost daily). Overall, \getval{pct_adoption_any} percent of students report using generative AI for academic purposes, with substantial variation in intensity: \getval{pct_freq_rarely} percent use it rarely, \getval{pct_freq_occasionally} percent occasionally, \getval{pct_freq_frequently} percent frequently, and \getval{pct_freq_veryfrequently} percent very frequently.
	
	An independent survey at Middlebury corroborates our adoption estimate. The student-led Zeitgeist \getval{zeitgeist_version} survey, conducted by \textit{The Middlebury Campus} in \getval{zeitgeist_period} with \getval{n_zeitgeist} respondents (\getval{pct_zeitgeist_rate} percent of the student body), asked ``How often do you use A.I.\ in your classes at Middlebury?''. \getval{pct_zeitgeist_adoption} percent reported some level of AI use, compared with \getval{pct_adoption_any_cmp} percent in our sample \citep{zeitgeist2025}. The close correspondence across surveys with different response rates and sampling periods suggests that selection bias is unlikely to substantially inflate our adoption rates.

	Middlebury's adoption rate aligns with patterns at other institutions. As of spring 2024, \getval{pct_harvard_adoption} percent of Harvard undergraduates used AI \citep{hirabayashi2024}, and a 2025 \textit{Inside Higher Ed} survey found that \getval{pct_ihe_adoption} percent of U.S.\ college students had used AI for coursework in the past year \citep{flaherty2025}---both close to our \getval{pct_adoption_any_cmp2} percent. In the U.K., \getval{pct_uk_adoption} percent of undergraduates had used AI for assessments \citep{freeman2025}. Earlier surveys show correspondingly lower rates. A fall 2023 \textit{BestColleges} survey found that \getval{pct_bestcolleges_adoption} percent of U.S.\ students had used AI on assignments or exams \citep{nam2023ai}, against about \getval{pct_midd_fall2023_adoption} percent at Middlebury by the same period. International data are similar: \getval{pct_ravselj_adoption} percent of higher-education students globally had ever used ChatGPT by late 2023--early 2024 \citep{ravselj.etal2025}; \getval{pct_sweden_adoption} percent of Swedish students had used ChatGPT by spring 2023 \citep{stohr.etal2024}; and \getval{pct_norway_adoption} percent of Norwegian students used AI at least occasionally \citep{carvajal2024}. The convergence across these samples suggests that near-universal AI adoption among university students is not specific to our context.

	These adoption rates far exceed those in the general population and workforce. Pew Research finds that \getval{pct_pew_adults_2025} percent of U.S.\ adults have ever used ChatGPT as of early 2025 \citep{mcclain2025}, and Gallup reports that \getval{pct_gallup_workers} percent of U.S.\ employees used AI at work at least a few times a year by mid-2025, with daily use at only \getval{pct_gallup_daily} percent \citep{gallup2025}. ChatGPT consumer logs show the platform had reached about \getval{pct_chatgpt_world} percent of the world's adult population by mid-2025 \citep{chatterji.etal2025}. \citet{bick.etal2026} and \citet{hartley.etal2026} estimate that \getval{pct_bick_low}--\getval{pct_bick_high} percent of the U.S.\ working-age population used generative AI for work in late 2024 and 2025. Both document substantial industry heterogeneity, with information services and technology at the top at \getval{pct_bick_tech_low}--\getval{pct_bick_tech_high} percent---still well below \getval{pct_adoption_round} percent at Middlebury. Even among workers in AI-exposed occupations, \citet{humlum.vestergaard2025} find adoption rates of only \getval{pct_humlum_adoption} percent.\footnote{The higher adoption rates in higher education may partly reflect demographic composition. Younger and more educated individuals consistently show greater AI adoption: \citet{mcclain2025} find that \getval{pct_pew_under30} percent of adults under 30 have used ChatGPT compared to \getval{pct_pew_overall} percent overall. \citet{bick.etal2026} document that workers aged \getval{age_range_young} are twice as likely to use AI at work as those aged \getval{age_range_old}, and college-educated workers are twice as likely to use AI as those without degrees. Similar age and education gradients appear in \citet{humlum.vestergaard2025} and \citet{liu2024generativeai}.}

	\subsection{Adoption by Student Characteristics and Field of Study}
	
	AI adoption varies considerably across demographic groups and academic disciplines (Figure~\ref{fig:ai_use} and Appendix~Table~\ref{tab:ai_freq_het}). Males report higher usage than females (\getval{pct_adoption_male_sec2} versus \getval{pct_adoption_female_sec2} percent). Usage differs by race and ethnicity: Black (\getval{pct_adoption_black} percent) and Asian (\getval{pct_adoption_asian} percent) students have the highest adoption rates, while white (\getval{pct_adoption_white} percent) and Hispanic (\getval{pct_adoption_latino} percent) students report lower usage. Students from private high schools use AI more than those from public schools (\getval{pct_adoption_private_hs} versus \getval{pct_adoption_public_hs} percent). Students with below-median GPAs report higher usage than their higher-achieving peers (\getval{pct_adoption_lowgpa_sec2} versus \getval{pct_adoption_highgpa_sec2} percent). Adoption varies most sharply by field of study: Natural Sciences leads at \getval{pct_adoption_natsci_sec2} percent and Social Sciences follows at \getval{pct_adoption_socsci} percent, while Languages (\getval{pct_adoption_languages_sec2} percent) and Literature (\getval{pct_adoption_literature_sec2} percent) show considerably lower adoption rates.
	
	To examine how student characteristics jointly relate to adoption, we estimate OLS regressions that include them all as covariates.\footnote{We exclude first-year GPA from these regressions because this variable is unavailable for first-year students.} Table~\ref{tab:ai_usage_correlates} reports estimates for four usage thresholds. Each column is the probability of meeting a progressively higher frequency threshold: any AI use (column 1), at least monthly use (column 2), at least weekly use (column 3), and daily use (column 4). 
	
	The regression results confirm these descriptive patterns. Holding other characteristics constant, males are \getval{pp_gender_gap_col1} pp more likely than females to use AI (column 1, $p < 0.05$), with this gender gap widening at higher usage frequencies (columns 2--4, all $p < 0.01$). Black and Asian students adopt AI at higher rates than white students, at \getval{pp_black_col1} and \getval{pp_asian_col1} pp respectively (column 1, both $p < 0.01$). Students from public high schools are \getval{pp_public_hs_col1} pp less likely to use AI than those from private schools, but this difference is not statistically significant.
	Field of study is the strongest predictor of adoption. Compared to Natural Sciences majors, students in Literature, Languages, and Humanities all use AI at lower rates, with differences reaching statistical significance at high usage frequencies (columns 3 and 4). Arts majors also show lower adoption on average, but the differences are not statistically significant. Social Sciences majors exhibit adoption rates similar to Natural Sciences majors across all frequency thresholds.	
	
	These heterogeneity patterns align with evidence from other settings. The gender gap at Middlebury---\getval{pp_gender_gap_cmp} pp higher for males---is consistent with a meta-analysis of \getval{n_otis_studies} studies by \citet{otis2024global}, who find that males are \getval{pp_otis_low}--\getval{pp_otis_high} pp more likely to use generative AI than females. The same gap shows up across educational studies of AI adoption \citep{nam2023ai, carvajal2024, stohr.etal2024, ravselj.etal2025}. Our finding that students with below-median GPAs adopt AI at higher rates aligns with \citet{carvajal2024}, who document higher adoption among students with lower admission grades. Whether greater adoption among lower-achieving students narrows or widens achievement gaps depends on whether AI enhances skill development or erodes it.
	
	Disciplinary differences at Middlebury mirror patterns documented elsewhere. \citet{stohr.etal2024} find that technology and engineering students use ChatGPT more than humanities students. \citet{nam2023ai} report that \getval{pct_nam_business} percent of business majors and \getval{pct_nam_stem} percent of STEM majors have used AI tools for coursework, versus \getval{pct_nam_humanities} percent of humanities majors. \citet{ravselj.etal2025} document comparable differences, with applied sciences students showing higher usage than arts and humanities students.
	
	These disciplinary differences persist into the workforce. \citet{bick.etal2026} find stark variation by college major: STEM graduates have the highest AI adoption rates (\getval{pct_bick_stem} percent), followed by Business/Economics graduates (\getval{pct_bick_business} percent); Liberal Arts graduates show much lower rates (\getval{pct_bick_liberal} percent). \citet{humlum.vestergaard2025} document similar patterns by occupation: roles requiring strong writing and technical skills---such as marketing specialists and journalists---have the highest adoption rates. Consistency across educational and professional contexts suggests that field-specific factors---particularly AI's applicability to different tasks---systematically shape adoption.
	
	\subsection{Timing of Generative AI Adoption}
	
	The speed of technology diffusion is a critical determinant of its economic and social impact \citep{david1990dynamo, hall2003adoption, stokey2021technology}. In educational contexts, rapid adoption can create or exacerbate inequalities between early and late adopters, particularly if the technology confers learning advantages \citep{world2016world}. To track adoption timing, we asked students when they first began using generative AI for academic purposes, with options ranging from ``This semester (\getval{semester_current})'' to ``Before Spring 2023'' (ChatGPT launched publicly in \getval{chatgpt_launch_date}).
	
\begin{figure}[H]
		\caption{The Evolution of Generative AI Adoption among Middlebury College Students} \label{fig:ai_use_time}
		\centering
		\includegraphics[width=.75\linewidth]{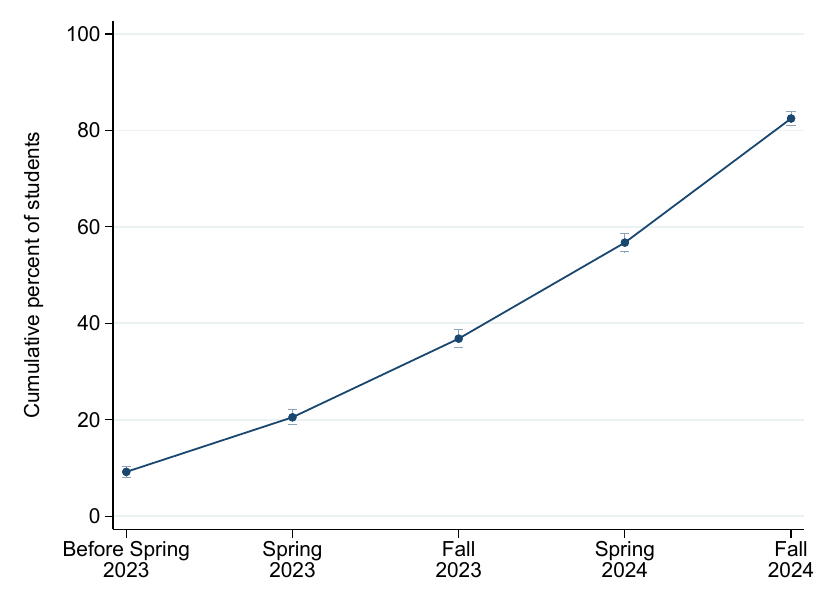}
		
		{\footnotesize
			\singlespacing \justify
			
			\textit{Notes:} This figure shows the cumulative percent of students who reported using generative AI tools for academic purposes over time. The data are based on retrospective self-reports collected in our December 2024 survey, where students were asked ``When did you first start using any form of Generative AI for academic purposes?'' Response options ranged from ``Before Spring 2023'' to ``This semester (Fall 2024).'' The $x$-axis represents academic semesters, while the $y$-axis represents the cumulative adoption rate. Vertical lines represent 95 percent confidence intervals calculated from the standard error of the proportion. \par
		}
		
	\end{figure}

	Students adopted generative AI at an extraordinary pace. Figure~\ref{fig:ai_use_time} shows that cumulative adoption grew from less than 10 percent before Spring 2023 to over \getval{pct_adoption_round2} percent by Fall 2024. Adoption accelerated over time, consistent with improvements in AI capabilities: among current users, \getval{pct_adopted_fall2024} percent adopted in Fall 2024 alone, \getval{pct_adopted_spring2024} percent in Spring 2024, and \getval{pct_adopted_fall2023} percent in Fall 2023.
	
	This pace far exceeds that observed in other populations. \citet{bick.etal2026} show that computers took over \getval{n_years_computers} years to reach \getval{pct_computer_benchmark} percent adoption among U.S.\ working-age adults, and the internet took about \getval{n_years_internet} years. Generative AI adoption in the broader population has also been slower: Pew Research found that just \getval{pct_pew_feb2024} percent of U.S.\ adults had ever used ChatGPT as of \getval{pew_feb_date}, up from \getval{pct_pew_jul2023} percent in \getval{pew_jul_date} \citep{mcclain2024}. Students reached \getval{pct_adoption_round} percent adoption in under two years, suggesting that academic settings accelerate AI diffusion more than other settings.
	
	To identify early versus late adopters, we examine how adoption timing varies across student characteristics (Table~\ref{tab:ai_adopt_correlates} and Appendix~Figure~\ref{fig:ai_cdf_dem}). Male students led adoption, with an \getval{pp_gender_earlyadopt} pp higher probability of using AI before Spring 2023 than females (column 1, $p < 0.01$)---a gap that persists across all periods (columns 2--5). Black and Asian students also adopted earlier than white students, though these differences reach statistical significance only in later periods (columns 4--5). Field of study is a strong predictor of adoption timing: Languages majors consistently lagged behind Natural and Social Sciences majors, with significantly lower adoption rates across nearly all periods.
	
	\subsection{Choice of Generative AI Models} \label{sec:models}
	
	Major AI companies operate on a freemium model: free versions coexist with premium subscriptions offering higher usage limits and more advanced models. This tiered structure creates potential for a new form of educational inequality---if paid versions confer academic advantages, students who cannot afford subscriptions may be systematically disadvantaged. We presented respondents with a list of free and paid versions of popular models and collected monthly subscription expenditures, with options ranging from no active subscription to more than \$\getval{usd_subscription_high} per month (ChatGPT Plus cost \$\getval{usd_chatgpt_monthly} per month at the time of our survey).
	
\begin{figure}[H]
		\caption{Adoption of Generative AI Models Among College Students} \label{fig:ai_models}
		\centering
		\includegraphics[width=.75\linewidth]{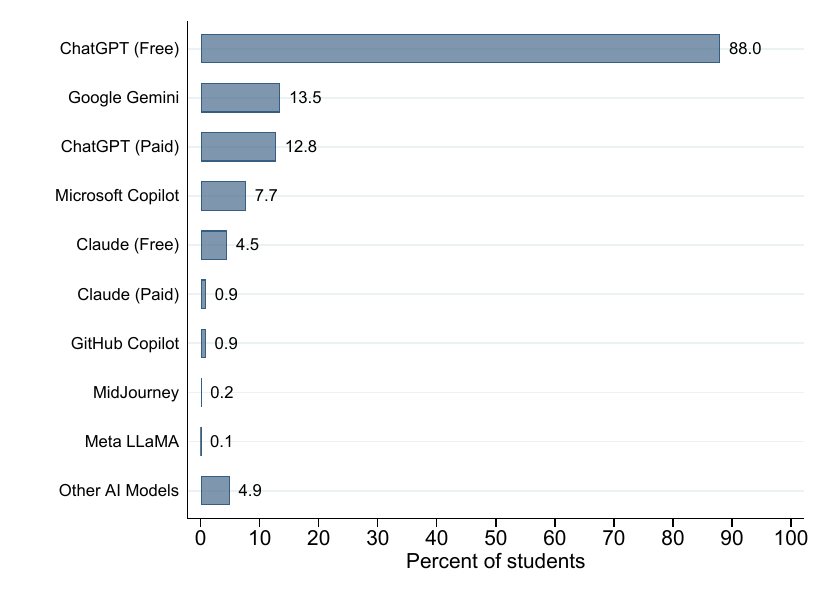}
		
		{\footnotesize
			\singlespacing \justify
			
			\textit{Notes:} This figure shows the adoption rates of various AI models as of Fall 2024. The horizontal axis shows the percent of students who reported using each tool, and the vertical axis lists the tools in descending order of adoption rates. \par
		}
		
	\end{figure}

	OpenAI's ChatGPT dominates AI usage among Middlebury students. Figure~\ref{fig:ai_models} shows that \getval{pct_chatgpt_free} percent of AI users rely on the free version, far more than any alternative. Google Gemini (\getval{pct_gemini} percent) and Microsoft Copilot (\getval{pct_copilot_use} percent) are distant competitors; other platforms each capture less than \getval{pct_other_models_ceiling} percent of users. This dominance mirrors patterns at other selective institutions: \citet{hirabayashi2024} find that over \getval{pct_harvard_chatgpt} percent of Harvard AI users report using ChatGPT. Similar patterns appear in other academic settings \citep{stohr.etal2024} and in the broader workforce \citep{bick.etal2026}.\footnote{ChatGPT's dominance is less pronounced in the general population. \citet{bick.etal2026} find that ChatGPT leads with \getval{pct_bick_chatgpt} percent adoption among U.S.\ adults, followed by Google Gemini at \getval{pct_bick_gemini} percent---a much smaller gap than we observe at Middlebury.}

	Despite near-universal AI adoption, only \getval{pct_pay_ai} percent of AI users pay for any AI service.\footnote{The share of students who report using paid ChatGPT (\getval{pct_pay_chatgpt} percent) is slightly larger than the share who report paying for any AI service (\getval{pct_pay_ai_cmp} percent). The gap may arise because some students access paid versions through shared accounts and do not personally pay.} The rate is similar to the \getval{pct_ravselj_pay} percent in \citet{ravselj.etal2025}'s multi-country survey but lower than the \getval{pct_harvard_pay} percent at Harvard \citep{hirabayashi2024}. The low payment rate suggests that for most students, the premium features---mainly higher usage limits and more advanced models---do not justify the cost. Payment patterns reveal disparities: males and Asian students are more likely to purchase AI subscriptions (Appendix~Table~\ref{tab:ai_model_correlates} and Appendix~Figure~\ref{fig:ai_pays_dem}), possibly reflecting differences in usage intensity.

	\section{Generative AI and the Production of Learning} \label{sec:task}
	
	\subsection{The Use of Generative AI across Academic Tasks}
	
	How is generative AI reshaping the traditional inputs to student learning? The educational production function includes time spent studying \citep[e.g.,][]{stinebrickner2008causal}, faculty instruction \citep[e.g.,][]{fairlie2014community}, peer interactions \citep[e.g.,][]{sacerdote2001}, and academic support services \citep[e.g.,][]{angrist2009incentives}. AI tools can complement or substitute for these and other inputs: asking AI to explain a concept may substitute for faculty office hours, and using it for proofreading may reduce the need for academic support services.

	To understand AI's role in the learning production function, we collected AI usage across ten academic tasks: proofreading, generating ideas, writing essays, editing drafts, coding assistance, creating images, explaining concepts, composing emails, summarizing materials, and finding information. For each task, students reported usage frequency on a \getval{n_scale_points} scale from never to daily. We supplement these data with open-ended responses.

	Students use generative AI across a wide range of academic tasks, with learning support and text processing showing the highest adoption. Figure~\ref{fig:ai_use_purpose}, Panel A shows that explaining concepts is the most common use case (\getval{pct_task_explain} percent of AI users), followed by summarizing texts (\getval{pct_task_summarize} percent), finding information (\getval{pct_task_findinfo} percent), and generating ideas (\getval{pct_task_genideas} percent). Writing assistance tasks---proofreading and editing essays---are also common (\getval{pct_task_proofread} and \getval{pct_task_editessays} percent, respectively). \getval{pct_task_coding} percent of AI users use it for coding help---a sizable share given that many degrees involve no programming. Only \getval{pct_task_writeessays} percent report using AI to write essays, suggesting reluctance to outsource the writing itself. The lowest adoption is for creating images (\getval{pct_task_images} percent), likely reflecting fewer academic use cases for this capability.
	
	An important limitation is that our survey relies on self-reported usage, which may introduce non-classical measurement error \citep{ling2025underreporting}---students might underreport uses they perceive as academically inappropriate. To assess validity, we compare our findings against two studies that analyze AI interaction logs: \citet{handa2025education}, who examine Claude usage among users with university email addresses, and \citet{ammari2025students}, who analyze ChatGPT logs from undergraduates at another U.S.\ university. Neither comparison is perfect: \citeauthor{handa2025education}'s data are conversation-level rather than student-level, and most students in our sample use ChatGPT rather than Claude (Section~\ref{sec:models}); \citeauthor{ammari2025students}'s data rely on students voluntarily sharing their logs. Both still offer useful benchmarks.
	
\begin{figure}[H]
		\caption{The Adoption of Generative AI among Middlebury College Students} \label{fig:ai_use}
		\centering
		\includegraphics[width=.54\linewidth]{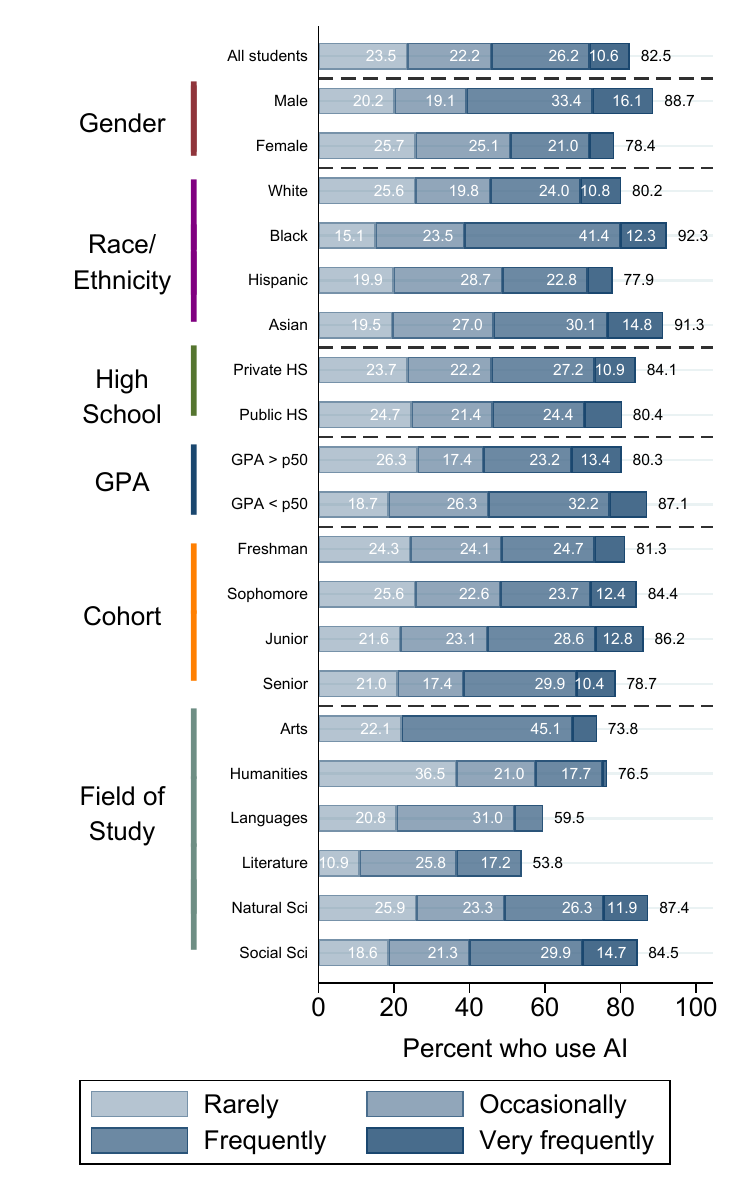}
		
		{\footnotesize
			\singlespacing \justify
			
			\textit{Notes:} This figure shows the fraction of students who report using AI during the academic semester, categorized by demographic characteristics, high school type, academic cohort, GPA, and field of study. Usage frequency is divided into four levels: ``Rarely'' (a few times a semester), ``Occasionally'' (a few times a month), ``Frequently'' (a few times a week), and ``Very Frequently'' (daily or almost daily).
			
			The category ``All students'' provides the baseline usage rate for the full sample. Gender categories are based on self-identification, with non-binary responses excluded due to a small sample size. ``Private HS'' refers to students who attended private high schools, while ``Public HS'' includes public institutions. ``Cohort'' denotes the student's academic year, ranging from first-year (``Freshman'') to fourth-year and beyond (``Senior''). GPA categories (``GPA $>$ p50'' and ``GPA $<$ p50'') split students into groups above or below the median first-year GPA, as self-reported on a 4.0 scale. See Appendix~\ref{app:fields} for the classification of majors into fields of study. \par
		}
	\end{figure}

	Our results are broadly consistent with Anthropic's data on Claude usage. \citet{handa2025education} report that the second-largest use case (\getval{pct_handa_technical} percent of conversations) involves ``technical explanations or solutions for academic assignments,'' while we find that \getval{pct_task_explain_cmp} percent of AI users use it for explaining concepts---a difference likely attributable to our student-level versus their conversation-level measurement. The most common Claude usage category involves ``designing practice questions, editing essays, or summarizing academic material'' (\getval{pct_handa_design} percent of conversations), consistent with our finding that \getval{pct_task_summarize_cmp} percent of AI users summarize texts and \getval{pct_task_editessays_cmp} percent edit essays. Disciplinary patterns also converge: \citet{handa2025education} find that computer science, natural sciences, and mathematics conversations are overrepresented, mirroring our finding that Natural Science majors show the highest AI adoption rates (Figure~\ref{fig:ai_use}).
	
	Further validation comes from \citet{ammari2025students}, who analyze ChatGPT logs from undergraduates. Although they use a different classification scheme---information seeking, content generation, language refinement, meta-cognitive engagement, and conversational repair---their findings align closely with ours. ``Concept explanation'' is the most common information-seeking behavior, consistent with our finding that explaining concepts is the most frequent use case. Information seeking and language refinement dominate over pure content generation, paralleling our results. Convergence across self-reported survey data, Anthropic's conversation logs, and \citeauthor{ammari2025students}'s ChatGPT logs---each using different measurement approaches and classification schemes---suggests that our results capture real patterns of student AI engagement rather than social desirability bias.
	
	\subsection{Automation versus Augmentation}
	
	Are students using generative AI primarily to \textit{augment} their learning or to \textit{automate} their coursework? This distinction matters for understanding AI's impact on human capital. Augmentation may enhance students' learning while maintaining active engagement; automation produces outputs that could be submitted with minimal student input, potentially undermining skill development.
	
\begin{figure}[H]
		\caption{Academic Uses of Generative AI} \label{fig:ai_use_purpose}
		
		\begin{center}
			\begin{subfigure}[t]{.8\textwidth}
				\caption*{Panel A. Across Common Academic Tasks}
				\centering
				\includegraphics[width=\textwidth]{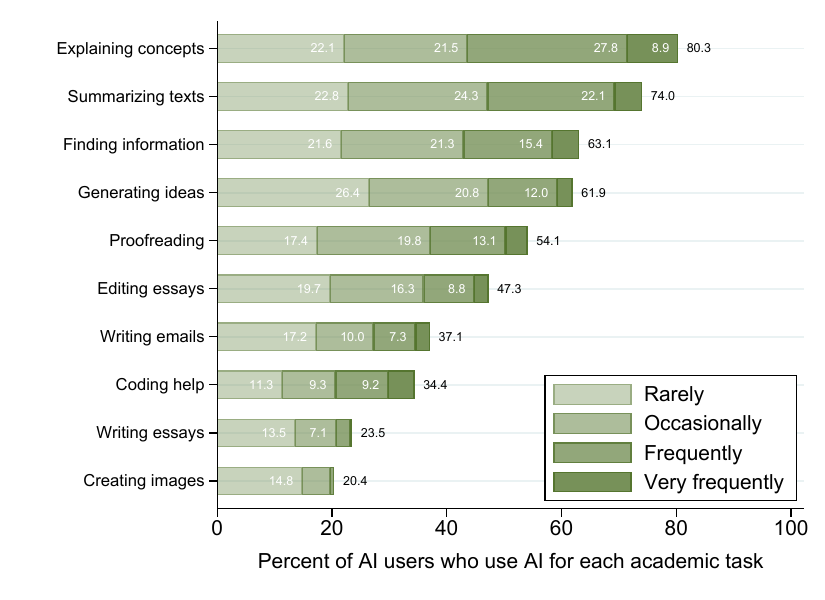}
			\end{subfigure}
			
			\begin{subfigure}[t]{.8\textwidth}
				\caption*{Panel B. Across Tasks that Augment versus Automate Student Effort}
				\centering
				\includegraphics[width=\textwidth]{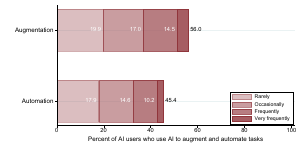}
			\end{subfigure} 
			
		\end{center}
		
		{\footnotesize\singlespacing \justify
			
			\textit{Notes:} This figure shows the percent of AI students who use generative AI for different academic tasks. For each task, usage frequency is divided into four levels: ``Rarely'' (a few times a semester), ``Occasionally'' (a few times a month), ``Frequently'' (a few times a week), and ``Very Frequently'' (daily or almost daily). The number at the end of each bar represents the total percent of students who use AI for that purpose at any frequency. Tasks are ordered by total usage, from highest to lowest. Results are based on responses to the question: ``For academic purposes, which of the following tasks do you typically use generative AI for?'' Sample includes all students who reported using AI during the academic semester. \par
		}
	\end{figure}

	Our main survey records AI use across ten academic tasks but does not classify them as augmentation or automation. To make this distinction, we conducted a supplementary survey of \getval{n_prolific_students} undergraduates and \getval{n_prolific_instructors} college instructors recruited through Prolific (see Appendix~\ref{app:augmentation_survey}). For each task, respondents classified whether AI acts more as a collaborator (works \textit{with} the student) or automates the task (does the work \textit{for} the student). We apply the combined modal classification to categorize tasks. Figure~\ref{fig:ai_use_purpose}, Panel B reports the share of AI users who use each category at various frequencies.

	Students use generative AI for both augmentation and automation, but augmentation is more common. Among AI users, the average usage rate across augmentation tasks is \getval{pct_avg_augmentation} percent, compared to \getval{pct_avg_automation} percent across automation tasks---a \getval{pp_aug_auto_gap} pp gap that suggests students find augmentation more valuable for day-to-day academic work. The pattern is consistent with AI conversation logs: \citet{handa2025economic} analyze workplace conversations with Claude and find that \getval{pct_handa_aug} percent involve augmentation and \getval{pct_handa_auto} percent involve automation.
	
	Qualitative evidence from open-ended responses reinforces the augmentation-automation framework (see Appendix~\ref{app:open_motiv} for additional results). For augmentation, many characterize AI as an ``on-demand tutor,'' especially valuable when traditional resources like office hours are unavailable. Non-native English speakers frequently mention using AI for proofreading to overcome language barriers; students in technical majors describe using it to debug code and understand error messages. Students automate mainly to save time: \getval{pct_openend_efficiency} percent of open-ended responses mention efficiency benefits (Appendix~Figure~\ref{fig:ai_keywords_motiv}). Students describe turning to AI during periods of overwhelming workload or looming deadlines, using it to generate initial drafts or complete routine assignments.\footnote{This pattern is consistent with anecdotal evidence suggesting that students rely on automation under time pressure, particularly in courses outside their major. See, for example: \href{https://www.newyorker.com/magazine/2025/07/07/the-end-of-the-english-paper}{``What Happens After A.I. Destroys College Writing?''} Hua Hsu, \textit{The New Yorker}, July 7, 2025.}
	
	The preference for augmentation over automation extends beyond Middlebury. In Appendix~Figure~\ref{fig:chatgpt_use_ravselj}, we re-analyze data from \citet{ravselj.etal2025}---a multi-country survey of higher education students---to examine how this balance varies by institutional quality.\footnote{While \citet{ravselj.etal2025} elicit a different set of academic tasks than our survey, we categorize them similarly. Augmentation tasks are study assistance, brainstorming, and coding assistance; automation tasks are proofreading, translating, summarizing, calculating, research assistance, personal assistance, and writing (academic, professional, and creative).} We classify universities into quintiles based on their \textit{Times Higher Education} World University Rankings. Worldwide, students who use AI show similar rates for augmentation (\getval{pct_ravselj_aug} percent) and automation (\getval{pct_ravselj_auto} percent) tasks---a much smaller gap than at Middlebury (Panel A). This aggregate pattern masks heterogeneity by institutional quality (Panel B). Top-quintile students show a slight preference for automation over augmentation; this gap closes through the middle quintiles, then reverses sharply at the bottom quintile, where augmentation reaches roughly \getval{pct_ravselj_bottom_aug} percent while automation stays near \getval{pct_ravselj_bottom_auto} percent.
	
	\subsection{Heterogeneity in Augmentation versus Automation Usage}
	
	To assess whether augmentation and automation patterns vary across student populations, we construct four measures: binary indicators for any use of each category, the proportion of tasks used in each category, Likert-scale intensity measures, and the difference between augmentation and automation (capturing relative preference). Table~\ref{tab:ai_autom_regs} presents regression estimates using these measures as outcomes.
	
	The balance between augmentation and automation varies by student characteristics. Males use AI more in both categories: \getval{pp_male_aug} pp more augmentation tasks and \getval{pp_male_auto} pp more automation tasks than females (columns 2 and 5, $p < 0.01$). Black students use AI more intensively for both augmentation (\getval{beta_black_aug} points, $p < 0.10$) and automation (\getval{beta_black_auto} points, $p < 0.05$), with no significant difference in their relative preference (columns 7--8).
	Asian students likewise use AI more in both categories than white students. Differences across fields are sizable: humanities, languages, and literature majors use AI less for both augmentation and automation than natural science majors.

	\section{Institutional Policies and AI Adoption} \label{sec:policy}
	
	\subsection{The Role of Institutional Policies in Shaping Student Behavior}
	
	Institutional policies shape technology adoption and diffusion \citep{acemoglu2025nobel}.\footnote{Middlebury has no college-wide AI policy; individual departments and faculty set their own.} To examine how policies affect behavior, we asked students to report their likelihood of using AI under scenarios ranging from complete prohibition to unrestricted use---a question made salient by the wide variation in institutional policies across colleges \citep{nolan2023chatgpt, xiao2023waiting}.
	
\begin{figure}[H]
		\caption{Student Reported Likelihood of Using AI under Different Policies} \label{fig:ai_policy}
		\centering
		\includegraphics[width=.75\linewidth]{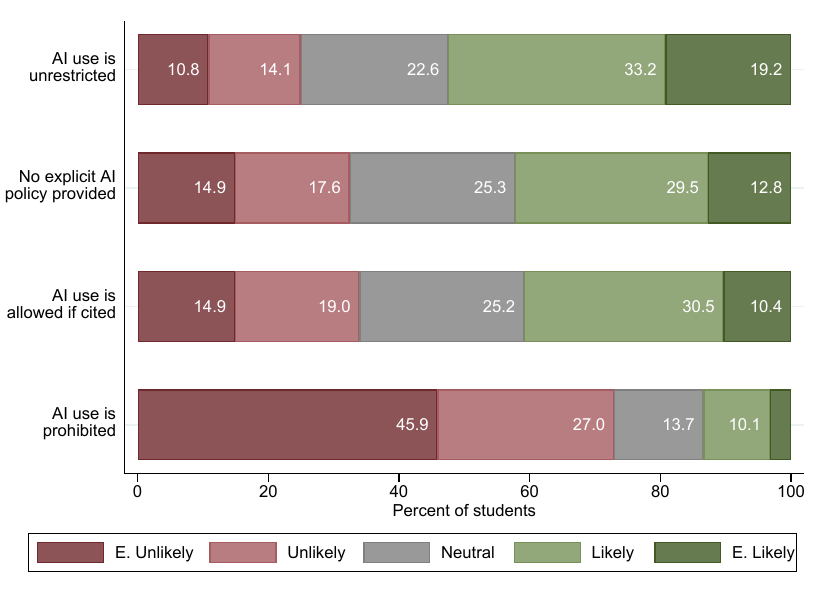}
		{\footnotesize
			\singlespacing \justify
			
			\textit{Notes:} This figure shows the percent of students who report different likelihoods of using AI under various policy scenarios. For each policy, responses are categorized on a five-point scale from ``Extremely unlikely to use AI'' to ``Extremely likely to use AI.'' The sample includes all survey respondents. The question asked was: ``How likely are you to use generative AI in a class with each of the following AI policies?'' \par
		}
	\end{figure}

	Institutional policies shape students' reported likelihood of using AI. Figure~\ref{fig:ai_policy} shows that under unrestricted use, \getval{pct_policy_unrestricted} percent of students report being likely or extremely likely to use AI. This share drops modestly when policies require citation (\getval{pct_policy_cite} percent) or when no explicit policy exists (\getval{pct_policy_noexplicit} percent). Explicit prohibition sharply changes intentions: only \getval{pct_policy_prohibited} percent say they would be likely or extremely likely to use AI when it is banned, while \getval{pct_policy_prohibited_unlikely} percent say they would be unlikely or extremely unlikely to do so (both figures include current users and non-users).\footnote{These policy magnitudes are broadly in line with findings from other contexts. \cite{carvajal2024} estimate that banning AI reduces usage by about \getval{pp_carvajal_overall} pp. In our survey, a ban reduces usage by \getval{pp_ban_overall} pp.} Institutional policies thus shape usage, though some students report they would use AI even when explicitly prohibited.\footnote{An important limitation of our analysis is the sequential nature of the policy questions, which may introduce bias through anchoring or contrast effects. If respondents anchor on their first response or evaluate subsequent scenarios relative to previous ones, the absolute magnitudes of our estimates may be biased. However, if these biases operate similarly across demographic groups, our analysis of between-group differences in policy responses remains informative. We are reassured by the similarity between our self-reported results and those of \cite{carvajal2024}, who employ a between-subject vignette experiment in which each respondent is randomly assigned to a single policy scenario, thereby avoiding sequential-response concerns.}

	Students do not respond to institutional policies uniformly. In Appendix~Table~\ref{tab:ai_policy_het}, we regress the likelihood of using AI under each policy scenario on student characteristics. Both the effect of prohibition and intended usage under prohibition vary by gender, race, and field of study. Black students reduce their intended use more sharply under a ban than white students, while Arts and Literature majors adjust only modestly---in part because they were less likely to use AI in the first place. Intended usage conditional on prohibition is similarly uneven: males remain substantially more likely than females to report they would still use AI, while Arts and Languages majors are among the least likely. Uniform policies thus produce disparate outcomes across the student body---both in the reductions they induce and in the residual usage they fail to deter.

	\subsection{Understanding of Institutional Policies and Resources}
	
	Information gaps and inattention can affect technology adoption decisions \citep{duflo.etal2011, hanna.etal2014}. Imperfect information about rules, available resources, or proper usage guidelines could lead to underadoption of beneficial technologies or inadvertent policy violations. To test for information gaps, we examine three dimensions of policy understanding: whether students find AI policies clear, whether they know about free access to premium AI tools, and whether they know how to cite AI when required.
	
\begin{figure}[H]
		\caption{Understanding of Generative AI Policies and Resources}\label{fig:ai_policies}
		
		\begin{center}
			\begin{subfigure}[t]{.8\textwidth}
				\caption*{Panel A. Understanding of AI Policies}
				\centering
				\includegraphics[width=\textwidth]{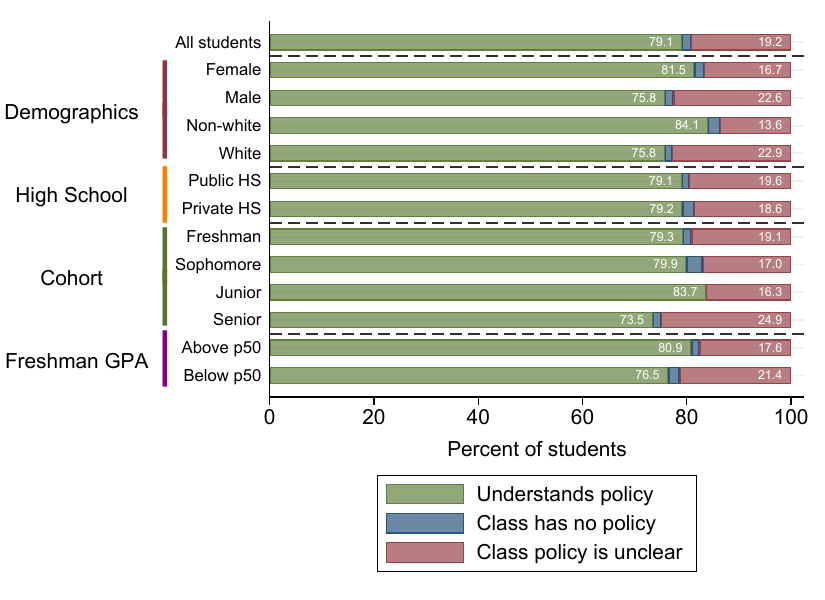}
			\end{subfigure}
			
			\begin{subfigure}[t]{.48\textwidth}
				\caption*{Panel B. Awareness of Copilot Access}
				\centering
				\includegraphics[width=\textwidth]{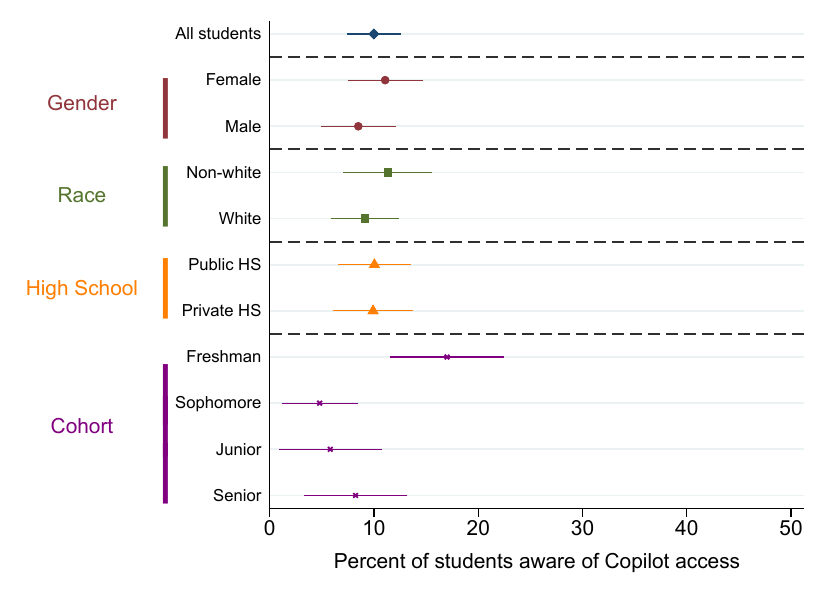}
			\end{subfigure} 
			\hfill		
			\begin{subfigure}[t]{.48\textwidth}
				\caption*{Panel C. Knowledge of Citation Requirements}
				\centering
				\includegraphics[width=\textwidth]{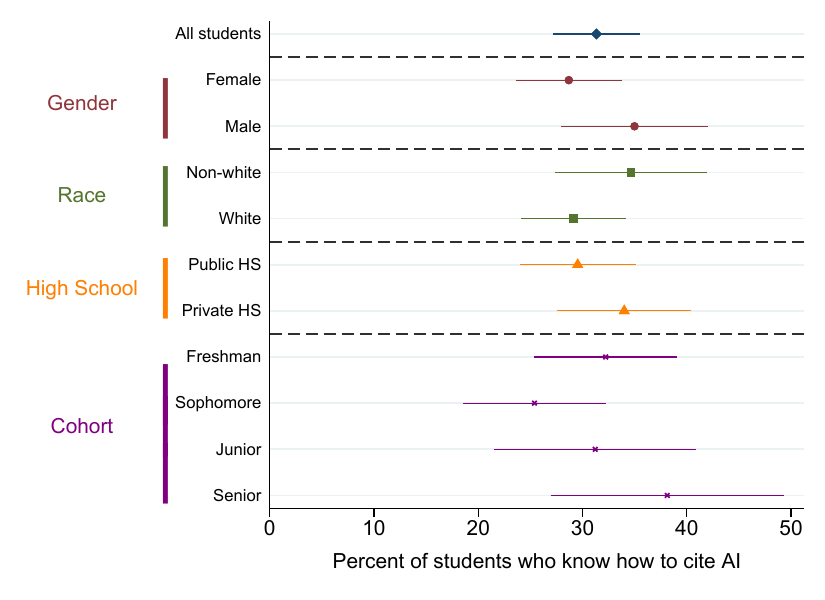}
			\end{subfigure} 
		\end{center}
		
		{\footnotesize\singlespacing \justify
			
			\textit{Notes:} This figure shows students' understanding of institutional AI policies and resources. Panel A displays the percent of students who report understanding AI policies in their classes, those who report having no explicit policy, and those who find policies unclear, broken down by demographic characteristics. Panel B shows the percent of students who are aware of their free access to Microsoft Copilot through Middlebury College. Panel C presents the percent of students who report knowing how to properly cite AI use in their academic work when required. For Panels B and C, horizontal lines represent 95 percent confidence intervals. Sample includes all survey respondents. \par
		}
	\end{figure}

	Most students understand AI policies, but gaps remain. Figure~\ref{fig:ai_policies}, Panel A shows that \getval{pct_understand_policy} percent of students understand when and where they can use AI in their classes; \getval{pct_policy_unclear_body} percent find policies unclear. Larger gaps appear elsewhere: only \getval{pct_copilot_aware_body} percent know they have free access to Microsoft Copilot through the college (Panel B), and just \getval{pct_cite_aware_body} percent know how to properly cite AI use (Panel C). Awareness varies by gender and race: females show better policy understanding than males (\getval{pct_understand_female} versus \getval{pct_understand_male} percent), and non-white students show higher awareness than white students across all three dimensions.\footnote{Similar patterns of limited awareness and confusion surrounding AI policies have been documented in other educational contexts. \citet{stohr.etal2024} find that only \getval{pct_stohr_policy} percent of Swedish students report that their teachers or universities have rules or guidelines on responsible AI use, suggesting widespread policy ambiguity.}
	
	Qualitative evidence from open-ended responses reinforces these frictions and reveals implementation challenges (Appendix~\ref{app:open_policy}). Students express frustration with vague guidelines and request specific examples of acceptable versus unacceptable uses. Many advocate for formal training, noting that knowing a policy exists differs from knowing how to integrate AI into their workflow. A recurring theme is blanket prohibitions as both ineffective and unfair, creating a prisoner's dilemma in which compliant students fall behind those who secretly violate the rules. Many responses call for a balanced approach---permitting AI use that supports learning while restricting uses that replace it---though the boundary between the two remains contested.	
	
	These findings point to two channels through which institutional interventions may shape AI adoption. Some policies operate through information and norms---clarifying acceptable use or raising awareness of available resources---and shift behavior by changing students' beliefs about appropriateness or their knowledge of options. Others operate through constraints and incentives---explicit prohibitions enforced through academic-integrity sanctions---and may shift behavior even with beliefs fixed, by changing AI use's costs and benefits. The information gaps we document suggest substantial scope for the first type of intervention; our policy-scenario results speak to the potential magnitude of the second.
	
	\section{Beliefs About AI's Educational Impact and Peer Usage} \label{sec:beliefs}

	\subsection{Student Beliefs About AI's Impact on Educational Outcomes}
	
	Beliefs about potential returns shape technology adoption \citep{foster2010microeconomics}. Students' perceptions of how AI affects their learning may influence both whether they adopt AI tools and how they use them. We asked students to evaluate AI's impact across four dimensions---understanding of course materials, overall learning ability, time management, and course grades---on a five-point scale from ``significantly reduces'' to ``significantly improves.''
	
\begin{figure}[H]
		\caption{Student Beliefs about the Impact of AI on their Academic Performance} \label{fig:ai_learning}
		\centering
		\includegraphics[width=.75\linewidth]{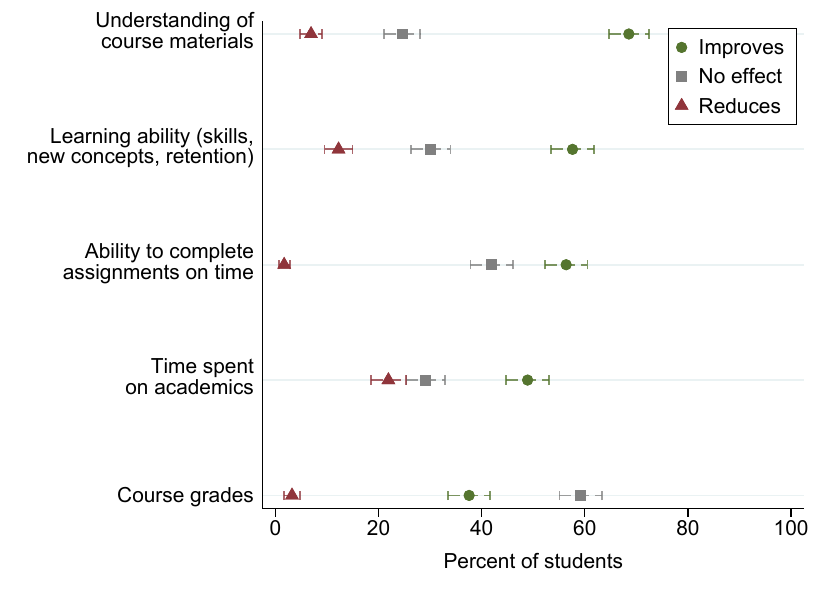}
		{\footnotesize
			\singlespacing \justify
			
			\textit{Notes:} This figure shows the percent of Middlebury students who believe that AI improves, reduces, or has no effect on different aspects of their academic experience. For each outcome, responses are categorized into three groups: ``Improves'' combines ``significantly improves'' and ``somewhat improves'' responses, ``Reduces'' combines ``significantly reduces'' and ``somewhat reduces'' responses, and ``No effect'' represents neutral responses. Sample includes all survey respondents who answered the question. ``Don't know'' responses are excluded. \par
		}
	\end{figure}

	Students generally view generative AI as beneficial, though perceived benefits vary across dimensions. Figure~\ref{fig:ai_learning} shows that \getval{pct_belief_understand} percent of students believe AI improves their understanding of course materials, and \getval{pct_belief_learn} percent report improvements in their ability to grasp concepts, retain information, or learn new skills. A similar \getval{pct_belief_ontime} percent report that AI improves their ability to complete assignments on time. Students are less confident about AI's impact on grades: only \getval{pct_belief_grades} percent believe it improves their grades, \getval{pct_belief_grades_noeffect} percent report no effect, and \getval{pct_belief_grades_negative} percent report negative effects.\footnote{Perceived benefits vary across student groups (Appendix~Table~\ref{tab:ai_learning_het}). Black students report the most positive perceptions, being \getval{pp_black_belief_learn} pp more likely than white students to believe AI improves learning ability. Male students consistently perceive greater benefits than females across most dimensions (\getval{pp_male_belief_range_low}-\getval{pp_male_belief_range_high} pp higher). Students in humanities, languages, and literature report less optimistic views than natural sciences majors, with literature majors \getval{pp_lit_belief_grades} pp less likely to believe AI improves grades.}
	This pattern suggests that students perceive AI as improving their learning process and workflow---through better understanding, skill development, and timely completion of work---but that these perceived benefits do not necessarily translate into better grades.\footnote{This pattern suggests that many students perceive a low elasticity of grades with respect to learning. While understanding this disconnect is beyond the scope of this paper, several mechanisms could explain it. Grade inflation may create ceiling effects that compress the grade distribution, limiting the scope for learning improvements to translate into higher grades. Alternatively, students may believe that effort and compliance with course requirements matter more for grades than deep understanding of the material---or that these inputs can compensate for gaps in learning.}
	
\begin{figure}[H]
		\caption{Relationship Between AI Adoption and Beliefs About AI's Academic Benefits}\label{fig:ai_adoption_beliefs}
		\begin{subfigure}[t]{.45\textwidth}
			\caption*{Panel A. Learning ability}
			\centering
			\includegraphics[width=\textwidth]{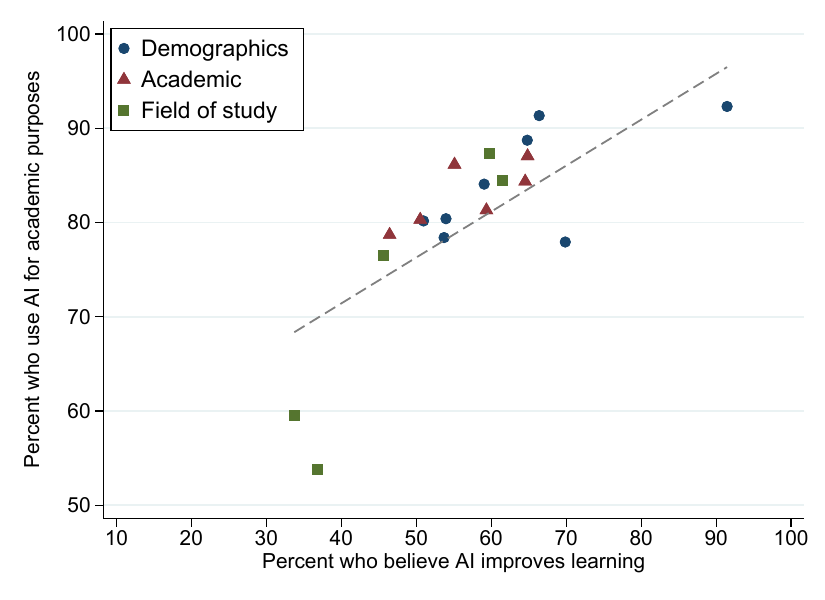}
		\end{subfigure}
		\hfill   
		\begin{subfigure}[t]{.45\textwidth}
			\caption*{Panel B. Understanding of course materials}
			\centering
			\includegraphics[width=\textwidth]{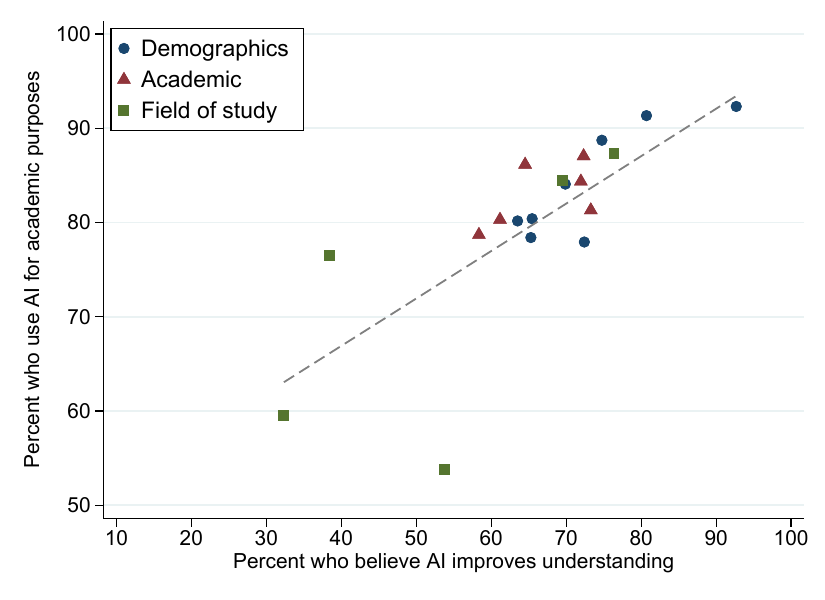}
		\end{subfigure} 
		\hfill \\~\\
		\begin{subfigure}[t]{.45\textwidth}
			\caption*{Panel C. Complete assignments on time}
			\centering
			\includegraphics[width=\textwidth]{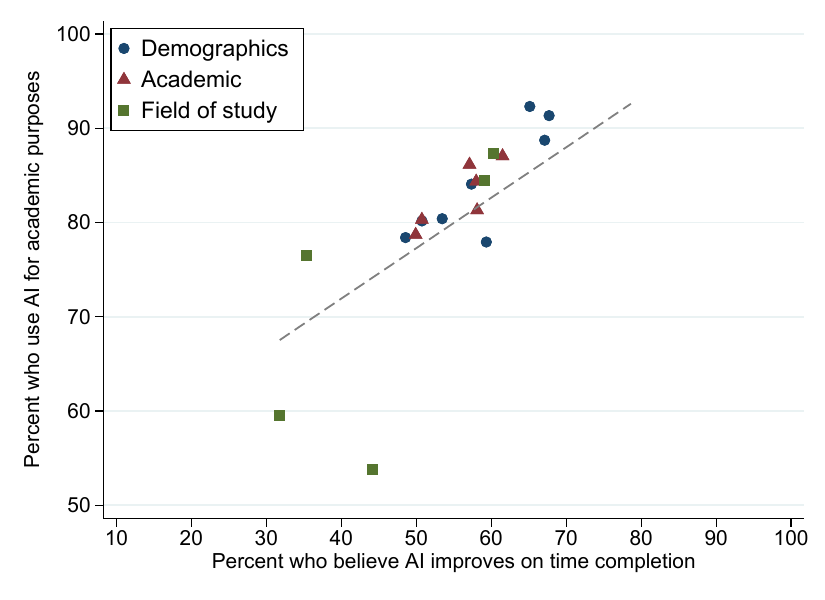}
		\end{subfigure}
		\hfill   
		\begin{subfigure}[t]{.45\textwidth}
			\caption*{Panel D. Course grades}
			\centering
			\includegraphics[width=\textwidth]{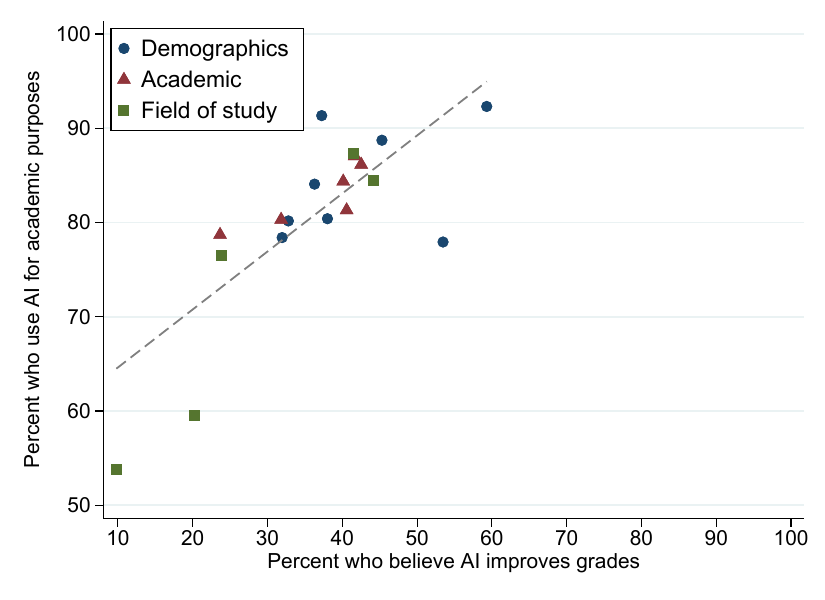}
		\end{subfigure} 
		\hfill    
		
		{\footnotesize\singlespacing \justify
			
			\textit{Notes:} This figure shows the relationship between AI adoption rates and beliefs that AI improves various academic outcomes across different student groups. Each panel plots the percentage who believe AI improves a specific outcome ($x$-axis) against the percent of students who use AI ($y$-axis). Points represent different student groups categorized by demographics (circles), academic characteristics (triangles), and field of study (squares). The dashed line shows the linear fit across all groups. Groups with fewer than ten students are not plotted. \par
		}
	\end{figure}

	Beliefs about AI's benefits are strongly correlated with adoption. Figure~\ref{fig:ai_adoption_beliefs} plots adoption rates against the share of students who believe AI improves a specific outcome, across student subgroups (e.g., males, white students, public-school students). Across all four academic dimensions, the relationship is positive, strong, and statistically significant. The correlation is strongest for beliefs about grades (Panel D): a \getval{pp_belief_change_unit} increase in the belief that AI improves grades is associated with a \getval{pp_slope_belief_grades} pp increase in AI adoption ($p < 0.01$). Similar relationships hold for beliefs about learning ability (\getval{pp_slope_belief_learn_sec4} pp, $p < 0.05$), understanding of course materials (\getval{pp_slope_belief_understand} pp, $p < 0.01$), and timely assignment completion (\getval{pp_slope_belief_ontime} pp, $p < 0.05$). These group-level correlations suggest that beliefs about AI's academic benefits---whether or not they reflect the actual benefits---shape adoption, though we cannot determine the direction of causation.
	
	These positive beliefs align with findings elsewhere. \citet{ravselj.etal2025} report that most students believe ChatGPT improves their general knowledge (\getval{pct_ravselj_general_know} percent) and specific knowledge (\getval{pct_ravselj_specific_know} percent)---remarkably similar to our findings on understanding of course materials (\getval{pct_belief_understand_cmp} percent) and learning ability (\getval{pct_belief_learn_cmp} percent). Similarly, \getval{pct_ravselj_deadlines} percent in \citet{ravselj.etal2025}'s sample believe ChatGPT helps meet deadlines, versus \getval{pct_belief_ontime_cmp} percent in our data. \citet{stohr.etal2024} find that \getval{pct_stohr_effective} percent of Swedish students believe AI makes them more effective learners, yet only \getval{pct_stohr_grades} percent believe it improves grades---mirroring our finding that perceived learning benefits exceed perceived grade effects.
	
	\subsection{Student Beliefs About Peer Use of Generative AI}

	Students' beliefs about peer AI use may shape adoption through social norms \citep[e.g.,][]{giaccherini2019behavioralist}, social learning \citep[e.g.,][]{foster1995learning, beaman2021can}, peer effects \citep[e.g.,][]{bailey2022peer}, and competitive pressure \citep[e.g.,][]{goehring2024technology}. We asked students to estimate the share of peers using generative AI for different purposes and under different policy environments.

\begin{figure}[H]
		\caption{Relationship Between Beliefs About AI Usage and Actual AI Usage} \label{fig:beliefs_corr}
		\centering
		\includegraphics[width=.75\linewidth]{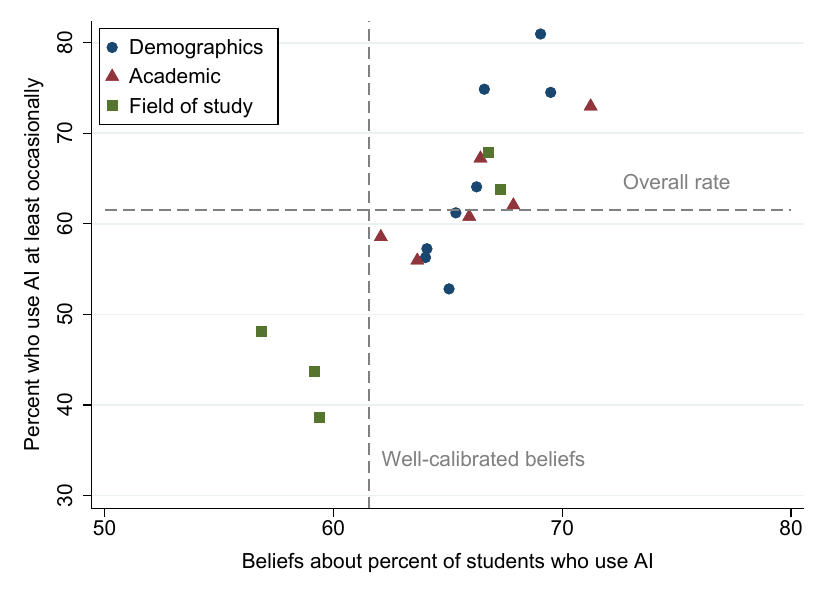}
		{\footnotesize
			\singlespacing \justify
			
			\textit{Notes:} This figure shows the relationship between students' beliefs about AI usage among their peers and actual AI usage rates across different demographic groups. Each point represents a different group of students (by demographics, academic characteristics, or field of study). The $y$-axis shows the percent of students in each group who use AI at least occasionally for academic purposes. The $x$-axis shows the mean belief within each group about what percent of Middlebury students use AI. The dashed horizontal line shows the overall true rate of AI use at least occasionally (\getval{pct_actual_peer_schoolwork} percent). Sample includes all survey respondents with at least ten observations per group. \par
		}
	\end{figure}

	On average, students believe that \getval{pct_belief_peer_schoolwork} percent of peers regularly use generative AI for schoolwork, with wide variation in individual estimates (Appendix~Figure~\ref{fig:ai_beliefs}). This average belief exceeds the actual rate of \getval{pct_actual_peer_schoolwork} percent (defined as use at least occasionally) by \getval{pp_schoolwork_overestimate} pp, implying slight overestimation. Yet this average masks substantial variation across student groups. Figure~\ref{fig:beliefs_corr} plots group-level beliefs ($x$-axis) against actual usage ($y$-axis) across demographic and academic groups. First, beliefs vary across groups---from \getval{pct_belief_cluster_low} to \getval{pct_belief_cluster_high} percent---yet nearly every group exceeds the overall true rate of \getval{pct_actual_peer_schoolwork} percent, so overestimation is near-universal. The largest overestimates are in lower-adoption groups such as females, white students, and majors in Humanities, Languages, and Literature. Second, beliefs track usage at the group level: groups with higher within-group adoption (ranging from \getval{pct_actual_range_low} to \getval{pct_actual_range_high} percent) report higher estimates of peer use, consistent with students inferring norms from their immediate social environment.

	Because beliefs track adoption closely, students' overestimation of peer AI use suggests that misperceived social norms may shape technology diffusion. If AI enhances learning, overestimation may accelerate beneficial adoption, and correcting misperceptions could slow it; if AI undermines skill development, the same overestimation may fuel overadoption, and corrections could safeguard against it.
	
	\section{Discussion} \label{sec:conclusion}
	
	This paper documents generative AI adoption at a selective U.S.\ college. We find rapid adoption, shifts in the educational production function through augmentation and automation, and a central role for beliefs and institutional policies in shaping AI use. Taken together, these findings point to three implications for institutional policy.

	First, we identify low-cost opportunities to make policy more effective by improving student information. The gaps we document in students' understanding of AI policies, citation practices, and available resources suggest that simple interventions---clear guidelines, illustrative examples of acceptable uses, and AI-literacy programs---can reduce unintentional integrity violations and support beneficial AI integration. Qualitative feedback confirms student demand for more explicit guidance.
	
	Second, our evidence challenges narratives that conflate widespread AI adoption with universal academic dishonesty.\footnote{See, for example: \href{https://nymag.com/intelligencer/article/openai-chatgpt-ai-cheating-education-college-students-school.html}{``Everyone Is Cheating Their Way Through College. ChatGPT has unraveled the entire academic project.''} John Herrman, \textit{New York Magazine}, May 7, 2025.} Although AI use is near-universal, students primarily use it to enhance learning through augmentation and selectively automate tasks when facing high time opportunity costs, not solely to circumvent academic effort. This distinction matters for policy: normalizing academic dishonesty as inevitable may shift social norms and encourage students who would otherwise use AI responsibly to engage in prohibited behaviors, believing ``everyone else is doing it.''

	Third, our findings caution against policy extremes \citep{merchant2024luddites, mcdonald.etal2025}. Blanket prohibitions risk disproportionately harming students who would benefit most from AI as a learning aid---if the perceived learning benefits translate into actual learning gains---while creating uneven compliance that places conscientious students at a disadvantage relative to rule-breakers. Unrestricted use based on revealed preference alone ignores potential market failures in educational settings. We document that students hold positive beliefs about AI's effects on learning and that these beliefs are strongly correlated with adoption. Yet the empirical evidence on AI's actual learning effects is mixed \citep[e.g.,][]{bastani.etal2025, contractor.reyes2025learning, desimone.tiberti2025, kestin.etal2025, kim.etal2025, lehmann.etal2025, lira.etal2025}. If students' beliefs are more optimistic than the true effects warrant, adoption based on those beliefs may harm learning. Permissive policies also risk creating competitive dynamics in which students feel compelled to adopt AI not for its learning benefits but to avoid falling behind in an educational ``arms race'' \citep{goehring2024technology}.

	\clearpage
	\begin{singlespace}
		\bibliographystyle{chicago}
		\bibliography{bib_adoption}
	\end{singlespace}

	\clearpage
	\section*{Figures and Tables}
	
	\clearpage

	\clearpage

	\clearpage

	\clearpage

	\clearpage
	
	\clearpage

	\clearpage
	\begin{table}[H]{\footnotesize
			\caption{Summary Statistics of Survey Participants}\label{tab:summ_stats}
			\vspace{-15pt}
			\begin{center}
				\begin{singlespace}
					\begin{tabular}{lccc}
						\midrule
						& \multicolumn{2}{c}{Survey Sample} & Admin records \\ \cmidrule{2-3} 
						& Unweighted & Weighted & \\
						& (1) & (2) & (3) \\ \midrule
						\multicolumn{4}{l}{\hspace{-1em} \textbf{Panel A. Demographics}} \\
						\input{results/summ_dem} \midrule
						
						\multicolumn{4}{l}{\hspace{-1em}\textbf{Panel B. Academic Characteristics}} \\
						\input{results/summ_aca} \midrule
						\multicolumn{4}{l}{\hspace{-1em}\textbf{Panel C. Field of Study}} \\
						\input{results/summ_maj} \midrule
						
						\input{results/summ_n} \midrule
						
					\end{tabular}
				\end{singlespace}
			\end{center}
			\begin{singlespace} 
				\justify \footnotesize \vspace{-1cm}
				\textit{Notes:} This table presents summary statistics from our survey of college students. Panel A reports demographic characteristics, including the proportion of participants identifying as male, female, white, Black, Hispanic, Asian, or who attended a private or public high school. Panel B provides academic characteristics, such as GPA (only available for non-freshmen), average weekly hours spent on academics, and academic year distribution (Freshman, Sophomore, Junior, and Senior). Note that in column 1--2, GPA refers to self-reported first-year GPA while in column 3 it is the overall GPA during Spring 2024. Panel C summarizes the distribution of participants across different fields of study. Major groups are mutually exclusive. \par
			\end{singlespace}
			
		}
	\end{table}

	\begin{table}[H]{\footnotesize
			\begin{center}
				\caption{Student Characteristics Associated with Frequency of Generative AI Use} \label{tab:ai_usage_correlates}
				\newcommand\w{2.4}
				\begin{tabular}{l@{}lR{\w cm}@{}L{0.45cm}R{\w cm}@{}L{0.45cm}R{\w cm}@{}L{0.45cm}R{\w cm}@{}L{0.45cm}}
					\midrule
					&& \multicolumn{8}{c}{Outcome: Uses AI during the semester with frequency of at least...}  \\\cmidrule{3-10} 
					&& A few times && A few times && A few times && Daily or almost \\  
					&& a semester && a month && a week && daily \\  
					&& (1)    && (2)     && (3)     && (4) \\  
					\midrule
					\input{results/ai_usage_regs}
					\midrule
					\input{results/ai_usage_N}
					\midrule
				\end{tabular}
			\end{center}
			\begin{singlespace} \vspace{-.5cm}
				\noindent \justify \textit{Notes:} This table assesses the relationship between AI adoption and student characteristics. We estimate: $$Y_{i} = \alpha + \beta X_i + \varepsilon_{i},$$ where $Y_i$ is a binary indicator of AI usage frequency threshold and $X_i$ is a vector of student characteristics including gender, race/ethnicity, high school type, cohort indicators, and academic division.				
				
				Each column uses a different threshold for AI usage frequency, categorized as: ``Rarely'' (a few times a semester), ``Occasionally'' (a few times a month), ``Frequently'' (a few times a week), and ``Very Frequently'' (daily or almost daily). Column 1 defines usage as any nonzero frequency; column 2 includes at least occasional use; column 3 includes frequent or higher use; and column 4 captures only very frequent use.
				
				The omitted categories are: Natural Sciences (and the small share of students with no declared or intended major) for academic division, white students for race/ethnicity, freshmen for cohort, female for gender, and private high school for school type. Heteroskedasticity-robust standard errors clustered at the student level in parentheses. Observations are weighted to adjust for sampling. {*} $p<0.10$, {*}{*} $p<0.05$, {*}{*}{*} $p<0.01$. \par
			\end{singlespace}   
		}
	\end{table}

	\begin{table}[H]{\footnotesize
			\begin{center}
				\caption{Student Characteristics Associated with Timing of Generative AI Adoption} \label{tab:ai_adopt_correlates}
				\newcommand\w{1.85}
				\begin{tabular}{l@{}lR{\w cm}@{}L{0.45cm}R{\w cm}@{}L{0.45cm}R{\w cm}@{}L{0.45cm}R{\w cm}@{}L{0.45cm}R{\w cm}@{}L{0.45cm}}
					\midrule
					&& \multicolumn{10}{c}{Outcome: Started using generative AI...}  \\\cmidrule{3-12} 
					&& Before && Spring 2023 && Fall 2023 && Spring 2024 && Fall 2024 \\  
					&& Spring 2023 && or before && or before && or before && or before  \\  
					&& (1)    && (2)     && (3)     && (4)  && (5) \\  
					\midrule
					\input{results/ai_adopt_regs}
					\midrule
					\input{results/ai_adopt_N}
					\midrule
				\end{tabular}
			\end{center}
			\begin{singlespace} \vspace{-.5cm}
				\noindent \justify \textit{Notes:} This table assesses the relationship between AI adoption and student characteristics. We estimate: $$Y_{i} = \alpha + \beta X_i + \varepsilon_{i},$$ where $Y_i$ is a binary indicator of AI adoption date and $X_i$ is a vector of student characteristics including gender, race/ethnicity, high school type, cohort indicators, and academic division.				
				Each column presents results for a different threshold of AI adoption. Column 1 shows the probability of adopting AI before Spring 2023; column 2 by Spring 2023; column 3 by Fall 2023; column 4 by Spring 2024; and column 5 by Fall 2024. The dependent variable in each regression is a binary indicator equal to one if the student had adopted AI by the specified time period.
				
				The omitted categories are: Natural Sciences (and the small share of students with no declared or intended major) for academic division, white students for race/ethnicity, freshmen for cohort, female for gender, and private high school for school type. Heteroskedasticity-robust standard errors clustered at the student level in parentheses. Observations are weighted to adjust for sampling. {*} $p<0.10$, {*}{*} $p<0.05$, {*}{*}{*} $p<0.01$. \par
			\end{singlespace}   
		}
	\end{table}

	\begin{landscape}
		\begin{table}[htpb]{\footnotesize
				\begin{center}
					\caption{Student Characteristics Associated with Task Augmentation and Automation} \label{tab:ai_autom_regs}
					\newcommand\w{1.55}
					\begin{tabular}{l@{}lR{\w cm}@{}L{0.45cm}R{\w cm}@{}L{0.45cm}R{\w cm}@{}L{0.45cm}R{\w cm}@{}L{0.45cm}R{\w cm}@{}L{0.45cm}R{\w cm}@{}L{0.45cm}R{\w cm}@{}L{0.45cm}R{\w cm}@{}L{0.45cm}}
						
						\midrule
						&& \multicolumn{6}{c}{Augmentation Tasks} & \multicolumn{6}{c}{Automation Tasks} & \multicolumn{4}{c}{Difference: Augm. - Autom.} \\ \cmidrule{3-7} \cmidrule{9-13} \cmidrule{15-18}
						&& Any $>0$ && Share $>0$ && Intensity 	&& Any $>0$ && Share $>0$ && Intensity && Share && Intensity \\
						&& (1) && (2) && (3) && (4) && (5) && (6) && (7) && (8) \\
						\midrule
						\input{results/ai_autom_regs}
						\midrule
						\input{results/ai_autom_N}
						\midrule
					\end{tabular}
				\end{center}
				\begin{singlespace} \vspace{-.5cm}
					\noindent \justify \textit{Notes:} This table reports estimated associations between student characteristics and their use of generative AI for academic tasks. In columns 1 and 4, the outcome is a dummy that equals one if a student reports using AI with any frequency for at least one augmentation or automation task, respectively. In columns 2 and 5, the outcome is the share of tasks within each category for which the student reports any use. In columns 3 and 6, the outcome is a continuous measure capturing average usage frequency for each task category, based on raw Likert-style responses. In columns 7 and 8, the outcome is the difference in average task share and usage intensity between augmentation and automation, respectively. Regressions are weighted and report robust standard errors clustered at the student level. $^{***}$ $p<0.01$, $^{**}$ $p<0.05$, $^*$ $p<0.10$.
					
				\end{singlespace}
				
			}
		\end{table}
	\end{landscape}
	
	\clearpage 
	\appendix
	\begin{center}
		\noindent {\LARGE \textbf{Appendix}}
	\end{center}
	\label{app:figs}

	\setcounter{table}{0}
	\setcounter{figure}{0}
	\setcounter{equation}{0}	
	\renewcommand{\thetable}{A\arabic{table}}
	\renewcommand{\thefigure}{A\arabic{figure}}
	\renewcommand{\theequation}{A\arabic{equation}}

	\section{Appendix Figures and Tables}

	\begin{figure}[H]
		\caption{Generative AI Usage Survey Design Overview}
		\label{fig:flow}
		\begin{center}
			\tikzstyle{block} = [rectangle, draw, fill=bluecomment!10, text width=16em, text centered, rounded corners, minimum height=3.5em, font=\small]
			\tikzstyle{sectionblock} = [rectangle, draw, fill=redcomment!10, text width=16em, text centered, rounded corners, minimum height=3.5em, font=\small]
			\tikzstyle{datablock} = [rectangle, draw, fill=greencomment!10, text width=14em, text centered, rounded corners, minimum height=3.5em, font=\small]
			\tikzstyle{beliefsblock} = [rectangle, draw, fill=yellowcomment!10, text width=14em, text centered, rounded corners, minimum height=3.5em, font=\small]
			\tikzstyle{subtext} = [font=\footnotesize\itshape]
			\tikzstyle{line} = [draw, -latex', thick]
			
			\begin{tikzpicture}[node distance = 1.6cm, auto, scale=0.55, transform shape]
				\node [draw, rectangle, fill=white, text width=14em, font=\small] at (10,-4) {
					\begin{tabular}{ll}
						\textbf{Key} & \\[0.2cm]
						\tikz\node [fill=bluecomment!10, minimum size=1em] {}; & Survey Structure \\[0.2cm]
						\tikz\node [fill=redcomment!10, minimum size=1em] {}; & Survey Sections \\[0.2cm]
						\tikz\node [fill=greencomment!10, minimum size=1em] {}; & Usage Data Collection \\[0.2cm]
						\tikz\node [fill=yellowcomment!10, minimum size=1em] {}; & Beliefs \& Perceptions
					\end{tabular}
				};
				
				\node [block, text width=22em, font=\normalsize\bfseries] (surveyheader) {\underline{AI Usage Survey at Middlebury College}};
				\node [subtext, below=0.1cm of surveyheader] {December 2024};
				
				\node [block, below=of surveyheader, node distance=2.2cm] (recruitment) {Participant Recruitment};
				\node [subtext, below=0.1cm of recruitment] {Campus-wide email invitations, framed as general technology use survey};
				
				\node [block, below=of recruitment, node distance=2.2cm] (incentives) {Participation Incentives};
				\node [subtext, below=0.1cm of incentives] {Amazon gift card lottery ranging from \$50-\$500};
				
				\node [sectionblock, below=of incentives, node distance=2.2cm] (demographics) {Section 1: Demographics \& Academic Information};
				
				\node [datablock, below left=1.8cm and 4cm of demographics] (demdata1) {Student Characteristics};
				\node [subtext, below=0.1cm of demdata1] {Gender, race/ethnicity, high school type};
				
				\node [datablock, below right=1.8cm and 4cm of demographics] (demdata2) {Academic Profile};
				\node [subtext, below=0.1cm of demdata2] {Year, major, self-reported GPA, study hours};
				
				\node [sectionblock, below=of demographics, node distance=5cm] (aiusage) {Section 2: Generative AI Usage Patterns};
				
				\node [datablock, below left=1.8cm and 4cm of aiusage] (usagedata1) {Adoption Metrics};
				\node [subtext, below=0.1cm of usagedata1] {Frequency, timing of first use, specific AI models used};
				
				\node [datablock, below right=1.8cm and 4cm of aiusage] (usagedata2) {Academic Applications};
				\node [subtext, below=0.1cm of usagedata2] {Tasks performed with AI, payment for premium services};
				
				\node [sectionblock, below=of aiusage, node distance=5cm] (perceptions) {Section 3: Perceptions \& Institutional Policies};
				
				\node [beliefsblock, below left=1.8cm and 4cm of perceptions] (percdata1) {Impact Perceptions};
				\node [subtext, below=0.1cm of percdata1] {Effects on learning, grades, time management};
				
				\node [beliefsblock, below right=1.8cm and 4cm of perceptions] (percdata2) {Policy Responses};
				\node [subtext, below=0.1cm of percdata2] {Usage likelihood under different policies};
				
				\node [beliefsblock, below=of perceptions, node distance=5cm] (peerbeliefs) {Peer Usage Beliefs};
				\node [subtext, below=0.1cm of peerbeliefs] {Estimates of AI adoption among peers for schoolwork and leisure};
				
				\node [block, below=of peerbeliefs, node distance=2.2cm] (openended) {Open-Ended Response Collection};
				\node [subtext, below=0.1cm of openended] {Motivations for AI use, policy feedback, suggestions for improvement};
				
				\path [line] (surveyheader) -- (recruitment);
				\path [line] (recruitment) -- (incentives);
				\path [line] (incentives) -- (demographics);
				\path [line] (demographics) -- (demdata1);
				\path [line] (demographics) -- (demdata2);
				\path [line] (demographics) -- (aiusage);
				\path [line] (aiusage) -- (usagedata1);
				\path [line] (aiusage) -- (usagedata2);
				\path [line] (aiusage) -- (perceptions);
				\path [line] (perceptions) -- (percdata1);
				\path [line] (perceptions) -- (percdata2);
				\path [line] (perceptions) -- (peerbeliefs);
				\path [line] (peerbeliefs) -- (openended);

			\end{tikzpicture}
		\end{center}
		{\footnotesize \textit{Note:} This figure illustrates the structure of the AI usage survey conducted at Middlebury College in December 2024. The survey collected information across three main sections: (1) demographic and academic background, (2) patterns of generative AI usage including adoption timing, frequency, and specific applications, and (3) perceptions of AI's impact on learning and responses to institutional policies.}
	\end{figure}

	\begin{figure}[H]
		\caption{Cumulative Generative AI Use by Student Characteristic}\label{fig:ai_cdf_dem}
		\centering
		\begin{subfigure}[t]{.48\textwidth}
			\caption*{Panel A. By Gender}
			\centering
			\includegraphics[width=\linewidth]{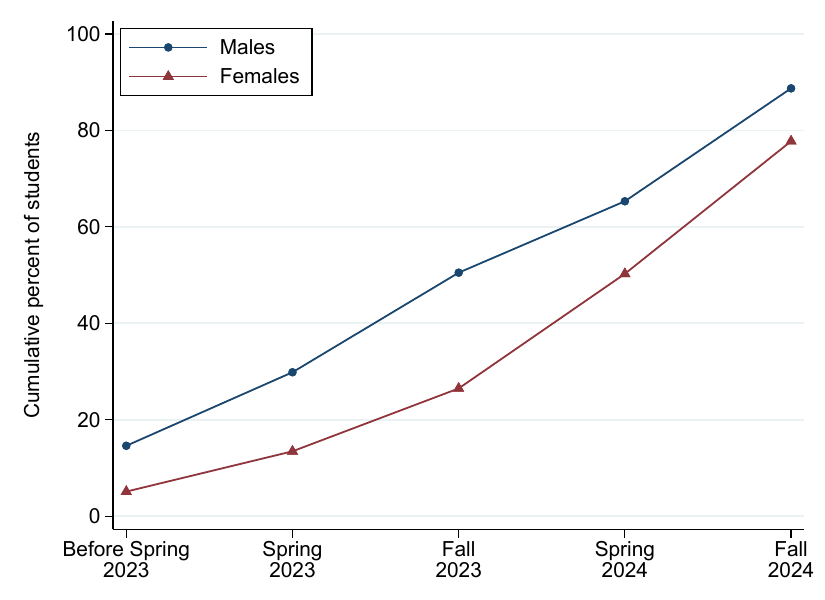}
		\end{subfigure}
		\hfill		
		\begin{subfigure}[t]{0.48\textwidth}
			\caption*{Panel B. By Race}
			\centering
			\includegraphics[width=\linewidth]{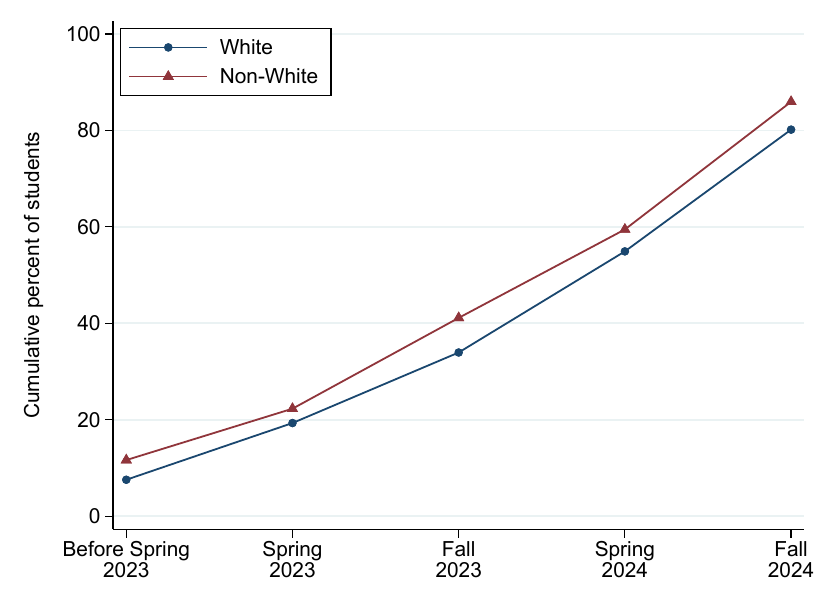}
		\end{subfigure}	
		\hfill		
		\begin{subfigure}[t]{.48\textwidth}
			\caption*{Panel C. By High School Type}
			\centering
			\includegraphics[width=\linewidth]{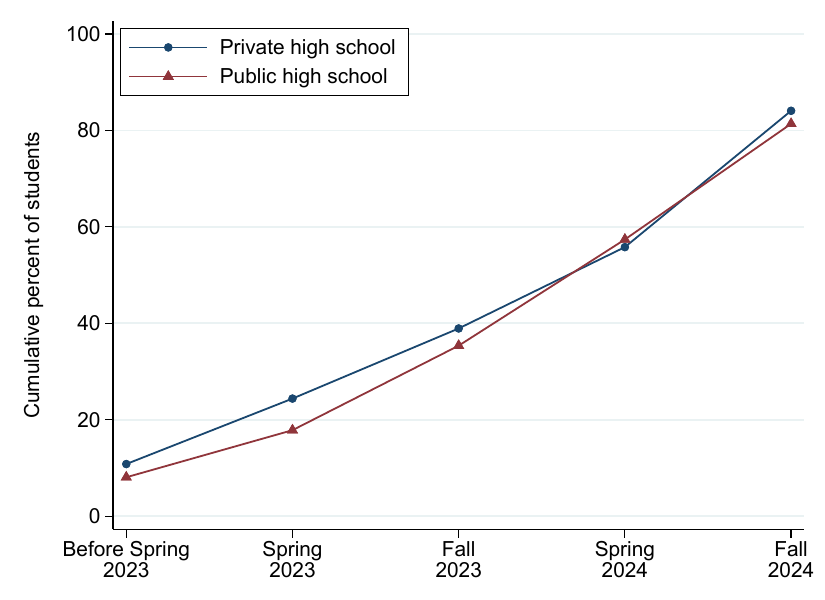}
		\end{subfigure}
		\hfill		
		\begin{subfigure}[t]{0.48\textwidth}
			\caption*{Panel D. By Freshman Year GPA}
			\centering
			\includegraphics[width=\linewidth]{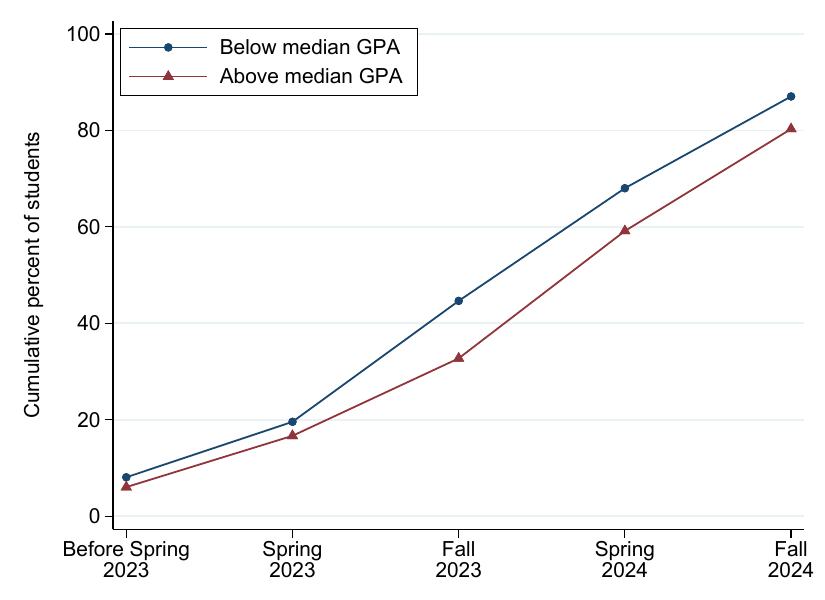}
		\end{subfigure}
		\hfill		
		\begin{subfigure}[t]{.48\textwidth}
			\caption*{Panel E. By Cohort}
			\centering
			\includegraphics[width=\linewidth]{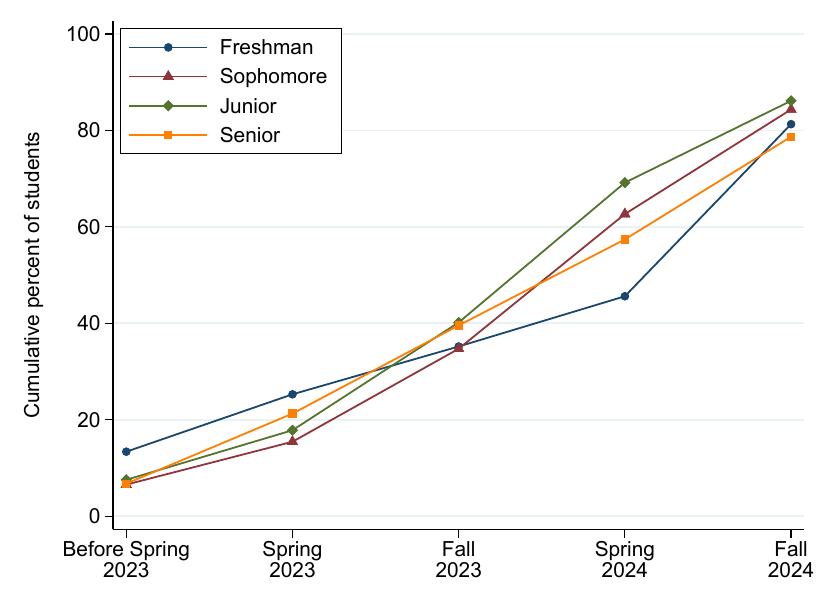}
		\end{subfigure}
		\hfill		
		\begin{subfigure}[t]{0.48\textwidth}
			\caption*{Panel F. By Field of Study}
			\centering
			\includegraphics[width=\linewidth]{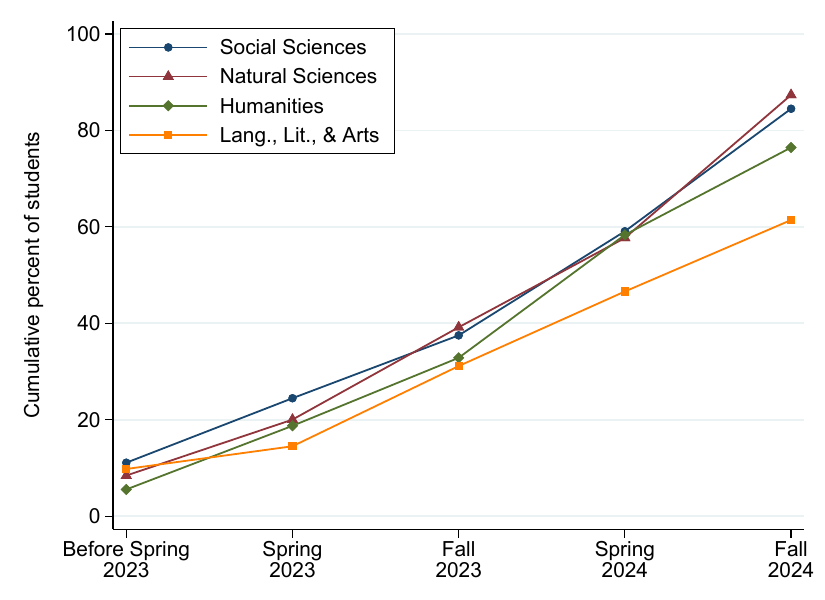}
		\end{subfigure}
		{\footnotesize
			\singlespacing \justify
			
			\textit{Notes:} This figure presents cumulative AI use based on different student characteristics. Each panel displays the cumulative distribution of AI use based on a specific characteristic: gender, race, school type, first-year GPA, cohort, or field of study. The cumulative percent of students is plotted against usage categories. The legends and colors correspond to subgroups within each demographic variable. \par
		}
	\end{figure}

	\clearpage
	\begin{figure}[H]
		\caption{Percent of Students Who Pay for Generative AI Tools} \label{fig:ai_pays_dem}
		\centering
		\includegraphics[width=.75\linewidth]{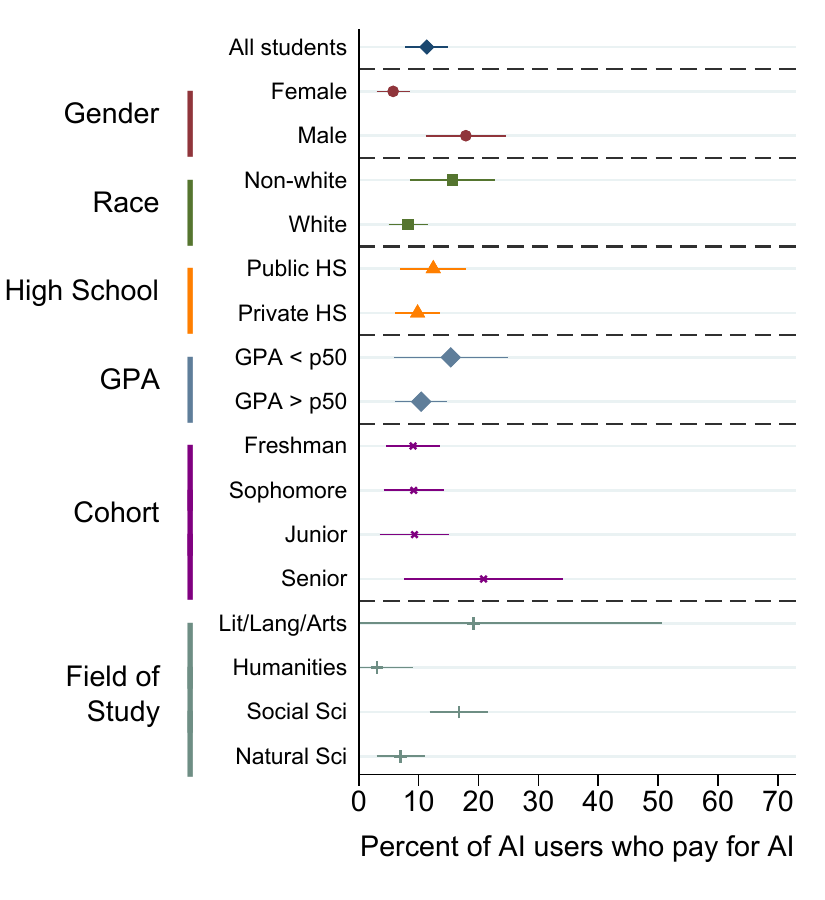}
		
		{\footnotesize
			\singlespacing
			\justify
			\emph{Notes:} This figure shows the percent of AI users who pay for AI tools (through any platform) across different demographic groups. Horizontal lines represent 95 percent confidence intervals calculated with heteroskedasticity-robust standard errors clustered at the student level. \par
		}
	\end{figure}

	\begin{figure}[htbp]
		\caption{Academic Uses of ChatGPT: Evidence from Global Survey} \label{fig:chatgpt_use_ravselj}
		
		\begin{center}
			
			\begin{subfigure}[t]{.8\textwidth}
				\caption*{Panel A. Across Tasks that Augment versus Automate Student Effort}
				\centering
				\includegraphics[width=\textwidth]{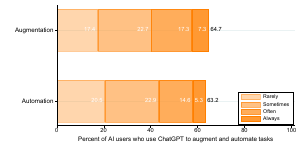} \vspace{.5cm}
			\end{subfigure} 
			
			\begin{subfigure}[t]{.7\textwidth}
				\caption*{Panel B. Across the College Quality Distribution}
				\centering
				\includegraphics[width=\textwidth]{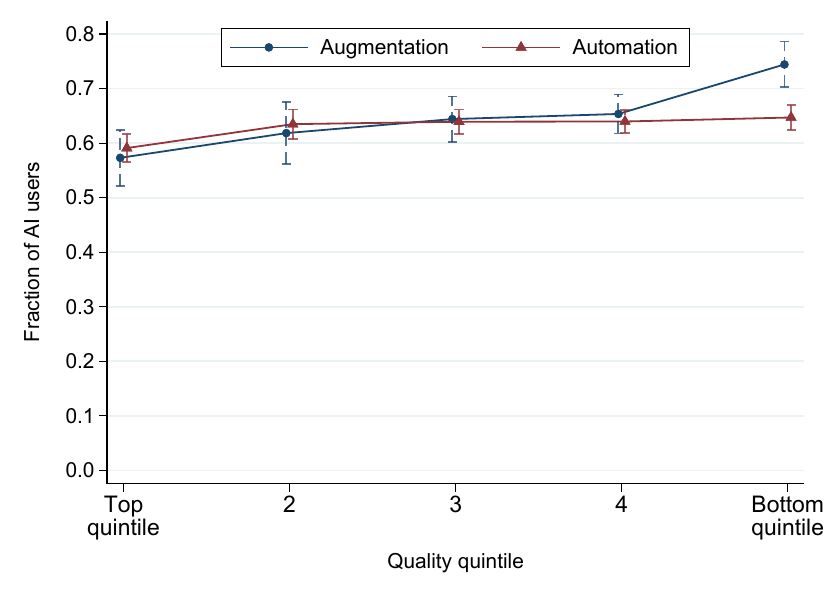}
			\end{subfigure} 
			
		\end{center} \vspace{-.5cm}
		
		{\footnotesize\singlespacing \justify \textit{Notes:} This figure shows the percent of students who use ChatGPT for different academic tasks based on data from \citet{ravselj.etal2025}. Panel A displays usage patterns across tasks categorized as augmenting (proofreading, translating, study assistance, research assistance) versus automating (academic writing, professional writing, creative writing, brainstorming, summarizing, calculating, coding assistance, personal assistance) student effort. Panel B shows usage patterns across university quality quintiles based on World University Rankings, with universities ranked in the top 20 percent (top quintile) showing slightly higher rates of augmentation relative to automation compared to bottom quintile institutions. The analysis includes universities with at least 30 student responses and excludes observations with missing usage data. \par
		}
	\end{figure}

	\clearpage
	\begin{landscape}

		\begin{figure}[htpb!]
			\caption{Student Beliefs about Generative AI Usage at Middlebury College}\label{fig:ai_beliefs}

			\begin{center}
				\begin{subfigure}[t]{.43\textwidth}
					\caption*{Panel A. For Schoolwork}
					\centering
					\includegraphics[width=\textwidth]{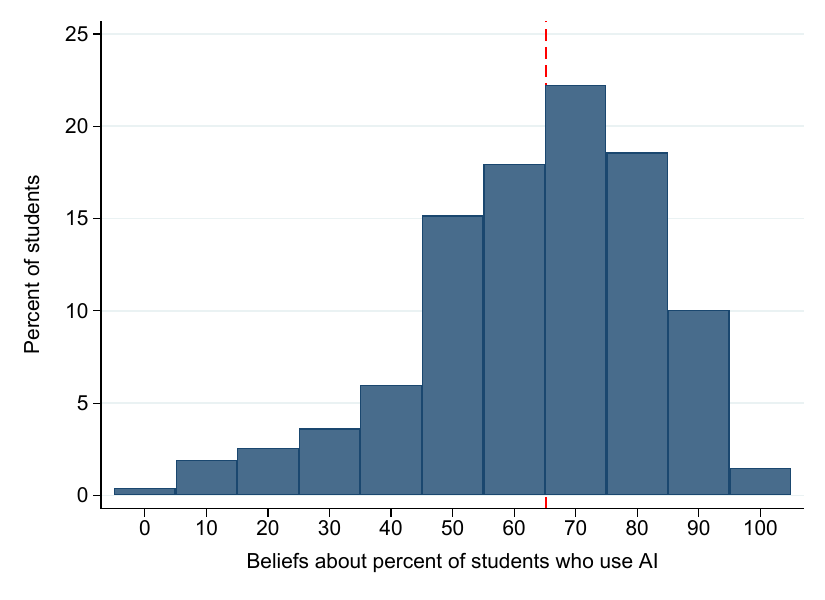}
				\end{subfigure}
				\hfill
				\begin{subfigure}[t]{.43\textwidth}
					\caption*{Panel B. For Leisure}
					\centering
					\includegraphics[width=\textwidth]{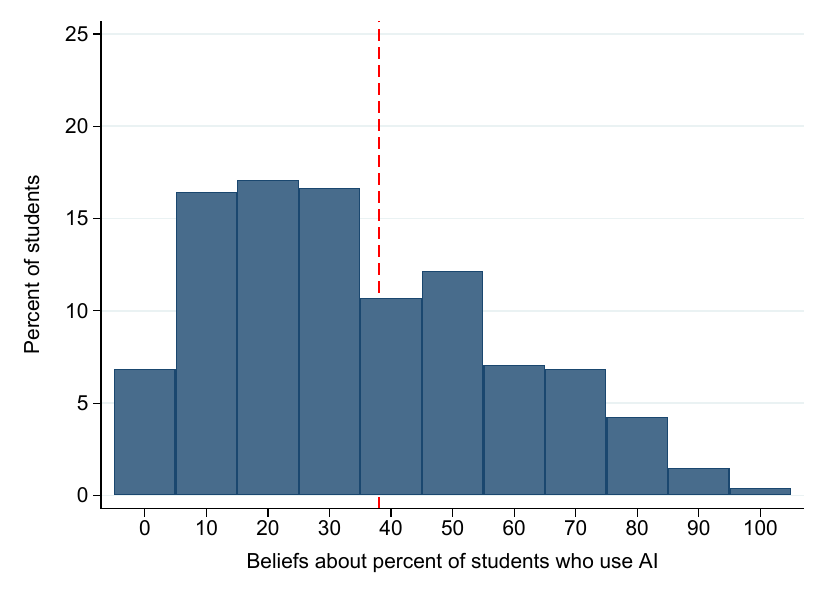}
				\end{subfigure}
				\hfill
				\begin{subfigure}[t]{.43\textwidth}
					\caption*{Panel C. For Any Purpose}
					\centering
					\includegraphics[width=\textwidth]{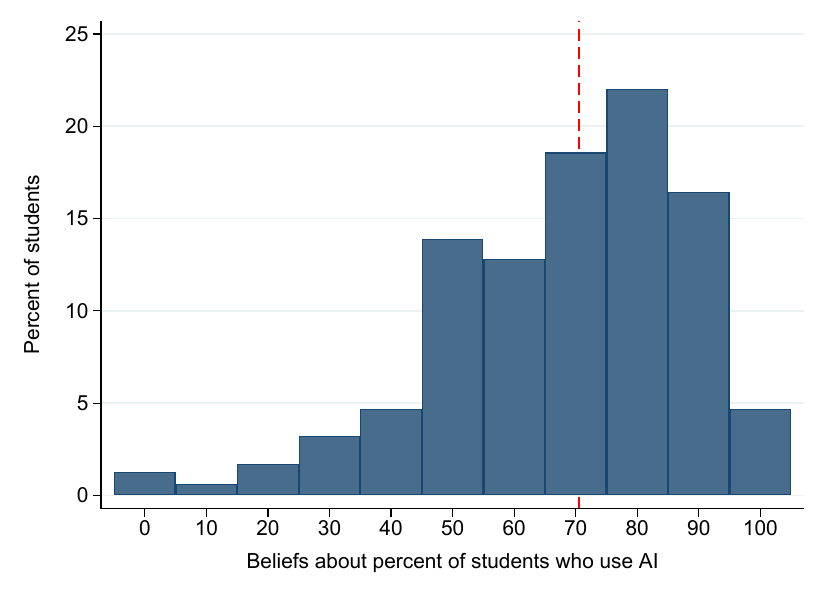}
				\end{subfigure}

				\vspace{1em}

				\begin{subfigure}[t]{.43\textwidth}
					\caption*{Panel D. Classes with No Policy}
					\centering
					\includegraphics[width=\textwidth]{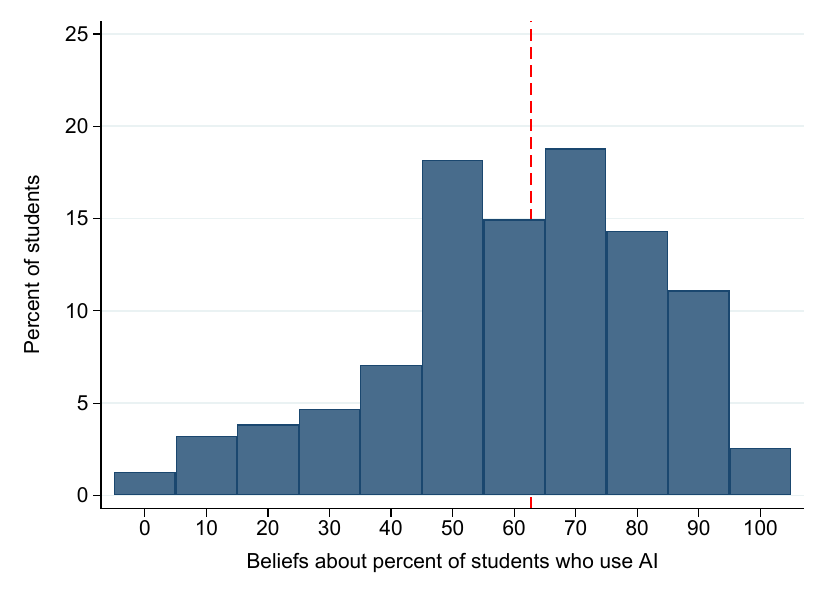}
				\end{subfigure}
				\hfill
				\begin{subfigure}[t]{.43\textwidth}
					\caption*{Panel E. Classes that Allow AI}
					\centering
					\includegraphics[width=\textwidth]{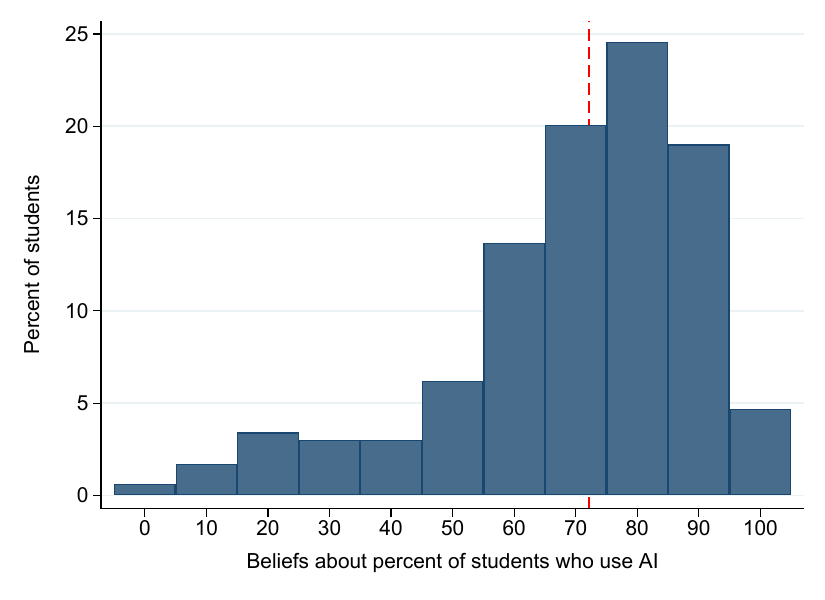}
				\end{subfigure}
				\hfill
				\begin{subfigure}[t]{.43\textwidth}
					\caption*{Panel F. Classes that Disallow AI}
					\centering
					\includegraphics[width=\textwidth]{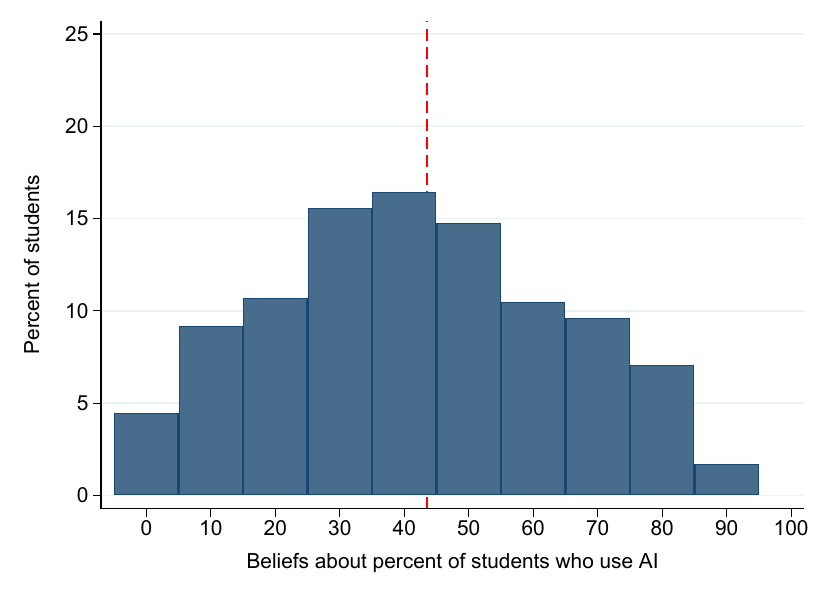}
				\end{subfigure}
			\end{center}

			{\footnotesize\singlespacing \justify
				\textit{Notes:} This figure shows the distribution of students' beliefs about generative AI usage among their peers at Middlebury College. Panels A--C display students' estimates of the percent of their peers who regularly use AI for schoolwork, leisure activities, and any purpose, respectively. Panels D--F show students' beliefs about AI usage in classes with different AI policies: those without an explicit policy (Panel D), those that allow AI use (Panel E), and those that prohibit AI use (Panel F). Each panel shows a histogram with bins of width ten percentage points (e.g., responses between 1--10 fall in the 10 bin, 11--20 in the 20 bin, etc.). The red dashed line indicates the mean response. Sample excludes respondents with missing values or who selected the default response for all six categories (which equals zero). \par
			}
		\end{figure}

	\end{landscape}

	\begin{table}[H]{\footnotesize
			\caption{Generative AI Usage Frequency by Student Characteristics}\label{tab:ai_freq_het}
			\vspace{-15pt}
			\begin{center}
				\begin{singlespace}
					\begin{tabular}{lccccc}
						\midrule
						&& \multicolumn{4}{c}{By Usage Frequency} \\ \cmidrule{3-6}
						& Any use & Rarely & Occasionally & Frequently & Very Frequently \\
						& (1) & (2) & (3) & (4) & (5) \\ \midrule						
						\multicolumn{5}{l}{\hspace{-1em} \textbf{Panel A. Demographics}} \\
						\input{results/ai_freq_dem} \midrule
						
						\multicolumn{5}{l}{\hspace{-1em}\textbf{Panel B. Academic Characteristics}} \\
						\input{results/ai_freq_aca} \midrule
						
						\multicolumn{5}{l}{\hspace{-1em}\textbf{Panel C. Field of Study}} \\
						\input{results/ai_freq_maj} \midrule
						
					\end{tabular}
				\end{singlespace}
			\end{center}
			\begin{singlespace} 
				\justify \footnotesize \vspace{-1cm}
				\textit{Notes:} This table presents the percent of students in each demographic group who report using AI at different frequencies during the academic semester. Each cell shows the percent of students within that group. Column 1 reports the total percent who use AI at any frequency. Columns 2 to 5 represent increasing usage frequencies: rarely (1–2 times per semester), occasionally (monthly), frequently (weekly), and very frequently (multiple times per week). Panel A reports percentages by demographic characteristics. Panel B shows percentages by academic characteristics. Panel C presents percentages by field of study. \par
			\end{singlespace}
		}
	\end{table}

	\begin{table}[H]{\footnotesize
			\begin{center}
				\caption{Student Characteristics Associated with Choice of Generative AI Models} \label{tab:ai_model_correlates}
				\newcommand\w{1.6}
				\begin{tabular}{l@{}lR{\w cm}@{}L{0.45cm}R{\w cm}@{}L{0.45cm}R{\w cm}@{}L{0.45cm}R{\w cm}@{}L{0.45cm}R{\w cm}@{}L{0.45cm}}
					\midrule
					&& \multicolumn{10}{c}{Outcome: =1 if student uses}  \\\cmidrule{3-12} 
					&& OpenAI's && Google's && Microsoft  && Other  && Pays for \\  					
					&& ChatGPT && Gemini && Copilot && Model  && GenAI \\  
					
					&& (1)    && (2)     && (3)     && (4)  && (5) \\  
					\midrule
					\input{results/ai_model_regs}
					\midrule
					\input{results/ai_model_N}
					\midrule
				\end{tabular}
			\end{center}
			\begin{singlespace} \vspace{-.5cm}
				\noindent \justify \textit{Notes:} This table assesses the relationship between AI model adoption and student characteristics. We estimate: $$Y_{i} = \alpha + \beta X_i + \varepsilon_{i},$$ where $Y_i$ is a binary indicator of AI model usage (columns 1--4) or payment for AI services (column 5), and $X_i$ is a vector of student characteristics including gender, race/ethnicity, high school type, cohort indicators, and academic division.
				
				Each column presents results for a different model or payment outcome. Column 1 shows usage of OpenAI's ChatGPT, column 2 Google Gemini, column 3 Microsoft Copilot, column 4 any other AI model, and column 5 whether the student pays for any generative AI service.				
				
				The omitted categories are Natural Sciences (and the small share of students with no declared or intended major) for academic division, white students for race/ethnicity, and freshmen for cohort. Heteroskedasticity-robust standard errors clustered at the student level in parentheses. Observations are weighted to adjust for sampling. {*} $p<0.10$, {*}{*} $p<0.05$, {*}{*}{*} $p<0.01$. \par
			\end{singlespace}   
		}
	\end{table}

	\begin{table}[H]{\footnotesize
			\begin{center}
				\caption{Student Characteristics Associated with AI Usage Likelihood Under Different Institutional Policies} \label{tab:ai_policy_het}
				\newcommand\w{1.95}
				\begin{tabular}{l@{}lR{\w cm}@{}L{0.45cm}R{\w cm}@{}L{0.45cm}R{\w cm}@{}L{0.45cm}R{\w cm}@{}L{0.45cm}R{\w cm}@{}L{0.45cm}}
					\midrule
					&& \multicolumn{10}{c}{Outcome: Would use generative AI in a given policy scenario...}  \\\cmidrule{3-12} 
					&& AI Use is && No Explicit && AI Allowed && AI Use is && Prohibition  \\
					&& Unrestricted && AI Policy && if Cited && Prohibited && Impact \\
					&& (1)    && (2)     && (3)     && (4)  && (5) \\  
					\midrule
					\input{results/ai_policy_regs}
					\midrule
					\input{results/ai_policy_N}
					\midrule
				\end{tabular}
			\end{center}
			\begin{singlespace} \vspace{-.5cm}
				\noindent \justify \textit{Notes:} This table examines how student characteristics relate to self-reported likelihood of using generative AI under different policy scenarios. Each column presents results for different policy scenarios. In columns 1--4, the dependent variable equals one if the student reports being ``likely'' or ``extremely likely'' to use AI under the specified policy, and zero otherwise. Column 5 represents the impact of moving from unrestricted use to complete prohibition.
				
				The omitted categories are: Natural Sciences (and the small share of students with no declared or intended major) for academic division, white students for race/ethnicity, freshmen for cohort, female for gender, and private high school for school type. Heteroskedasticity-robust standard errors clustered at the student level in parentheses. Observations are weighted to adjust for sampling. {*} $p<0.10$, {*}{*} $p<0.05$, {*}{*}{*} $p<0.01$. \par
			\end{singlespace}   
		}
	\end{table}

	\begin{table}[H]{\footnotesize
			\begin{center}
				\caption{Student Characteristics Associated with Perceived Learning Benefits } \label{tab:ai_learning_het}
				\newcommand\w{1.85}
				\begin{tabular}{l@{}lR{\w cm}@{}L{0.45cm}R{\w cm}@{}L{0.45cm}R{\w cm}@{}L{0.45cm}R{\w cm}@{}L{0.45cm}R{\w cm}@{}L{0.45cm}}
					\midrule
					&& \multicolumn{10}{c}{Outcome: Believes that generative AI improves...}  \\\cmidrule{3-12} 
					&& Learning && Understand && Course && Assignment && Time on \\
					&& Ability && Materials && Grades && Completion && Academics \\  
					&& (1)    && (2)     && (3)     && (4)  && (5) \\  
					\midrule
					\input{results/ai_improves_regs}
					\midrule
					\input{results/ai_improves_N}
					\midrule
				\end{tabular}
			\end{center}
			\begin{singlespace} \vspace{-.5cm}
				\noindent \justify \textit{Notes:} This table assesses the relationship between AI adoption and student characteristics. Each column presents results for beliefs about different academic outcomes: learning ability (e.g., ability to grasp concepts, retain information, or learn new skills) in column 1, understanding of course materials in column 2, course grades in column 3, ability to complete assignments on time in column 4, and time spent on academics in column 5. The dependent variable in each regression equals one if the student believes AI ``somewhat improves'' or ``significantly improves'' the outcome, and zero if they believe it has no effect, reduces, or significantly reduces the outcome. ``Don't know'' responses are excluded.
				
				The omitted categories are: Natural Sciences (and the small share of students with no declared or intended major) for academic division, white students for race/ethnicity, freshmen for cohort, female for gender, and private high school for school type. Heteroskedasticity-robust standard errors clustered at the student level in parentheses. Observations are weighted to adjust for sampling. {*} $p<0.10$, {*}{*} $p<0.05$, {*}{*}{*} $p<0.01$. \par
			\end{singlespace}   
		}
	\end{table}
	
	\section{Empirical Appendix} \label{app:empirical}
	
	\setcounter{table}{0}
	\setcounter{figure}{0}
	\setcounter{equation}{0}	
	\renewcommand{\thetable}{B\arabic{table}}
	\renewcommand{\thefigure}{B\arabic{figure}}
	\renewcommand{\theequation}{B\arabic{equation}}

	\subsection{Robustness to Alternative Weighting Schemes} \label{app:weights}

	This appendix examines how our main findings depend on the choice of poststratification weights. Our benchmark results weight observations by major to match the distribution of fields of study in the student population. We present results using three alternatives: (i) weighting by gender and cohort, (ii) weighting by race, and (iii) unweighted estimates.

	The gender-by-cohort weights use \getval{n_weight_cells_gc} cells (two genders $\times$ \getval{n_cohorts} cohorts). Population cohort totals come from administrative enrollment records. Gender shares within each cohort come from administrative data on expected graduation terms, which give a gender-by-graduation-year cross-tabulation for the classes of 2025, 2026, and 2027. Freshmen (Class of 2028) do not yet have assigned graduation terms; we use the overall gender ratio from the Common Data Set (\getval{pct_cds_female} percent female). The race weights use \getval{n_weight_cells_race} cells (white, Black, Hispanic, Asian) with population counts from the Common Data Set. Students outside these four categories or who prefer not to say receive missing weights and are excluded from the race-weighted estimates. All weights are normalized to sum to total enrollment.

	Appendix~Tables~\ref{tab:weight_adoption}--\ref{tab:weight_beliefs} show that our main findings are robust across these specifications. The largest variation is for coding assistance under gender-by-cohort weights: the rate shifts from \getval{pct_coding_benchmark} percent (benchmark) to \getval{pct_coding_gcweight} percent, reflecting the correlation between gender composition and coding-related majors. Small-cell disciplines show wider variation: Arts usage ranges from \getval{pct_arts_low} to \getval{pct_arts_high} percent across weighting schemes, reflecting the small number of Arts respondents.

	\begin{table}[H]{\footnotesize
		\caption{Generative AI Adoption Rates by Weighting Scheme}
		\label{tab:weight_adoption}
		\begin{center}
			\begin{singlespace}
				\begin{tabular}{lcccc}
					\midrule
					& \multicolumn{4}{c}{Weighting Scheme:} \\\cmidrule{2-5}
					& Benchmark & Gender $\times$ Cohort & Race & Unweighted \\
					& (by major) &  & & \\
					& (1) & (2) & (3) & (4) \\ \midrule
					\input{results/weight_adoption}
					\midrule
				\end{tabular}
			\end{singlespace}
		\end{center}
		\begin{singlespace}
			\justify \footnotesize \vspace{-0.5cm}
			\textit{Notes:} This table presents the percent of students who report using generative AI for academic purposes during the semester, across different poststratification weighting schemes. Column 1 shows our benchmark estimates using weights based on major. Column 2 uses weights based on the joint distribution of gender and cohort. Column 3 uses weights based on race/ethnicity. Column 4 shows unweighted estimates. The sample in column 3 excludes students who report a race/ethnicity outside the four categories used for weighting (white, Black, Hispanic, Asian) or who prefer not to say. \par
		\end{singlespace}
		}
	\end{table}

	\begin{table}[H]{\footnotesize
		\caption{Academic Uses of Generative AI by Weighting Scheme}
		\label{tab:weight_tasks}
		\begin{center}
			\begin{singlespace}
				\begin{tabular}{lcccc}
					\midrule
					& \multicolumn{4}{c}{Weighting Scheme:} \\\cmidrule{2-5}
					& Benchmark & Gender $\times$ Cohort & Race & Unweighted \\
					& (by major) &  & & \\
					& (1) & (2) & (3) & (4) \\ \midrule
					\input{results/weight_tasks}
					\midrule
				\end{tabular}
			\end{singlespace}
		\end{center}
		\begin{singlespace}
			\justify \footnotesize \vspace{-0.5cm}
			\textit{Notes:} This table presents the percent of AI users who report using generative AI for each academic task, across different poststratification weighting schemes. Tasks are grouped into augmentation (explaining concepts, generating ideas, editing drafts, and coding assistance) and automation (summarizing texts, finding information, proofreading, writing emails, writing essays, and creating images) based on the combined survey classification described in Appendix~\ref{app:augmentation_survey}. The bottom rows show the average usage rate across tasks within each category. Column definitions follow Appendix~Table~\ref{tab:weight_adoption}. Sample restricted to students who report using AI during the academic semester. \par
		\end{singlespace}
		}
	\end{table}

	\begin{table}[H]{\footnotesize
		\caption{Student Beliefs About AI's Academic Impact by Weighting Scheme}
		\label{tab:weight_beliefs}
		\begin{center}
			\begin{singlespace}
				\begin{tabular}{lcccc}
					\midrule
					& \multicolumn{4}{c}{Weighting Scheme:} \\\cmidrule{2-5}
					& Benchmark & Gender $\times$ Cohort & Race & Unweighted \\
					& (by major) &  & & \\
					& (1) & (2) & (3) & (4) \\ \midrule
					\input{results/weight_beliefs}
					\midrule
				\end{tabular}
			\end{singlespace}
		\end{center}
		\begin{singlespace}
			\justify \footnotesize \vspace{-0.5cm}
			\textit{Notes:} This table presents the percent of students who believe that generative AI ``somewhat improves'' or ``significantly improves'' each aspect of their academic experience, across different poststratification weighting schemes. Students who responded ``I don't know'' are excluded. Column definitions follow Appendix~Table~\ref{tab:weight_adoption}. \par
		\end{singlespace}
		}
	\end{table}

	\clearpage
	\subsection{Field of Study Classifications} \label{app:fields}

	This appendix details the classification of majors into broad fields of study:
	
	\begin{itemize}
		\item \textbf{Natural Sciences:} Includes Biology, Chemistry, Computer Science, Earth and Climate Sciences/Geology, Environmental Studies, Mathematics, Molecular Biology \& Biochemistry, Neuroscience, Physics, and Statistics.

		\item \textbf{Social Sciences:} Includes Anthropology, Economics, Education, Geography, International \& Global Studies, International Politics \& Economics, Political Science, Psychology, and Sociology.

		\item \textbf{Humanities:} Includes American Studies, Architectural Studies, Art History \& Museum Studies, Black Studies, Classical Studies, History, History of Art \& Architecture, Philosophy, and Religion.

		\item \textbf{Literature:} Includes Comparative Literature, English/English \& American Literatures, and Literary Studies.

		\item \textbf{Languages:} Includes Arabic, Chinese, French \& Francophone Studies, German, Japanese Studies, Russian, and Spanish.

		\item \textbf{Arts:} Includes Film \& Media Culture, Music, Studio Art, and Theatre.

		\item \textbf{Has not declared major:} Students who had not yet declared a major at the time of the survey are grouped by their intended field of study, as reported in the survey.
	\end{itemize}

	\clearpage
	\subsection{Task-Specific Use of Generative AI}
	
	The aggregate patterns in Section~\ref{sec:task} may mask heterogeneity across student groups. Appendix~Table~\ref{tab:ai_purpose_correlates} reports regression estimates of how student characteristics predict AI usage for each of our ten academic tasks.
	
	Usage patterns differ by student characteristics. Males show higher adoption across most applications, with the largest gaps for finding information (\getval{pp_male_findinfo} pp; $p < 0.01$), summarizing texts (\getval{pp_male_summarize} pp; $p < 0.01$), and creating images (\getval{pp_male_images} pp; $p < 0.01$). Black students show \getval{pp_black_findinfo} pp higher usage for finding information than white students ($p < 0.05$) and higher adoption of writing assistance tools (\getval{pp_black_editing} pp for editing text, $p < 0.05$; \getval{pp_black_emails} pp for writing emails, $p < 0.10$). Hispanic students show higher usage for generating ideas (\getval{pp_latino_ideas} pp; $p < 0.01$) and writing emails (\getval{pp_latino_emails} pp; $p < 0.01$), while Asian students show higher adoption for writing emails (\getval{pp_asian_emails} pp; $p < 0.01$) and explaining concepts (\getval{pp_asian_explain} pp, $p < 0.05$). Public high school students report lower usage for concept explanation (\getval{pp_public_explain} pp; $p < 0.10$) and writing emails (\getval{pp_public_emails} pp; $p < 0.10$), but higher usage for proofreading (\getval{pp_public_proofread} pp; $p < 0.10$).
	
	Field of study is a strong predictor. Arts majors show lower adoption across multiple tasks than Natural Science students, with gaps of \getval{pp_arts_findinfo} pp for finding information ($p < 0.05$), \getval{pp_arts_ideas} pp for generating ideas ($p < 0.05$), and \getval{pp_arts_coding} pp for coding assistance ($p < 0.01$). Humanities majors are \getval{pp_hum_ideas} pp less likely to use AI for generating ideas ($p < 0.01$) and \getval{pp_hum_coding} pp less likely for coding assistance ($p < 0.01$). Languages majors show the largest differences: \getval{pp_lang_summarize} pp less likely to use AI for summarizing texts ($p < 0.01$), \getval{pp_lang_ideas} pp less likely for generating ideas ($p < 0.01$), and \getval{pp_lang_coding} pp less likely for coding assistance ($p < 0.01$). Social science students show higher usage for summarizing texts (\getval{pp_socsci_summarize} pp; $p < 0.05$) but lower usage for coding assistance (\getval{pp_socsci_coding_abs} pp; $p < 0.01$).

	\begin{landscape}
		\begin{table}[htpb]{\footnotesize
				\begin{center}
					\caption{Student Characteristics Associated with Task-Specific Use of Generative AI} \label{tab:ai_purpose_correlates}
					\newcommand\w{1.2}
					\begin{tabular}{l@{}lR{\w cm}@{}L{0.45cm}R{\w cm}@{}L{0.45cm}R{\w cm}@{}L{0.45cm}R{\w cm}@{}L{0.45cm}R{\w cm}@{}L{0.45cm}R{\w cm}@{}L{0.45cm}R{\w cm}@{}L{0.45cm}R{\w cm}@{}L{0.45cm}R{\w cm}@{}L{0.45cm}R{\w cm}@{}L{0.45cm}}
						\midrule
						&& \multicolumn{18}{c}{=1 if student uses generative AI with any frequency during the academic semester to...} \\\cmidrule{3-22}
						&& Explain && Summ. && Find && Gen. && Proof- && Edit && Write && Code && Write && Create \\
						&& Cnpts.  && Texts && Info. && Ideas && read  && Text && Emails && Assist. && Essays && Images \\
						&& (1) && (2) && (3) && (4) && (5) 	&& (6) && (7) && (8) && (9) && (10) \\
						\midrule
						\input{results/ai_purpose_regs}
						\midrule
						\input{results/ai_purpose_N}
						\midrule
					\end{tabular}
				\end{center}
				\begin{singlespace} \vspace{-.5cm}
					\noindent \justify \textit{Notes:} This table reports estimated associations between student characteristics and use of AI for specific academic tasks. Each column shows the result for a different academic task. The omitted categories are: Natural Sciences (and the small share of students with no declared or intended major) for academic division, white students for race/ethnicity, freshmen for cohort, female for gender, and private high school for school type. Regressions are weighted and use heteroskedasticity-robust standard errors clustered at the student level. $^{***}$, $^{**}$, and $^*$ indicate significance at the 1\%, 5\%, and 10\% levels, respectively. \par
				\end{singlespace}   
			}
		\end{table}
		
	\end{landscape}

	\section{Supplementary Survey on Task Classification} \label{app:augmentation_survey}
	
	\setcounter{table}{0}
	\setcounter{figure}{0}
	\setcounter{equation}{0}	
	\renewcommand{\thetable}{C\arabic{table}}
	\renewcommand{\thefigure}{C\arabic{figure}}
	\renewcommand{\theequation}{C\arabic{equation}}
	
	Our main survey records how frequently students use AI for \getval{n_tasks} academic tasks but does not classify these tasks as augmentation or automation. This classification is essential for the analysis in Section~\ref{sec:task}, in which we examine whether students primarily use AI to complement their own effort or to replace it. To make the distinction, we designed a supplementary survey to elicit independent classifications from undergraduate students and college instructors outside our main sample. This appendix describes the survey design, reports the classification results, and examines their robustness.

	\subsection{Sample and Recruitment}

	We recruited respondents through Prolific in \getval{prolific_date}. The student sample has \getval{n_prolific_students} enrolled U.S.\ undergraduates; the instructor sample has \getval{n_prolific_instructors} college instructors, including faculty, lecturers, and teaching assistants. We paid respondents \$\getval{usd_prolific_pay} for a survey that took about \getval{min_prolific_duration} minutes. Respondents had to be U.S.-based and either currently enrolled as an undergraduate or currently teaching at the college level.

	\subsection{Survey Design}

	The survey presented respondents with the same ten academic tasks measured in our Middlebury survey: explaining concepts, summarizing texts, finding information, generating ideas, proofreading, editing drafts, composing emails, coding assistance, writing essays, and creating images. We asked respondents to imagine working on a college assignment and to evaluate how using AI would change the nature of each task (see Appendix~\ref{app:classification_instrument} for survey instrument).

	We elicited classifications along multiple dimensions, each capturing a different aspect of how AI interacts with student effort and learning. All respondents evaluated three classification dimensions---for/with, effort, and learning. The student sample additionally rated the importance of each task. To mitigate order effects and respondent fatigue, we randomized the order in which dimensions appeared: the three classification dimensions were presented in random order relative to each other, and in the student survey the importance block appeared either before or after them.

	The \textit{for/with} dimension is our primary classification variable. For each task, respondents indicated whether AI acts more as a collaborator that ``works with'' the student (augmentation) or automates the task by ``doing it for'' the student (automation). This binary classification maps directly onto the augmentation--automation framework in Section~\ref{sec:task}.

	The \textit{effort} dimension asked respondents to assess whether AI use reduces, increases, or has no effect on the cognitive effort required for each task. This three-point scale captures whether AI substitutes for or complements student effort, independent of how that effort maps onto learning.

	The \textit{learning} dimension asked whether AI use for each task increases, decreases, or has no effect on student learning. This dimension captures respondents' beliefs about the educational consequences of AI use, which may differ from their beliefs about effort. A task could reduce effort while increasing learning (e.g., AI explains a concept the student would have struggled with alone) or reduce both (e.g., AI writes an essay the student never engages with).

	We asked only the student sample to rate the \textit{importance} of each task for their own learning on a \getval{scale_importance_low}--\getval{scale_importance_high} slider scale, since instructors may weigh task importance differently from students.

	Each dimension's block also included a ranking task in which respondents ordered all \getval{n_tasks} tasks from one end of the dimension to the other (e.g., from ``most works with'' to ``most does for''). These rankings provide an ordinal measure that complements the categorical classifications. The survey concluded with questions about respondents' own AI usage frequency, both overall and for each of the \getval{n_tasks} tasks.

	\subsection{Main Classification Results}

	We classify each task using the combined modal response across both samples. For each task, we compute the share of pooled respondents (students and instructors) who classify AI as ``working with'' versus ``doing for'' the student, and assign the task to the majority category.

	Appendix~Table~\ref{tab:class_forwith} presents the results. The combined classification categorizes \getval{n_aug_tasks} tasks as augmentation---explaining concepts, generating ideas, editing essays, and coding help---and \getval{n_auto_tasks} tasks as automation---summarizing texts, finding information, proofreading, writing emails, writing essays, and creating images. Students and instructors agree on the classification of \getval{n_tasks_agreed} out of \getval{n_tasks} tasks. The two samples disagree on two tasks: students classify both finding information and editing essays as augmentation, while instructors classify both as automation. In both cases, the combined classification follows the pooled majority.

\begin{figure}[H]
		\caption{Student vs.\ Instructor For/With Rankings}\label{fig:class_rankings}
		\centering
		\includegraphics[width=.7\linewidth]{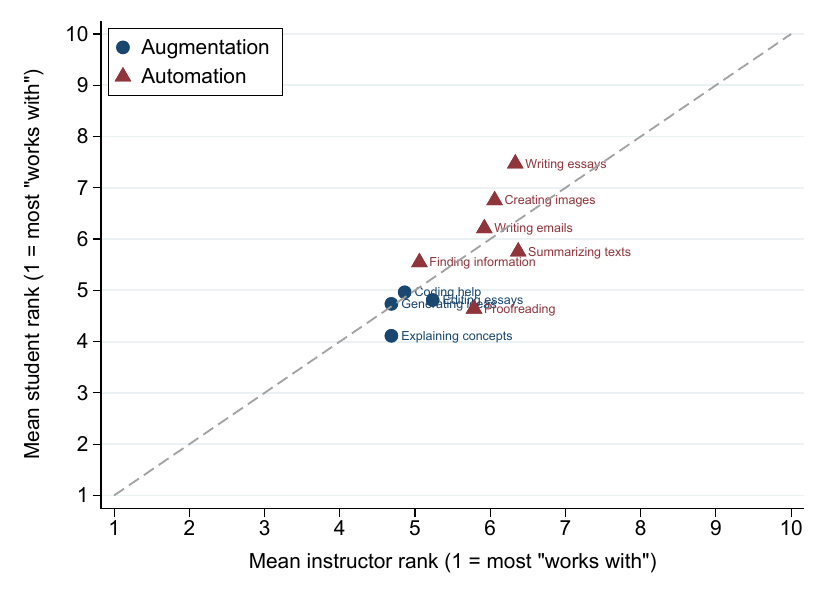}
		{\footnotesize \singlespacing \justify

			\textit{Notes:} This scatterplot shows mean for/with rankings from students ($y$-axis) and instructors ($x$-axis) for each of the ten academic tasks. Rankings range from 1 (most ``works with'') to 10 (most ``does for''). Circles denote tasks classified as augmentation in the combined survey classification; triangles denote automation tasks. The dashed line is the 45-degree line. \par
		}
	\end{figure}

	Appendix~Figure~\ref{fig:class_rankings} plots mean for/with rankings from students against those from instructors. Tasks cluster near the 45-degree line, confirming broad agreement on which tasks are more collaborative versus more automated. The correlation between rankings is high, with explaining concepts and generating ideas consistently ranked as the most ``with'' tasks and writing essays and creating images as the most ``for'' tasks.

	\begin{table}[H]{\footnotesize
		\caption{For/With Classification: Students vs.\ Instructors} \label{tab:class_forwith}
		\begin{center}
			\begin{tabular}{lccccccc}
				\midrule
				& \multicolumn{2}{c}{Students} & \multicolumn{2}{c}{Instructors} & \multicolumn{2}{c}{Overall} & Survey \\
				\cmidrule(lr){2-3} \cmidrule(lr){4-5} \cmidrule(lr){6-7}
				Task & \% With & \% For & \% With & \% For & \% With & \% For & Classification \\
				\midrule
				\input{results/class_forwith_table}
				\midrule
			\end{tabular}
		\end{center}
		\begin{singlespace}
			\justify \footnotesize \vspace{-1cm}
			\textit{Notes:} This table shows the percentage of respondents who classify AI as ``working with'' or ``doing for'' the student for each academic task. The student sample includes 56 undergraduates and the instructor sample includes 51 instructors at U.S.\ colleges and universities. The ``Survey'' column shows the combined modal classification (averaging across both samples). \par
		\end{singlespace}
		}
	\end{table}

	\subsection{Additional Dimensions: Effort, Learning, and Importance}

	The effort and learning dimensions provide evidence that is broadly consistent with the for/with classification. Appendix~Table~\ref{tab:class_effort} shows that for most tasks, both students and instructors report that AI reduces cognitive effort. The reduction is largest for tasks classified as automation: large majorities of respondents report that AI reduces the effort required for proofreading, writing emails, and creating images. Among augmentation tasks, the picture is more mixed. Explaining concepts and generating ideas show smaller effort reductions, and a nontrivial share of respondents report that AI increases the effort required for coding help---consistent with the view that debugging and understanding code with AI requires active engagement.

	Appendix~Table~\ref{tab:class_learning} presents results for the learning dimension. Respondents generally view augmentation tasks as more likely to increase learning than automation tasks. Explaining concepts stands out: a majority of both students and instructors report that AI use increases learning for this task. By contrast, respondents most often rate automation tasks---such as writing emails, proofreading, and creating images---as having no effect on learning. Writing essays is the task respondents most often rate as decreasing learning, consistent with the concern that outsourcing core writing undermines skill development.

	Appendix~Table~\ref{tab:class_importance} shows that students rate augmentation tasks as more important to their learning than automation tasks. Explaining concepts, generating ideas, and coding help receive the highest mean importance ratings, while creating images and writing emails receive the lowest. This pattern is consistent with students valuing tasks in which AI complements their effort over those in which AI substitutes for it.

	\begin{table}[H]{\footnotesize
		\caption{Effort Dimension: Students vs.\ Instructors} \label{tab:class_effort}
		\begin{center}
			\begin{tabular}{llcccccc}
				\midrule
				& & \multicolumn{3}{c}{Students} & \multicolumn{3}{c}{Instructors} \\
				\cmidrule(lr){3-5} \cmidrule(lr){6-8}
				Task & Classification &  $\downarrow$ Effort & No effect & $\uparrow$ Effort & $\downarrow$ Effort & No effect  & $\uparrow$ Effort \\
				\midrule
				\input{results/class_effort_table}
				\midrule
			\end{tabular}
		\end{center}
		\begin{singlespace}
			\justify \footnotesize \vspace{-0.5cm}
			\textit{Notes:} This table shows the percentage of respondents who report that AI use for each task reduces, has no effect on, or increases the student's cognitive effort. Classification column reflects the combined survey classification. \par
		\end{singlespace}
		}
	\end{table}

	\begin{table}[H]{\footnotesize
		\caption{Learning Dimension: Students vs.\ Instructors} \label{tab:class_learning}
		\begin{center}
			\begin{tabular}{llcccccc}
				\midrule
				& & \multicolumn{3}{c}{Students} & \multicolumn{3}{c}{Instructors} \\
				\cmidrule(lr){3-5} \cmidrule(lr){6-8}
				Task & Class. & $\downarrow$ Learning & No effect & $\uparrow$ Learning & $\downarrow$ Learning & No effect &  $\uparrow$ Learning \\
				\midrule
				\input{results/class_learning_table}
				\midrule
			\end{tabular}
		\end{center}
		\begin{singlespace}
			\justify \footnotesize \vspace{-0.5cm}
			\textit{Notes:} This table shows the percentage of respondents who report that AI use for each task decreases, has no effect on, or increases student learning. Classification column reflects the combined survey classification. \par
		\end{singlespace}
		}
	\end{table}

	\begin{table}[H]{\footnotesize
		\caption{Task Importance: Student Sample} \label{tab:class_importance}
		\begin{center}
			\begin{tabular}{llcc}
				\midrule
				Task & Classification & Mean importance  & Mean rank \\
				  &    & (0--100) & (1--10) \\				\midrule
				\input{results/class_importance_table}
				\midrule
			\end{tabular}
		\end{center}
		\begin{singlespace}
			\justify \footnotesize \vspace{-0.5cm}
			\textit{Notes:} This table shows mean importance ratings (0--100 slider) and mean importance rankings for each academic task, from the student sample only. Tasks are sorted by mean rank. Classification column reflects the combined survey classification. Summary rows show averages across augmentation and automation tasks. \par
		\end{singlespace}
		}
	\end{table}

	\subsection{Robustness: AI Users Only}

	Students and instructors who rarely or never use AI may classify tasks differently from those with firsthand experience. Appendix~Table~\ref{tab:class_aiusers} restricts the classification sample to respondents who report using AI at least occasionally. This restriction changes only one classification: finding information flips from automation to augmentation, consistent with AI users having a more interactive experience with this task. All other classifications remain unchanged.

	\begin{table}[H]{\footnotesize
		\caption{For/With Classification: AI Users Only} \label{tab:class_aiusers}
		\begin{center}
			\begin{tabular}{lcccccc}
				\midrule
				& \multicolumn{2}{c}{Overall (full)} & \multicolumn{2}{c}{Overall (AI users)} & Full & AI users \\
				\cmidrule(lr){2-3} \cmidrule(lr){4-5}
				Task & \% With & \% For & \% With & \% For & Classification & Classification \\
				\midrule
				\input{results/class_forwith_aiusers_table}
				\midrule
			\end{tabular}
		\end{center}
		\begin{singlespace}
			\justify \footnotesize \vspace{-0.5cm}
			\textit{Notes:} This table replicates the classification analysis restricting to respondents who report using AI more than rarely. The first two columns show the overall (student and instructor combined) percentages from the full sample; the next two columns show the corresponding percentages among AI users only. Tasks in bold indicate changes in classification relative to the full sample. \par
		\end{singlespace}
		}
	\end{table}

	\clearpage
	\section{Qualitative Evidence on Student Perspectives on AI Use} \label{app:open_motiv}
	
	\setcounter{table}{0}
	\setcounter{figure}{0}
	\setcounter{equation}{0}	
	\renewcommand{\thetable}{D\arabic{table}}
	\renewcommand{\thefigure}{D\arabic{figure}}
	\renewcommand{\theequation}{D\arabic{equation}}
	
\begin{figure}[H]
		\caption{Word Cloud of Student Motivations for Generative AI Use}\label{fig:ai_wordcloud_motivation}
		\centering
		\includegraphics[width=.8\linewidth]{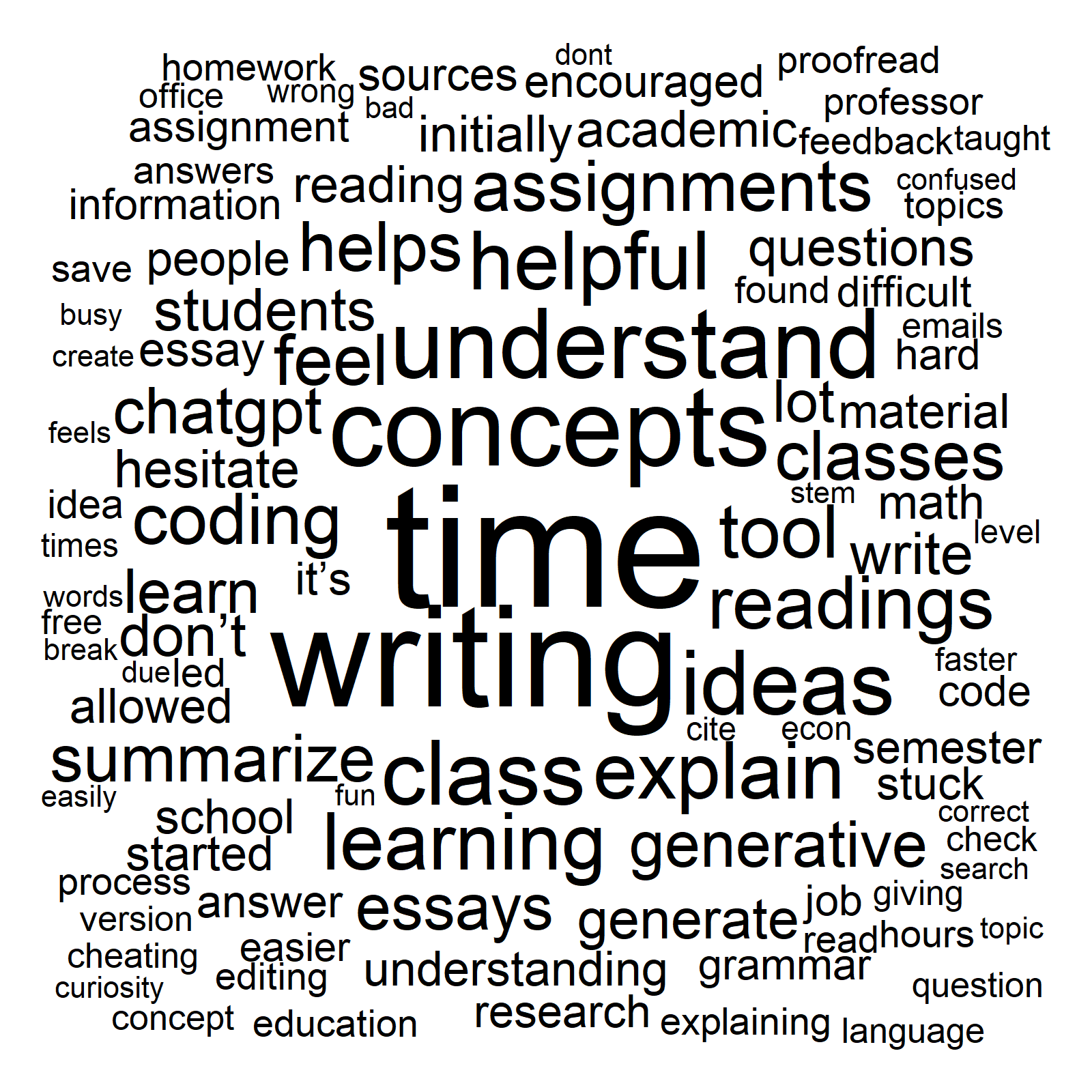}
		{
			\footnotesize \singlespacing \justify
			
			\textit{Notes:} Word cloud displaying words that appear at least five times in \getval{n_openend_motiv} student responses after removing common English stop words and the word ``AI''. Text size is proportional to word frequency. The visualization is based on responses to the question: ``Please describe the factors that have personally influenced your use of generative AI in your academic work. What initially led you to try it, what has motivated you to use it or caused you to hesitate?' \par
		}
	\end{figure}

	This section analyzes student responses to an open-ended question about their use of generative AI: ``Please describe the factors that have personally influenced your use of generative AI in your academic work. What initially led you to try it, what has motivated you to use it or caused you to hesitate?'' \getval{pct_openend_response_motiv} percent of respondents answered this optional question. Appendix~Figure~\ref{fig:ai_wordcloud_motivation} presents a word cloud of the most frequent words in the responses.

	\subsection{Validating the Open-Ended Response Measure} 
	
\begin{figure}[H]
		\caption{Relationship Between AI Sentiment and AI Adoption}\label{fig:vader}
		\centering
		\begin{subfigure}[t]{.48\textwidth}
			\caption*{Panel A. Ever Used Generative AI}
			\centering
			\includegraphics[width=\linewidth]{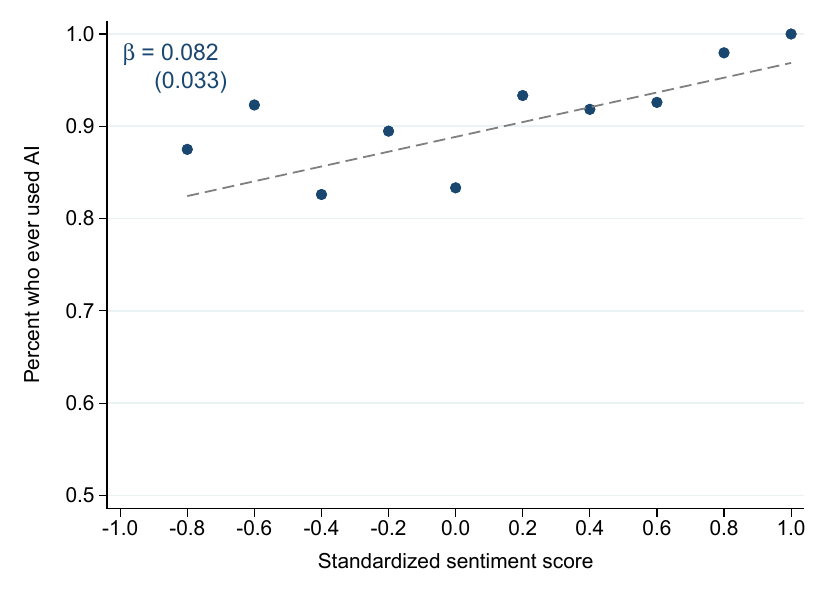}
		\end{subfigure}
		\hfill		
		\begin{subfigure}[t]{0.48\textwidth}
			\caption*{Panel B. Uses AI for Academic Purposes}
			\centering
			\includegraphics[width=\linewidth]{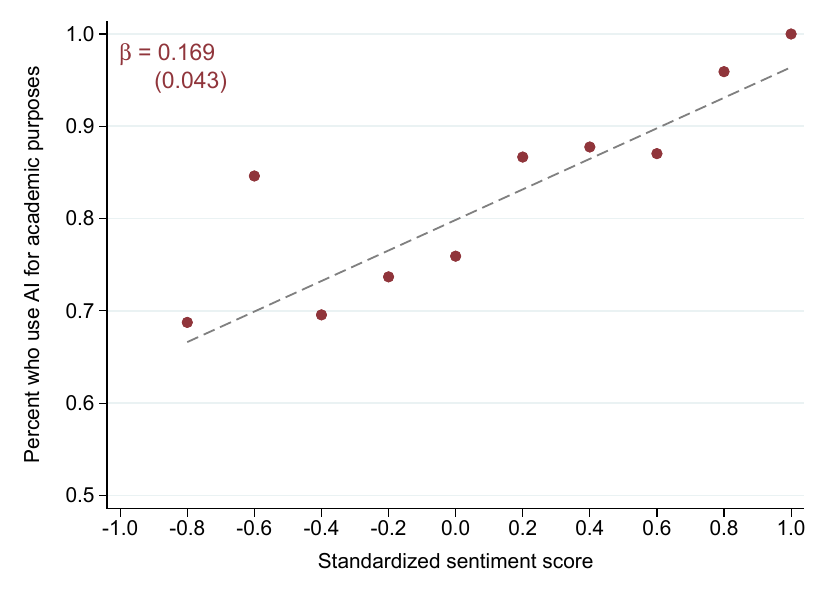}
		\end{subfigure}	
		{\footnotesize
			\singlespacing \justify
			
			\textit{Notes:} This figure presents the relationship between AI sentiment and AI adoption. Panel A shows the proportion of respondents who have ever used generative AI, while Panel B shows the proportion who use AI for academic purposes. Each point represents the mean adoption rate for respondents within sentiment score bins of width 0.2. Sentiment scores are standardized compound scores computed using \cite{hutto2014vader}'s VADER algorithm applied to responses to an open-ended question about generative AI. Positive values indicate positive sentiment and negative values indicate negative sentiment. The dashed lines show OLS best-fit lines estimated on the microdata, with coefficients and standard errors (in parentheses) displayed in the top-left corner of each panel. \par
		}
	\end{figure}

	We first validate our open-ended response measure. We use VADER (Valence Aware Dictionary and sEntiment Reasoner), a lexicon-based sentiment analysis tool \citep{hutto2014vader}, and test whether sentiment scores correlate with actual AI adoption. If the responses capture real attitudes, students with more positive sentiment should adopt AI at higher rates. We test this prediction using two measures: whether students have ever used generative AI, and whether they currently use AI for academic purposes. Appendix~Figure~\ref{fig:vader} presents binned scatterplots of AI adoption against standardized sentiment scores.
	
	Sentiment toward generative AI strongly predicts adoption. Students with negative sentiment scores show adoption rates of \getval{pct_vader_neg_ever_low}--\getval{pct_vader_neg_ever_high} percent for ever using AI and \getval{pct_vader_neg_acad_low}--\getval{pct_vader_neg_acad_high} percent for academic use, while those with positive scores reach nearly \getval{pct_vader_pos_ever} percent for general use and \getval{pct_vader_pos_acad_low}--\getval{pct_vader_pos_acad_high} percent for academic purposes. The relationship is stronger for academic AI use (Panel B), where $\hat{\beta} = \getval{beta_vader_acad}$ is twice as large as for general adoption.\footnote{This stronger association is expected: our open-ended question asked about academic AI use, so sentiment scores are most relevant in that domain.} The open-ended responses thus capture meaningful variation in attitudes that maps to behavior.
	
	\subsection{How Students use Generative AI} 
	
\begin{figure}[H]
		\caption{Frequency of Keywords in Student Motivations for AI Use}\label{fig:ai_keywords_motiv}
		\centering
		\includegraphics[width=.8\linewidth]{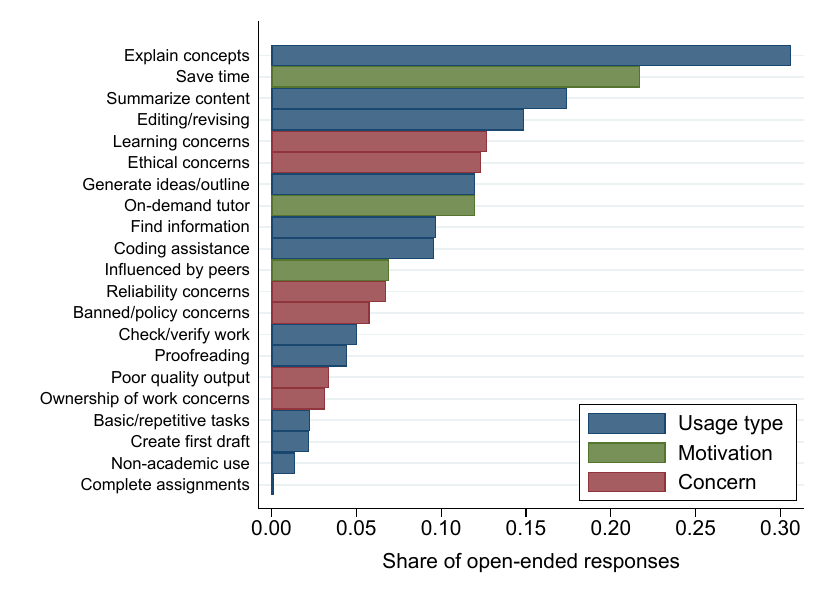}
		{
			\footnotesize \singlespacing \justify
			
			\textit{Notes:} The figure shows the share of open-ended responses that mentioned different themes related to AI use. The responses come from the question ``Please describe the factors that have personally influenced your use of generative AI in your academic work. What initially led you to try it, what has motivated you to use it or caused you to hesitate?'' Color coding indicates the category of each theme. Usage type refers to how students use AI tools. Motivation captures what drove students to try AI. Concerns include mentions of course policies and academic integrity, individual reservations about AI use, worries about AI's impact on education, and AI's technical limitations. \par
		}
		
	\end{figure}

	To analyze these responses systematically, we classified each one using keywords. For example, if a student mentioned saving time, we tagged the response as ``time-saver''; if a student expressed concern about learning, we tagged it as ``negative learning.'' We allowed multiple keywords per response. Appendix~Figure~\ref{fig:ai_keywords_motiv} shows the frequency of keywords. The responses reveal how students use AI, what motivates them, and what causes some to hesitate.
	
	The most common use is as an explanatory tool. Nearly \getval{pct_openend_explain} percent of responses mentioned using AI to understand course material, asking it to break down concepts from readings and lectures. One student reported: ``I can ask AI to explain concepts to me that I have a hard time grasping. [...] I can keep asking 'simplify' or 'break down even more.''' Students also use AI to summarize dense readings, which they say helps manage heavy course loads.
	
	Students use AI throughout different stages of the writing process. Some use AI to generate initial drafts that serve as starting points. One student explained: ``Helps me get started with a base for most of my essays. It feels easier to edit something already written and make it my own than to write from scratch.'' Others use AI more narrowly for brainstorming when stuck on specific problems. As one student noted: ``I use it if I am feeling stuck to push me to the right direction (whether a mathematical problem or an essay idea).'' Many also report using AI as an editing tool to improve grammar, sentence structure, and overall writing flow. This editing use is especially common among non-native English speakers. As one student explained: ``English is not my first language and it frustrates me sometimes that I cannot find the best way to phrase a certain idea and AI is a useful tool to have to find alternate expressions.''
	
	Students also mentioned task-specific uses. In coding-intensive courses, students use AI for debugging and understanding programming concepts. Others use it for administrative tasks like formatting citations and drafting emails. Many treat AI as an enhanced search engine: ``It has significantly reduced the time it takes to conduct research on new topics and ideas, and helps me by giving me a thorough selection of sources to use for projects of any kind.''
	
	\subsection{Why Students Adopt AI}
	
	Saving time was the most cited reason for adoption. Nearly \getval{pct_openend_timesaving} percent of responses mentioned using AI to complete work more efficiently, viewing it as a way to manage demanding course loads. Students particularly embrace AI for mechanical or administrative tasks---one noted using it for ``Writing emails quick and creating resume/ cover letter templates.'' But AI assistance extends beyond grunt work. Some use it to ``spend less time doing assignments and homework,'' especially when they view the work as peripheral: ``when I come across work I deem as ineffective, I want to spend as little time as possible doing it.''
	
	Having an on-demand tutor was another key motivation. Around \getval{pct_openend_tutor} percent of responses described using AI as a ``tutor'' when other resources were unavailable. As one student noted, ``I use it as a last resort (if there are no office hours, after looking up videos, etc.) if I need extra help. I'd like to think that the way I use it is similar to going to office hours or TA hours.''
	
	Peer influence also drove AI adoption. Some students reported feeling pressure to use AI to remain competitive with their classmates. One student explained: ``I noticed others use it, are getting better grades than me, and they say they learn better with the help of AI, so I gave it a try.'' Others worried about being at a competitive disadvantage: ``Other people were using it and told me about it. I felt like I would be at a disadvantage if I wasn't also using it.''
	
	\subsection{Concerns and Limitations}
	
	The most frequent concern was negative impacts on learning. One student who initially used AI extensively reported: ``In the past, I have simply plugged and chugged homework assignments into ChatGPT and submitted it. Those assignments feedback from teachers was positive and I was getting good grades, but I definitely felt that my own learning outcomes to be significantly worse.'' Others viewed AI use as fundamentally incompatible with their educational goals: ``my task is as a humanities student is to think, not calculate; why should I let AI do the thinking for me? It would defeat the purpose of pursuing my education.''
	
	Many students described ethical uncertainty about where to draw the line. One noted: ``I never use it to explicitly write something because that feels like overt cheating, but sometimes I hesitate when it completely solves Econ problems. I understand how it does it, and it helps me to learn, but it still sometimes feels a little morally gray.'' Another expressed similar ambivalence: ``I tend to only use it when [it] will save me time in a moral way.'' Closely related was the desire to maintain ownership of their work: ``I don't have interest in using generative AI for my academic work because I want my work to reflect my own ideas.'' Another noted: ``It usually would not even occur to me to turn to AI to substitute writing because I want to take credit for my work, and using AI seems to diminish that.''
	
	Technical limitations also deterred some students. Concerns about inaccurate outputs (``hallucinations'') and poor quality were common, especially for creative or complex analytical work. One student noted: ``In my poetry class we were instructed to use it to come up with poems and they were awful, so that kinda turned me away from using it to do my work for me.''
	
	\subsection{Discussion}
	
	Two overarching themes emerge from the responses.

	First, students vary substantially in how they incorporate AI into their academic lives, and this heterogeneity depends largely on what they perceive as ``appropriate'' uses. For tasks they view as core---writing essays, solving problems---many hesitate. One student put it clearly: ``Most of my work is writing or reading. If I'm not doing the writing, what is the purpose of me taking the class?'' Yet students draw different boundaries between central tasks and grunt work. Some use AI extensively, viewing their role as a manager who provides direction while AI handles implementation. Others restrict it to brainstorming, editing, or drafting emails. Still others avoid AI entirely for academic work, often for ethical reasons. Even among users, adoption patterns reflect individual trade-offs between time savings, learning goals, and integrity concerns.

	Second, efficiency and learning are in tension. The time-saving benefits are easy to observe and quantify. But they are unlikely to be a free lunch: some may come at the cost of spending less time with material that requires deeper engagement. As one student noted: ``There may be a negative effect in that it eliminates much of the 'struggle' in learning.'' Yet having an on-demand tutor can also improve learning: ``It can explain concepts to me in a way that is tailored to my learning style.'' This tension suggests that AI's impact on learning depends not on whether students use it---almost all do---but on how they use it.
	

	\clearpage
	\section{Qualitative Evidence on Student Views of AI Policies} \label{app:open_policy}
	
	\setcounter{table}{0}
	\setcounter{figure}{0}
	\setcounter{equation}{0}	
	\renewcommand{\thetable}{E\arabic{table}}
	\renewcommand{\thefigure}{E\arabic{figure}}
	\renewcommand{\theequation}{E\arabic{equation}}
	
\begin{figure}[H]
		\caption{Word Cloud of Student Feedback on Generative AI Policies}\label{fig:ai_wordcloud_policy}
		\centering
		\includegraphics[width=.8\linewidth]{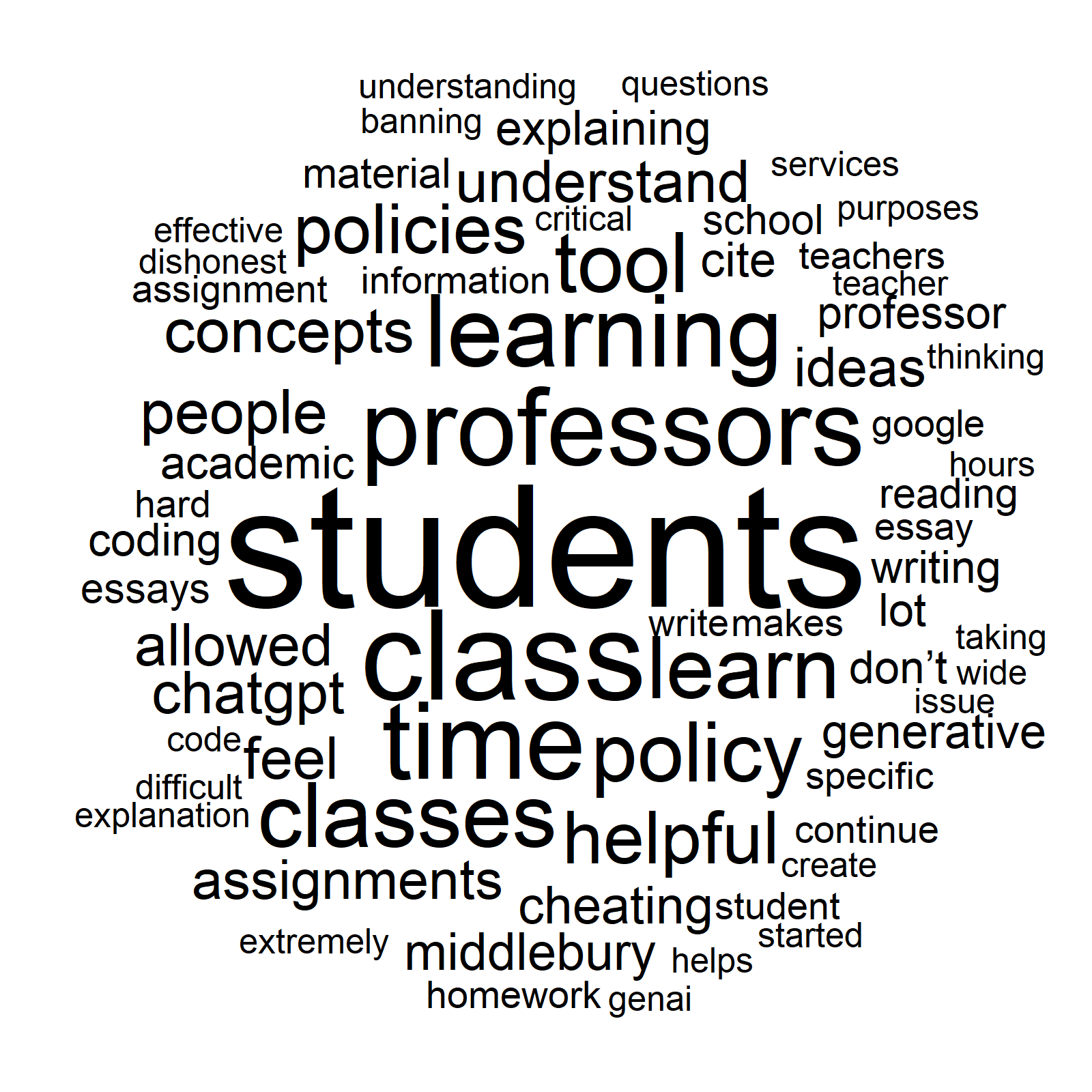}
		{
			\footnotesize \singlespacing \justify
			
			\textit{Notes:} Word cloud displaying words that appear at least five times in \getval{n_openend_policy} student responses after removing common English stop words and the word ``AI''. Text size is proportional to word frequency. The visualization is based on responses to the question: ``Do you have any specific feedback or suggestions about Middlebury's generative AI policies, resources, or support services?'' \par
		}
	\end{figure}

\begin{figure}[H]
		\caption{Frequency of Keywords in Student Feedback on AI Policies}\label{fig:ai_keywords_policy}
		\centering
		\includegraphics[width=.8\linewidth]{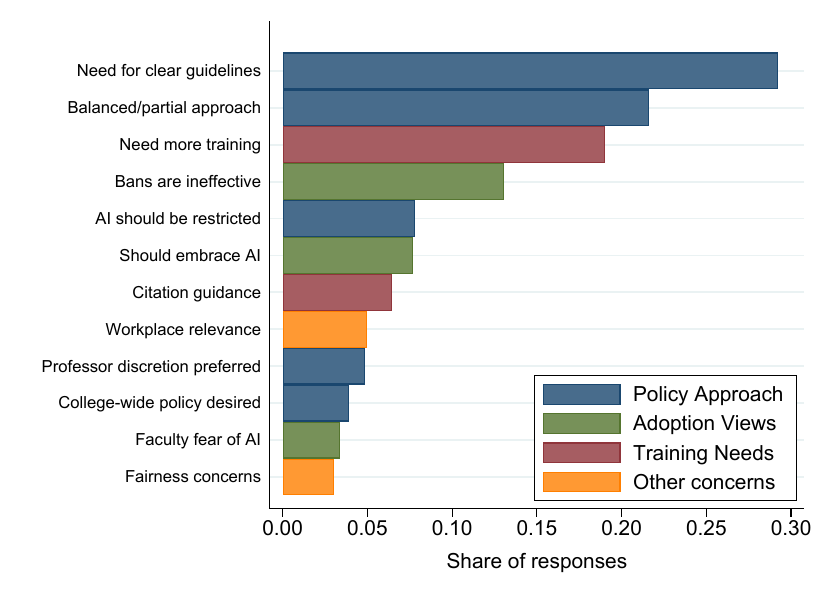}
		{
			\footnotesize \singlespacing \justify
			
			\textit{Notes:} The figure shows the share of open-ended responses that mentioned different themes related to Middlebury's AI policies. The responses come from the question ``Do you have any specific feedback or suggestions about Middlebury's generative AI policies, resources, or support services?'' Color coding indicates the category of each theme. Policy Approach captures suggestions about how AI should be regulated at the college. Adoption Views reflect positions on whether and how AI should be integrated into academic work. Training Needs indicates requests for guidance and support. Other Concerns include issues of workplace relevance and fairness. \par
		}
		
	\end{figure}

	This section analyzes student responses to an open-ended question about Middlebury's AI policies: ``Do you have any specific feedback or suggestions about Middlebury's generative AI policies, resources, or support services?'' Appendix~Figure~\ref{fig:ai_wordcloud_policy} presents a word cloud of the most frequent words; Appendix~Figure~\ref{fig:ai_keywords_policy} shows keyword frequencies.

	\subsection{Polarized Views on Generative AI Policy Approaches}
	
	Students expressed sharply different views about AI policies. Some advocated embracing the technology: ``The tool is there, there is supply and there is demand. Don't fight another war on drugs. Don't live in a fake reality.'' Others called for restrictions, arguing that ``the use of generative AI is dishonest and corrosive'' and that it ``prohibits these organic processes and divorces students from true learning.''
	
	The most common position, however, called for a balanced approach. Students distinguished between uses that enhance learning (concept explanation) and those that substitute for it (generating entire essays). One student wrote: ``AI also can really be helpful at explaining a textbook problem that doesn't make sense, or guiding slightly with homework, or creating study materials, or editing/tightening up your prose. All of those things are good, and universities should figure out how to maximize AI use for those reasons and to minimize students just feeding their problem sets into ChatGPT.''
	
	A recurring theme was the perceived futility of blanket bans. One student noted: ``I don't think anyone really cares what the policy of any given class is. If professors want people to not use it, they need to structure assessments in a way that will discourage use.'' Another compared AI bans to restricting internet use: ``AI policies seem to be totally irrelevant. It's like telling people they can't use the internet as a resource for the class.'' The perceived ineffectiveness of bans creates fairness concerns: ``I think if it is banned in a class, that should be enforced (and right now it absolutely is not)... As with any form of cheating, those who don't cheat are put at a disadvantage.''
	
	\subsection{Need for Clear Guidelines}
	
	Students repeatedly asked for clarity. Many said they did not know what counted as acceptable use in a given course: ``I think it should be more clear whether we can use it and how and how to cite it since most professors rarely mention it at all.'' Beyond knowing the rules, students wanted to understand why they existed: ``I think that Professor's should be very specific about what is allowed and their reasoning behind their policy.''

	A related frustration was inconsistency across courses. As one student put it: ``Sometimes its confusing when one class allows it and another doesn't and the other encourages it and so on so if there was a school wide or department wide policy that could help.'' Others pushed back against uniform rules, noting that AI is more useful in some fields than others: ``GenAI is more effective in some classes/majors than others. Making sure professors understand how students use GenAI and how useful GenAI is in their class (given the course structure, nature of assignments/material, etc.) is very important for the class policy.''
	
	\subsection{Training and Support Services}
	
	Many students called for formal training on generative AI use. One proposed ``a workshop that teaches you to effectively use GenAI without violating the honor code.'' Others wanted help distinguishing productive from counterproductive uses: ``I think it could be useful to develop some sort of training. How do we use AI in a way that actually benefits our learning? I tried out some things on my own but I feel that I need more guidance.'' The underlying concern was not whether to use AI, but how to use it well.

	Students frequently connected this desire for training to workplace preparedness. One noted that ``As the world uses more and more AI, I think it is an important tool that students should know how to leverage.'' Another framed the issue more bluntly: ``The moment us students leave campus, we will be using it in the professional world, and when used in combination with one's own skills, it is merely a tool to maximize efficiency.'' For many respondents, learning to use AI well is not a nice-to-have but something their college education should prepare them for.

	\subsection{Discussion}

	The picture that emerges from these \getval{n_openend_policy} responses is one of unresolved contradictions. Most students have already made AI part of their workflow, using it for brainstorming, editing, and concept clarification. They distinguish that use from having ChatGPT write an essay, and they want policies that reflect the distinction. Blanket bans, in their view, do not. As one student noted, such bans ``seem to be totally irrelevant'' when enforcement is inconsistent and compliance puts rule-followers at a disadvantage. Students also disagree about who should set the rules. Some want a campus-wide standard so they know what to expect; others argue that an economics professor and a creative writing instructor face such different questions about AI that a uniform rule makes no sense. What nearly everyone agrees on is that the status quo fails students: policies vary by course, often go unstated, and rarely come with any explanation of the reasoning behind them.

	\clearpage

	\setcounter{table}{0}
	\setcounter{figure}{0}
	\setcounter{equation}{0}	
	\renewcommand{\thetable}{F\arabic{table}}
	\renewcommand{\thefigure}{F\arabic{figure}}
	\renewcommand{\theequation}{F\arabic{equation}}
	
	\clearpage			
	\includepdf[scale=1,pages=1,pagecommand=\section{Survey Instrument} \label{app:survey}]{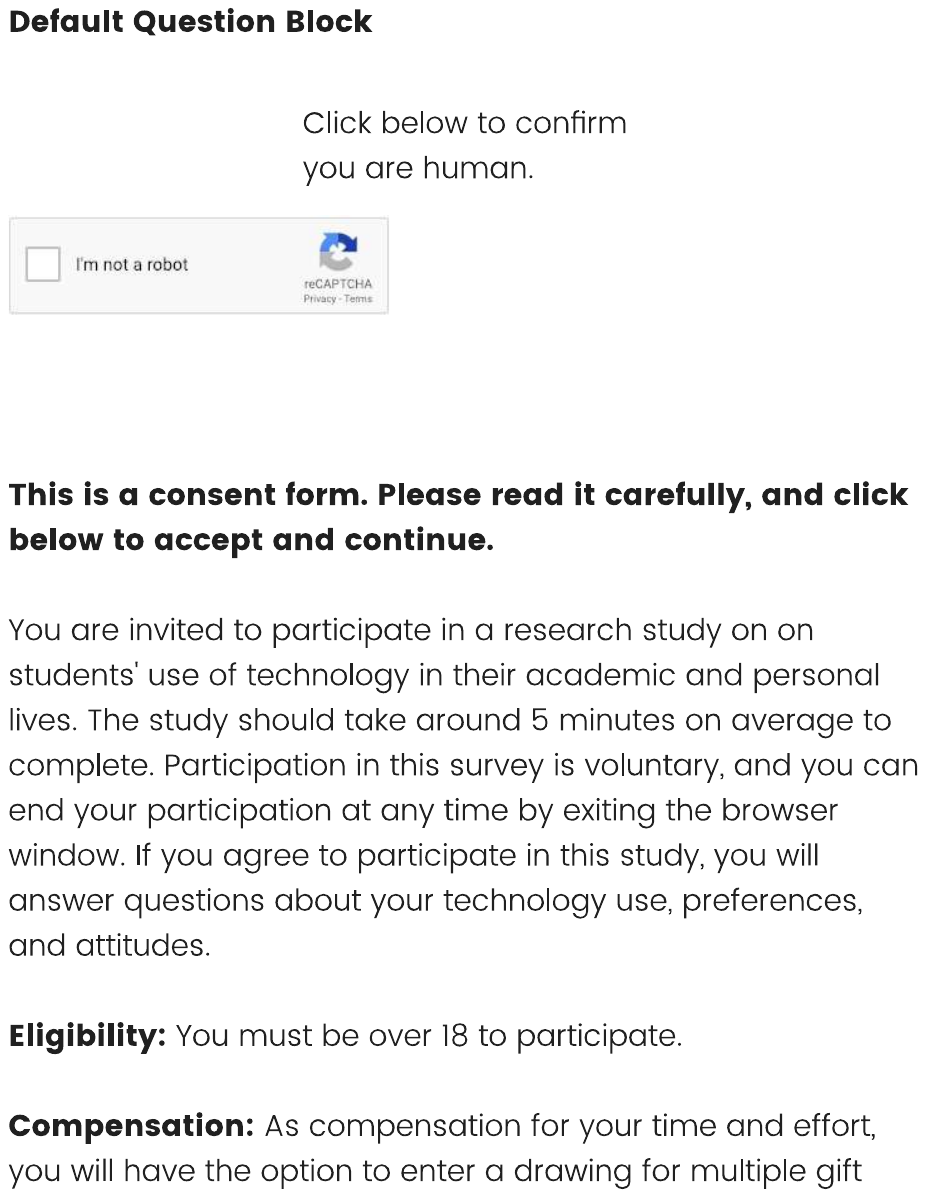}
	\includepdf[pages=2-, pagecommand={\thispagestyle{plain}}]{assets/ai-survey-midd.pdf}

	\setcounter{table}{0}
	\setcounter{figure}{0}
	\setcounter{equation}{0}
	\renewcommand{\thetable}{G\arabic{table}}
	\renewcommand{\thefigure}{G\arabic{figure}}
	\renewcommand{\theequation}{G\arabic{equation}}

	\clearpage
	\includepdf[scale=1,pages=1,pagecommand={\section{Classification Survey Instrument} \label{app:classification_instrument}}]{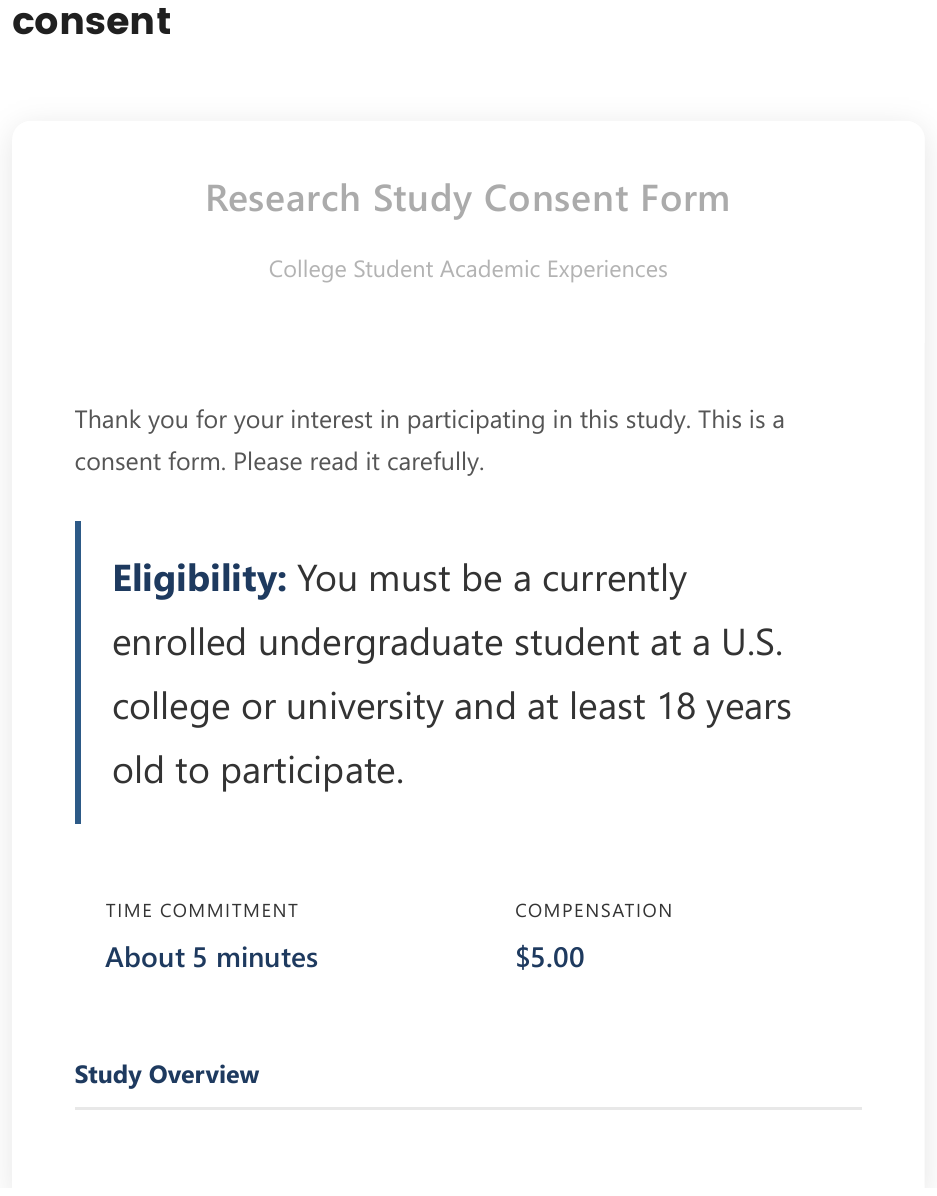}
	\includepdf[pages=2-, pagecommand={\thispagestyle{plain}}]{assets/classification-survey-instrument.pdf}

\end{document}

%% file: results/summ_dem.tex
Male &0.446 &0.433 &0.463 \\
Female &0.508 &0.516 &0.533 \\
White &0.618 &0.603 &0.538 \\
Black &0.036 &0.033 &0.052 \\
Hispanic &0.099 &0.106 &0.124 \\
Asian &0.155 &0.162 &0.073 \\
Private high school &0.420 &0.399 &-- \\
Public high school &0.543 &0.556 &-- \\

%% file: results/summ_aca.tex
GPA &3.740 &3.736 &3.670 \\
Hours spent on academics per week &17.899 &17.889 &-- \\
Freshman &0.311 &0.355 &0.255 \\
Sophomore &0.273 &0.272 &0.257 \\
Junior &0.202 &0.179 &0.201 \\
Senior &0.213 &0.194 &0.287 \\

%% file: results/summ_maj.tex
Arts &0.017 &0.028 &0.023 \\
Humanities &0.071 &0.091 &0.073 \\
Languages &0.025 &0.025 &0.025 \\
Literature &0.047 &0.040 &0.025 \\
Natural Sciences &0.336 &0.390 &0.244 \\
Social Sciences &0.468 &0.385 &0.243 \\
Has not declared major &0.035 &0.041 &0.364 \\

%% file: results/summ_n.tex
N (\# degrees) &43 &43 &49 \\
N (\# students) &634 &2,760 &2,760 \\

%% file: results/ai_usage_regs.tex
Male&&0.103&\sym{***}&0.164&\sym{***}&0.205&\sym{***}&0.092&\sym{***}\\
&&(0.033&)&(0.043&)&(0.043&)&(0.030&)\\
Black&&0.136&\sym{***}&0.222&\sym{**}&0.186&&0.026&\\
&&(0.047&)&(0.095&)&(0.117&)&(0.076&)\\
Latino&&$-$0.004&&0.051&&$-$0.026&&$-$0.023&\\
&&(0.060&)&(0.073&)&(0.065&)&(0.038&)\\
Asian&&0.114&\sym{***}&0.173&\sym{***}&0.097&&0.052&\\
&&(0.038&)&(0.056&)&(0.059&)&(0.041&)\\
Public HS&&$-$0.041&&$-$0.060&&$-$0.043&&$-$0.014&\\
&&(0.033&)&(0.043&)&(0.042&)&(0.026&)\\
Sophomores&&0.051&&0.056&&0.075&&0.061&\sym{*}\\
&&(0.043&)&(0.055&)&(0.052&)&(0.035&)\\
Juniors&&0.071&&0.119&\sym{**}&0.125&\sym{**}&0.066&\sym{*}\\
&&(0.045&)&(0.060&)&(0.062&)&(0.037&)\\
Seniors&&0.012&&0.054&&0.104&\sym{*}&0.040&\\
&&(0.048&)&(0.061&)&(0.059&)&(0.039&)\\
Arts&&$-$0.180&&$-$0.177&&0.054&&$-$0.078&\\
&&(0.116&)&(0.153&)&(0.158&)&(0.098&)\\
Humanities&&$-$0.121&\sym{*}&$-$0.230&\sym{***}&$-$0.207&\sym{***}&$-$0.112&\sym{***}\\
&&(0.069&)&(0.084&)&(0.075&)&(0.028&)\\
Languages&&$-$0.239&\sym{*}&$-$0.164&&$-$0.245&\sym{***}&$-$0.094&\sym{***}\\
&&(0.134&)&(0.137&)&(0.067&)&(0.029&)\\
Literature&&$-$0.326&\sym{***}&$-$0.177&\sym{*}&$-$0.185&\sym{**}&$-$0.097&\sym{***}\\
&&(0.095&)&(0.101&)&(0.083&)&(0.023&)\\
Social Sci.&&$-$0.036&&0.032&&0.055&&0.028&\\
&&(0.036&)&(0.047&)&(0.048&)&(0.033&)\\

%% file: results/ai_usage_N.tex
Mean Dep. Var.&&0.825&&0.589&&0.368&&0.106&\\
R$-$squared&&0.084&&0.087&&0.106&&0.057&\\
\(N\) (Students)&&616&&616&&616&&616&\\

%% file: results/ai_adopt_regs.tex
Male&&0.087&\sym{***}&0.152&\sym{***}&0.220&\sym{***}&0.131&\sym{***}&0.086&\sym{**}\\
&&(0.024&)&(0.035&)&(0.041&)&(0.043&)&(0.035&)\\
Black&&$-$0.020&&0.033&&0.057&&0.184&\sym{*}&0.157&\sym{***}\\
&&(0.058&)&(0.100&)&(0.108&)&(0.100&)&(0.047&)\\
Latino&&$-$0.013&&0.032&&0.034&&$-$0.044&&$-$0.022&\\
&&(0.037&)&(0.054&)&(0.063&)&(0.070&)&(0.063&)\\
Asian&&0.037&&0.024&&0.129&\sym{**}&0.151&\sym{***}&0.136&\sym{***}\\
&&(0.039&)&(0.049&)&(0.058&)&(0.057&)&(0.038&)\\
Public HS&&$-$0.013&&$-$0.056&&$-$0.084&\sym{**}&$-$0.014&&$-$0.068&\sym{**}\\
&&(0.024&)&(0.035&)&(0.041&)&(0.043&)&(0.034&)\\
Sophomores&&$-$0.056&\sym{*}&$-$0.071&\sym{*}&0.029&&0.177&\sym{***}&0.039&\\
&&(0.030&)&(0.042&)&(0.050&)&(0.055&)&(0.045&)\\
Juniors&&$-$0.051&&$-$0.068&&0.074&&0.242&\sym{***}&0.065&\\
&&(0.034&)&(0.049&)&(0.061&)&(0.059&)&(0.047&)\\
Seniors&&$-$0.063&\sym{*}&$-$0.009&&0.070&&0.135&\sym{**}&0.018&\\
&&(0.032&)&(0.052&)&(0.058&)&(0.060&)&(0.049&)\\
Arts&&0.023&&$-$0.130&&0.138&&0.059&&$-$0.161&\\
&&(0.102&)&(0.117&)&(0.125&)&(0.125&)&(0.113&)\\
Humanities&&$-$0.019&&0.003&&$-$0.079&&$-$0.021&&$-$0.109&\\
&&(0.036&)&(0.066&)&(0.082&)&(0.085&)&(0.071&)\\
Languages&&$-$0.037&\sym{*}&$-$0.135&\sym{***}&$-$0.324&\sym{***}&$-$0.287&\sym{**}&$-$0.235&\sym{*}\\
&&(0.020&)&(0.037&)&(0.051&)&(0.124&)&(0.133&)\\
Literature&&0.082&&0.084&&$-$0.075&&$-$0.103&&$-$0.300&\sym{***}\\
&&(0.071&)&(0.093&)&(0.096&)&(0.103&)&(0.095&)\\
Social Sci.&&0.019&&0.032&&$-$0.038&&0.021&&$-$0.025&\\
&&(0.027&)&(0.038&)&(0.044&)&(0.047&)&(0.038&)\\

%% file: results/ai_adopt_N.tex
Mean Dep. Var.&&0.089&&0.199&&0.357&&0.551&&0.801&\\
R$-$squared&&0.043&&0.061&&0.095&&0.081&&0.075&\\
\(N\) (Students)&&633&&633&&633&&633&&633&\\

%% file: results/ai_autom_regs.tex
Male&&0.052&\sym{**}&0.080&\sym{***}&0.237&\sym{***}&0.029&&0.095&\sym{***}&0.272&\sym{***}&$-$0.016&&$-$0.035&\\
&&(0.026&)&(0.028&)&(0.076&)&(0.027&)&(0.024&)&(0.059&)&(0.023&)&(0.055&)\\
Black&&$-$0.009&&0.053&&0.446&\sym{**}&$-$0.008&&0.109&&0.477&\sym{**}&$-$0.056&&$-$0.031&\\
&&(0.080&)&(0.068&)&(0.220&)&(0.083&)&(0.069&)&(0.199&)&(0.052&)&(0.155&)\\
Latino&&0.064&\sym{*}&0.074&\sym{*}&0.241&\sym{**}&0.049&&0.088&\sym{**}&0.277&\sym{***}&$-$0.014&&$-$0.036&\\
&&(0.034&)&(0.042&)&(0.118&)&(0.035&)&(0.043&)&(0.104&)&(0.041&)&(0.091&)\\
Asian&&0.055&\sym{*}&0.066&\sym{*}&0.180&\sym{*}&0.058&\sym{*}&0.050&\sym{*}&0.156&\sym{**}&0.016&&0.025&\\
&&(0.031&)&(0.035&)&(0.095&)&(0.031&)&(0.028&)&(0.069&)&(0.030&)&(0.074&)\\
Public HS&&$-$0.003&&$-$0.028&&$-$0.018&&$-$0.019&&$-$0.005&&$-$0.007&&$-$0.024&&$-$0.011&\\
&&(0.029&)&(0.028&)&(0.076&)&(0.028&)&(0.024&)&(0.058&)&(0.023&)&(0.056&)\\
Sophomores&&0.033&&0.020&&0.059&&$-$0.002&&0.003&&0.018&&0.017&&0.041&\\
&&(0.032&)&(0.034&)&(0.091&)&(0.033&)&(0.030&)&(0.070&)&(0.027&)&(0.065&)\\
Juniors&&0.049&&0.099&\sym{**}&0.232&\sym{**}&$-$0.020&&0.033&&0.119&&0.067&\sym{**}&0.113&\\
&&(0.038&)&(0.039&)&(0.104&)&(0.041&)&(0.035&)&(0.086&)&(0.033&)&(0.076&)\\
Seniors&&$-$0.013&&0.081&\sym{**}&0.313&\sym{***}&0.017&&0.018&&0.150&\sym{*}&0.063&\sym{*}&0.163&\sym{*}\\
&&(0.042&)&(0.040&)&(0.115&)&(0.031&)&(0.030&)&(0.081&)&(0.033&)&(0.083&)\\
Arts&&0.001&&$-$0.377&\sym{***}&$-$0.840&\sym{***}&$-$0.104&&$-$0.030&&0.042&&$-$0.347&\sym{***}&$-$0.882&\sym{***}\\
&&(0.069&)&(0.075&)&(0.159&)&(0.118&)&(0.065&)&(0.166&)&(0.084&)&(0.197&)\\
Humanities&&$-$0.096&&$-$0.199&\sym{***}&$-$0.561&\sym{***}&0.025&&$-$0.073&\sym{*}&$-$0.188&\sym{*}&$-$0.126&\sym{***}&$-$0.373&\sym{***}\\
&&(0.063&)&(0.049&)&(0.115&)&(0.048&)&(0.041&)&(0.099&)&(0.046&)&(0.095&)\\
Languages&&$-$0.146&&$-$0.279&\sym{***}&$-$0.582&\sym{***}&$-$0.015&&$-$0.045&&$-$0.157&&$-$0.234&\sym{***}&$-$0.424&\sym{***}\\
&&(0.143&)&(0.094&)&(0.211&)&(0.110&)&(0.055&)&(0.121&)&(0.061&)&(0.133&)\\
Literature&&$-$0.103&&$-$0.160&\sym{**}&$-$0.474&\sym{***}&$-$0.076&&$-$0.062&&$-$0.097&&$-$0.098&\sym{*}&$-$0.377&\sym{***}\\
&&(0.094&)&(0.079&)&(0.164&)&(0.100&)&(0.058&)&(0.144&)&(0.056&)&(0.121&)\\
Social Sci.&&$-$0.001&&$-$0.014&&0.054&&0.039&&0.061&\sym{**}&0.233&\sym{***}&$-$0.076&\sym{***}&$-$0.179&\sym{***}\\
&&(0.029&)&(0.030&)&(0.086&)&(0.029&)&(0.027&)&(0.066&)&(0.024&)&(0.059&)\\

%% file: results/ai_autom_N.tex
Mean Dep. Var.&&0.917&&0.560&&2.159&&0.916&&0.454&&1.885&&0.106&&0.274&\\
R$-$squared&&0.041&&0.127&&0.133&&0.026&&0.090&&0.147&&0.098&&0.107&\\
\(N\) (Students)&&515&&515&&515&&515&&515&&515&&515&&515&\\

%% file: results/ai_freq_dem.tex
Male	&88.7	&20.2	&19.1	&33.4	&16.1	\\
Female	&78.4	&25.7	&25.1	&21.0	&6.6	\\
White	&80.2	&25.6	&19.8	&24.0	&10.8	\\
Black	&92.3	&15.1	&23.5	&41.4	&12.3	\\
Hispanic	&77.9	&19.9	&28.7	&22.8	&6.6	\\
Asian	&91.3	&19.5	&27.0	&30.1	&14.8	\\
Private HS	&84.1	&23.7	&22.2	&27.2	&10.9	\\
Public HS	&80.4	&24.7	&21.4	&24.4	&9.9	\\

%% file: results/ai_freq_aca.tex
GPA > p50	&80.3	&26.3	&17.4	&23.2	&13.4	\\
GPA < p50	&87.1	&18.7	&26.3	&32.2	&9.9	\\
Freshman	&81.3	&24.3	&24.1	&24.7	&8.1	\\
Sophomore	&84.4	&25.6	&22.6	&23.7	&12.4	\\
Junior	&86.2	&21.6	&23.1	&28.6	&12.8	\\
Senior	&78.7	&21.0	&17.4	&29.9	&10.4	\\

%% file: results/ai_freq_maj.tex
Arts	&73.8	&22.1	&0.0	&45.1	&6.6	\\
Humanities	&76.5	&36.5	&21.0	&17.7	&1.3	\\
Languages	&59.5	&20.8	&31.0	&7.7	&0.0	\\
Literature	&53.8	&10.9	&25.8	&17.2	&0.0	\\
Natural Sciences	&87.4	&25.9	&23.3	&26.3	&11.9	\\
Social Sciences	&84.5	&18.6	&21.3	&29.9	&14.7	\\

%% file: results/ai_model_regs.tex
Male&&0.023&&0.083&\sym{**}&0.011&&$-$0.007&&0.103&\sym{***}\\
&&(0.020&)&(0.036&)&(0.027&)&(0.033&)&(0.029&)\\
Black&&$-$0.017&&0.002&&$-$0.012&&$-$0.013&&$-$0.043&\\
&&(0.055&)&(0.095&)&(0.061&)&(0.076&)&(0.050&)\\
Latino&&0.027&&$-$0.030&&0.033&&$-$0.012&&0.020&\\
&&(0.026&)&(0.050&)&(0.051&)&(0.051&)&(0.043&)\\
Asian&&0.029&&$-$0.074&\sym{**}&$-$0.007&&$-$0.065&\sym{**}&0.134&\sym{***}\\
&&(0.019&)&(0.035&)&(0.033&)&(0.032&)&(0.044&)\\
Public HS&&$-$0.032&\sym{*}&0.004&&0.017&&$-$0.010&&$-$0.003&\\
&&(0.017&)&(0.033&)&(0.026&)&(0.030&)&(0.028&)\\
Sophomores&&0.011&&0.015&&$-$0.064&\sym{*}&0.008&&0.025&\\
&&(0.028&)&(0.041&)&(0.033&)&(0.036&)&(0.034&)\\
Juniors&&0.044&\sym{**}&0.072&&$-$0.065&\sym{*}&0.059&&0.028&\\
&&(0.020&)&(0.053&)&(0.038&)&(0.044&)&(0.038&)\\
Seniors&&0.026&&0.058&&$-$0.068&\sym{**}&0.054&&0.102&\sym{**}\\
&&(0.026&)&(0.050&)&(0.034&)&(0.045&)&(0.045&)\\
Arts&&$-$0.082&&$-$0.086&&0.002&&$-$0.067&\sym{**}&0.342&\sym{*}\\
&&(0.088&)&(0.102&)&(0.090&)&(0.032&)&(0.190&)\\
Humanities&&$-$0.032&&$-$0.101&\sym{*}&$-$0.078&\sym{***}&0.015&&$-$0.047&\\
&&(0.041&)&(0.058&)&(0.023&)&(0.057&)&(0.040&)\\
Languages&&0.041&\sym{*}&$-$0.078&&$-$0.072&\sym{***}&0.047&&$-$0.063&\\
&&(0.022&)&(0.071&)&(0.023&)&(0.137&)&(0.039&)\\
Literature&&$-$0.026&&$-$0.141&\sym{***}&$-$0.084&\sym{***}&0.040&&$-$0.064&\sym{**}\\
&&(0.058&)&(0.031&)&(0.023&)&(0.080&)&(0.029&)\\
Social Sci.&&$-$0.006&&$-$0.042&&$-$0.005&&0.068&\sym{**}&0.084&\sym{***}\\
&&(0.020&)&(0.040&)&(0.031&)&(0.034&)&(0.032&)\\

%% file: results/ai_model_N.tex
Mean Dep. Var.&&0.962&&0.135&&0.077&&0.110&&0.114&\\
R$-$squared&&0.026&&0.039&&0.030&&0.029&&0.137&\\
\(N\) (Students)&&516&&516&&516&&516&&516&\\

%% file: results/ai_policy_regs.tex
Male&&0.134&\sym{***}&0.190&\sym{***}&0.129&\sym{***}&0.107&\sym{***}&0.026&\\
&&(0.045&)&(0.045&)&(0.045&)&(0.033&)&(0.045&)\\
Black&&0.261&\sym{***}&0.233&\sym{**}&0.082&&0.020&&0.242&\sym{**}\\
&&(0.097&)&(0.114&)&(0.119&)&(0.083&)&(0.116&)\\
Latino&&$-$0.039&&$-$0.004&&$-$0.037&&$-$0.056&&0.017&\\
&&(0.074&)&(0.072&)&(0.072&)&(0.042&)&(0.072&)\\
Asian&&0.097&&0.091&&$-$0.004&&0.022&&0.075&\\
&&(0.060&)&(0.059&)&(0.059&)&(0.044&)&(0.059&)\\
Public HS&&$-$0.016&&$-$0.037&&$-$0.065&&$-$0.019&&0.002&\\
&&(0.044&)&(0.044&)&(0.044&)&(0.031&)&(0.044&)\\
Sophomores&&0.023&&0.062&&0.023&&0.017&&0.006&\\
&&(0.056&)&(0.055&)&(0.057&)&(0.039&)&(0.058&)\\
Juniors&&0.011&&0.103&&0.068&&0.035&&$-$0.024&\\
&&(0.065&)&(0.063&)&(0.064&)&(0.045&)&(0.064&)\\
Seniors&&0.023&&0.042&&$-$0.029&&$-$0.009&&0.032&\\
&&(0.063&)&(0.063&)&(0.062&)&(0.045&)&(0.060&)\\
Arts&&$-$0.479&\sym{***}&$-$0.384&\sym{**}&$-$0.303&\sym{**}&0.011&&$-$0.490&\sym{***}\\
&&(0.149&)&(0.158&)&(0.131&)&(0.127&)&(0.053&)\\
Humanities&&$-$0.079&&$-$0.177&\sym{**}&$-$0.181&\sym{**}&$-$0.068&&$-$0.011&\\
&&(0.087&)&(0.082&)&(0.083&)&(0.046&)&(0.088&)\\
Languages&&$-$0.115&&$-$0.256&\sym{**}&$-$0.219&\sym{**}&0.004&&$-$0.119&\\
&&(0.139&)&(0.113&)&(0.101&)&(0.071&)&(0.140&)\\
Literature&&$-$0.275&\sym{***}&$-$0.220&\sym{***}&$-$0.178&\sym{**}&$-$0.057&&$-$0.219&\sym{**}\\
&&(0.089&)&(0.084&)&(0.090&)&(0.041&)&(0.088&)\\
Social Sci.&&0.019&&0.009&&0.042&&0.085&\sym{**}&$-$0.067&\\
&&(0.049&)&(0.049&)&(0.049&)&(0.035&)&(0.050&)\\

%% file: results/ai_policy_N.tex
Mean Dep. Var.&&0.525&&0.422&&0.409&&0.134&&0.391&\\
R$-$squared&&0.067&&0.088&&0.064&&0.058&&0.038&\\
\(N\) (Students)&&599&&599&&599&&599&&599&\\

%% file: results/ai_improves_regs.tex
Male&&0.122&\sym{***}&0.101&\sym{**}&0.140&\sym{***}&0.027&&0.181&\sym{***}\\
&&(0.045&)&(0.041&)&(0.045&)&(0.048&)&(0.045&)\\
Black&&0.361&\sym{***}&0.244&\sym{***}&0.283&\sym{**}&$-$0.056&&0.103&\\
&&(0.065&)&(0.055&)&(0.129&)&(0.128&)&(0.110&)\\
Latino&&0.179&\sym{**}&0.069&&0.203&\sym{***}&0.020&&0.076&\\
&&(0.070&)&(0.064&)&(0.077&)&(0.076&)&(0.072&)\\
Asian&&0.125&\sym{**}&0.149&\sym{***}&0.060&&0.025&&0.136&\sym{**}\\
&&(0.061&)&(0.052&)&(0.059&)&(0.064&)&(0.060&)\\
Public HS&&$-$0.067&&$-$0.067&\sym{*}&0.014&&$-$0.052&&$-$0.054&\\
&&(0.045&)&(0.041&)&(0.045&)&(0.047&)&(0.045&)\\
Sophomores&&0.089&&0.026&&0.029&&$-$0.133&\sym{**}&0.041&\\
&&(0.055&)&(0.050&)&(0.057&)&(0.059&)&(0.058&)\\
Juniors&&0.008&&$-$0.027&&0.075&&$-$0.148&\sym{**}&0.043&\\
&&(0.067&)&(0.059&)&(0.066&)&(0.067&)&(0.065&)\\
Seniors&&$-$0.083&&$-$0.098&&$-$0.106&\sym{*}&$-$0.149&\sym{**}&$-$0.048&\\
&&(0.063&)&(0.060&)&(0.060&)&(0.066&)&(0.064&)\\
Arts&&0.113&&$-$0.084&&$-$0.254&\sym{**}&0.128&&0.099&\\
&&(0.152&)&(0.146&)&(0.118&)&(0.208&)&(0.128&)\\
Humanities&&$-$0.115&&$-$0.373&\sym{***}&$-$0.149&\sym{*}&$-$0.025&&$-$0.245&\sym{***}\\
&&(0.090&)&(0.083&)&(0.078&)&(0.089&)&(0.086&)\\
Languages&&$-$0.186&&$-$0.385&\sym{***}&$-$0.107&&$-$0.143&&$-$0.201&\sym{*}\\
&&(0.127&)&(0.125&)&(0.108&)&(0.130&)&(0.121&)\\
Literature&&$-$0.222&\sym{**}&$-$0.224&\sym{**}&$-$0.314&\sym{***}&$-$0.085&&$-$0.138&\\
&&(0.102&)&(0.103&)&(0.074&)&(0.115&)&(0.119&)\\
Social Sci.&&0.015&&$-$0.085&\sym{*}&0.030&&$-$0.000&&$-$0.022&\\
&&(0.049&)&(0.044&)&(0.051&)&(0.051&)&(0.050&)\\

%% file: results/ai_improves_N.tex
Mean Dep. Var.&&0.576&&0.685&&0.376&&0.489&&0.564&\\
R$-$squared&&0.087&&0.120&&0.091&&0.032&&0.082&\\
\(N\) (Students)&&561&&565&&543&&560&&560&\\

%% file: results/weight_adoption.tex
Overall	&	82.5	&	84.3	&	83.6	&	83.8	\\
Male	&	88.7	&	88.9	&	89.4	&	89.4	\\
Female	&	78.4	&	80.3	&	80.4	&	80.1	\\
\addlinespace
White	&	80.2	&	83.6	&	82.8	&	82.8	\\
Black	&	92.3	&	95.0	&	91.3	&	91.3	\\
Hispanic	&	77.9	&	80.4	&	80.0	&	80.0	\\
Asian	&	91.3	&	90.2	&	89.8	&	89.8	\\
\addlinespace
Private HS	&	84.1	&	86.5	&	85.8	&	85.6	\\
Public HS	&	80.4	&	82.1	&	81.2	&	81.8	\\
\addlinespace
GPA $\geq$ median	&	80.3	&	82.6	&	82.7	&	82.4	\\
GPA $<$ median	&	87.1	&	88.2	&	88.7	&	88.0	\\
\addlinespace
Natural Sciences	&	87.4	&	86.8	&	85.6	&	86.4	\\
Social Sciences	&	84.5	&	89.6	&	89.0	&	89.0	\\
Humanities	&	76.5	&	76.7	&	79.9	&	77.3	\\
Literature	&	53.8	&	50.5	&	51.7	&	50.0	\\
Languages	&	59.5	&	52.2	&	54.0	&	53.3	\\
Arts	&	73.8	&	57.6	&	58.3	&	63.6	\\
\midrule
\(N\) (Students)	&	616	&	589	&	560	&	616	\\

%% file: results/weight_tasks.tex
\multicolumn{5}{l}{\hspace{-1em} \textit{Augmentation tasks}} \\
\hspace{1em} Explaining concepts	&	85.4	&	85.7	&	86.5	&	85.9	\\
\hspace{1em} Generating ideas	&	66.0	&	68.8	&	69.6	&	67.7	\\
\hspace{1em} Editing essay drafts	&	51.5	&	52.2	&	52.8	&	52.6	\\
\hspace{1em} Coding assistance	&	37.0	&	45.3	&	41.2	&	42.1	\\
\addlinespace
\multicolumn{5}{l}{\hspace{-1em} \textit{Automation tasks}} \\
\hspace{1em} Summarizing texts	&	79.5	&	80.3	&	79.9	&	80.3	\\
\hspace{1em} Finding information	&	67.1	&	68.8	&	69.6	&	67.9	\\
\hspace{1em} Proofreading	&	58.5	&	59.3	&	59.4	&	58.6	\\
\hspace{1em} Writing emails	&	40.2	&	41.5	&	41.3	&	41.5	\\
\hspace{1em} Writing essays	&	26.1	&	27.1	&	25.4	&	26.9	\\
\hspace{1em} Creating images	&	21.8	&	21.6	&	19.2	&	20.6	\\
\addlinespace
\midrule
Augmentation (average)	&	59.6	&	62.6	&	62.1	&	61.6	\\
Automation (average)	&	48.3	&	49.1	&	48.6	&	48.7	\\
\midrule
\(N\) (Students)	&	510	&	510	&	510	&	510	\\

%% file: results/weight_beliefs.tex
Learning ability	&	57.6	&	59.2	&	60.2	&	59.2	\\
Understanding of course materials	&	68.5	&	68.5	&	69.2	&	68.7	\\
Course grades	&	37.6	&	39.6	&	41.0	&	40.1	\\
Ability to complete assignments on time	&	56.4	&	57.6	&	55.9	&	57.3	\\
Time spent on academics	&	48.9	&	47.3	&	47.5	&	48.8	\\
\midrule
\(N\) (Students)	&	560	&	537	&	511	&	560	\\

%% file: results/ai_purpose_regs.tex
Male&&0.076&\sym{**}&0.120&\sym{***}&0.162&\sym{***}&0.049&&0.042&&0.097&\sym{**}&0.041&&0.097&\sym{**}&0.044&&0.163&\sym{***}\\
&&(0.037&)&(0.042&)&(0.046&)&(0.046&)&(0.050&)&(0.049&)&(0.046&)&(0.043&)&(0.044&)&(0.039&)\\
Black&&0.114&&0.042&&0.265&\sym{***}&$-$0.113&&0.148&&0.302&\sym{***}&0.222&\sym{*}&$-$0.091&&0.093&&$-$0.117&\\
&&(0.078&)&(0.111&)&(0.101&)&(0.121&)&(0.125&)&(0.117&)&(0.118&)&(0.100&)&(0.122&)&(0.076&)\\
Latino&&0.158&\sym{***}&0.064&&0.084&&0.214&\sym{***}&0.072&&0.079&&0.198&\sym{**}&$-$0.156&\sym{***}&$-$0.014&&0.122&\\
&&(0.050&)&(0.066&)&(0.076&)&(0.064&)&(0.083&)&(0.080&)&(0.083&)&(0.059&)&(0.073&)&(0.075&)\\
Asian&&0.112&\sym{**}&0.034&&0.014&&0.038&&0.080&&0.061&&0.247&\sym{***}&0.054&&0.006&&$-$0.081&\sym{*}\\
&&(0.045&)&(0.056&)&(0.061&)&(0.061&)&(0.064&)&(0.064&)&(0.061&)&(0.058&)&(0.054&)&(0.042&)\\
Public HS&&$-$0.089&\sym{**}&$-$0.057&&$-$0.007&&$-$0.001&&0.090&\sym{*}&0.025&&$-$0.074&&$-$0.048&&0.000&&0.021&\\
&&(0.038&)&(0.042&)&(0.046&)&(0.046&)&(0.049&)&(0.049&)&(0.045&)&(0.042&)&(0.041&)&(0.038&)\\
Sophomores&&0.040&&0.066&&0.040&&0.024&&0.027&&$-$0.079&&$-$0.018&&0.093&\sym{*}&$-$0.006&&$-$0.091&\sym{*}\\
&&(0.047&)&(0.054&)&(0.059&)&(0.058&)&(0.062&)&(0.061&)&(0.057&)&(0.052&)&(0.054&)&(0.047&)\\
Juniors&&0.051&&$-$0.009&&0.063&&0.085&&0.075&&$-$0.010&&0.052&&0.272&\sym{***}&0.053&&$-$0.036&\\
&&(0.052&)&(0.060&)&(0.063&)&(0.062&)&(0.069&)&(0.069&)&(0.065&)&(0.061&)&(0.062&)&(0.053&)\\
Seniors&&0.020&&0.058&&$-$0.022&&0.060&&0.046&&$-$0.041&&0.011&&0.284&\sym{***}&$-$0.039&&0.057&\\
&&(0.054&)&(0.055&)&(0.065&)&(0.065&)&(0.068&)&(0.067&)&(0.063&)&(0.063&)&(0.055&)&(0.058&)\\
Arts&&$-$0.122&&0.043&&$-$0.409&\sym{**}&$-$0.424&\sym{**}&0.188&&$-$0.370&\sym{***}&0.219&&$-$0.592&\sym{***}&$-$0.048&&$-$0.176&\sym{*}\\
&&(0.127&)&(0.118&)&(0.169&)&(0.179&)&(0.172&)&(0.130&)&(0.136&)&(0.096&)&(0.117&)&(0.096&)\\
Humanities&&$-$0.120&&$-$0.143&&$-$0.138&&$-$0.333&\sym{***}&$-$0.044&&$-$0.025&&$-$0.120&&$-$0.318&\sym{***}&$-$0.060&&0.069&\\
&&(0.078&)&(0.091&)&(0.087&)&(0.087&)&(0.096&)&(0.091&)&(0.077&)&(0.072&)&(0.072&)&(0.079&)\\
Languages&&$-$0.377&\sym{**}&$-$0.497&\sym{***}&0.010&&$-$0.425&\sym{***}&0.072&&0.043&&$-$0.036&&$-$0.356&\sym{***}&$-$0.055&&0.239&\\
&&(0.160&)&(0.131&)&(0.174&)&(0.144&)&(0.185&)&(0.187&)&(0.166&)&(0.119&)&(0.110&)&(0.176&)\\
Literature&&$-$0.159&&0.029&&$-$0.165&&0.029&&$-$0.026&&$-$0.191&\sym{*}&$-$0.060&&$-$0.319&\sym{***}&$-$0.133&\sym{*}&$-$0.017&\\
&&(0.113&)&(0.115&)&(0.122&)&(0.116&)&(0.124&)&(0.109&)&(0.126&)&(0.101&)&(0.080&)&(0.091&)\\
Social Sci.&&0.008&&0.099&\sym{**}&0.017&&0.048&&0.087&\sym{*}&0.057&&0.089&\sym{*}&$-$0.170&\sym{***}&0.110&\sym{**}&$-$0.035&\\
&&(0.040&)&(0.045&)&(0.050&)&(0.049&)&(0.053&)&(0.053&)&(0.050&)&(0.047&)&(0.049&)&(0.041&)\\

%% file: results/ai_purpose_N.tex
Mean Dep. Var.&&0.803&&0.740&&0.631&&0.619&&0.541&&0.473&&0.371&&0.344&&0.235&&0.204&\\
R$-$squared&&0.076&&0.089&&0.072&&0.105&&0.028&&0.053&&0.089&&0.159&&0.037&&0.083&\\
\(N\) (Students)&&515&&515&&515&&515&&515&&515&&515&&515&&515&&515&\\

%% file: results/class_forwith_table.tex
Explaining concepts &  64.3 &  30.4 &  58.8 &  41.2 &  61.5 &  35.8 & AUG \\
Summarizing texts &  35.7 &  58.9 &   9.8 &  90.2 &  22.8 &  74.5 & AUTO \\
Finding information &  50.0 &  44.6 &  47.1 &  52.9 &  48.5 &  48.8 & AUTO \\
Generating ideas &  60.7 &  33.9 &  52.9 &  47.1 &  56.8 &  40.5 & AUG \\
Proofreading &  46.4 &  48.2 &  29.4 &  70.6 &  37.9 &  59.4 & AUTO \\
Editing essays &  58.9 &  35.7 &  41.2 &  58.8 &  50.0 &  47.2 & AUG \\
Writing emails &  37.5 &  57.1 &  35.3 &  64.7 &  36.4 &  60.9 & AUTO \\
Coding help &  58.9 &  35.7 &  64.7 &  35.3 &  61.8 &  35.5 & AUG \\
Writing essays &  37.5 &  57.1 &  31.4 &  68.6 &  34.5 &  62.8 & AUTO \\
Creating images &  28.6 &  66.1 &  37.3 &  62.7 &  33.0 &  64.4 & AUTO \\

%% file: results/class_effort_table.tex
Explaining concepts & AUG &  71.4 &  21.4 &   3.6 &  76.5 &  13.7 &   9.8 \\
Summarizing texts & AUTO &  85.7 &   5.4 &   5.4 &  90.2 &   3.9 &   5.9 \\
Finding information & AUTO &  76.8 &  12.5 &   7.1 &  82.4 &  11.8 &   5.9 \\
Generating ideas & AUG &  73.2 &  21.4 &   1.8 &  84.3 &   7.8 &   7.8 \\
Proofreading & AUTO &  76.8 &  16.1 &   3.6 &  82.4 &   7.8 &   9.8 \\
Editing essays & AUG &  76.8 &  12.5 &   7.1 &  84.3 &  13.7 &   2.0 \\
Writing emails & AUTO &  69.6 &  23.2 &   3.6 &  86.3 &  13.7 &   0.0 \\
Coding help & AUG &  82.1 &  12.5 &   1.8 &  78.4 &  17.6 &   3.9 \\
Writing essays & AUTO &  87.5 &   5.4 &   3.6 &  80.4 &   9.8 &   9.8 \\
Creating images & AUTO &  76.8 &   7.1 &  12.5 &  80.4 &  13.7 &   5.9 \\
\midrule
\textit{All tasks} & &  77.7 &  13.8 &   5.0 &  82.5 &  11.4 &   6.1 \\
\textit{AUG tasks} & &  75.9 &  17.0 &   3.6 &  80.9 &  13.2 &   5.9 \\
\textit{AUTO tasks} & &  78.9 &  11.6 &   6.0 &  83.7 &  10.1 &   6.2 \\

%% file: results/class_learning_table.tex
Explaining concepts & AUG &  23.2 &   7.1 &  64.3 &  35.3 &   5.9 &  58.8 \\
Summarizing texts & AUTO &  26.8 &  25.0 &  42.9 &  56.9 &  15.7 &  27.5 \\
Finding information & AUTO &  32.1 &  23.2 &  39.3 &  49.0 &  13.7 &  37.3 \\
Generating ideas & AUG &  50.0 &  19.6 &  25.0 &  52.9 &  19.6 &  27.5 \\
Proofreading & AUTO &  32.1 &  23.2 &  39.3 &  39.2 &  41.2 &  19.6 \\
Editing essays & AUG &  44.6 &  16.1 &  33.9 &  52.9 &  31.4 &  15.7 \\
Writing emails & AUTO &  53.6 &  30.4 &  10.7 &  45.1 &  33.3 &  21.6 \\
Coding help & AUG &  37.5 &  16.1 &  41.1 &  35.3 &  19.6 &  45.1 \\
Writing essays & AUTO &  73.2 &   7.1 &  14.3 &  84.3 &   3.9 &  11.8 \\
Creating images & AUTO &  37.5 &  46.4 &  10.7 &  31.4 &  54.9 &  13.7 \\
\midrule
\textit{All tasks} & &  41.1 &  21.4 &  32.1 &  48.2 &  23.9 &  27.8 \\
\textit{AUG tasks} & &  38.8 &  14.7 &  41.1 &  44.1 &  19.1 &  36.8 \\
\textit{AUTO tasks} & &  42.6 &  25.9 &  26.2 &  51.0 &  27.1 &  21.9 \\

%% file: results/class_importance_table.tex
Explaining concepts & AUG &  67.8 &   3.8 \\
Finding information & AUTO &  67.6 &   4.0 \\
Generating ideas & AUG &  64.9 &   4.3 \\
Writing essays & AUTO &  69.5 &   4.5 \\
Proofreading & AUTO &  57.5 &   5.2 \\
Summarizing texts & AUTO &  54.7 &   5.6 \\
Editing essays & AUG &  64.0 &   5.8 \\
Coding help & AUG &  45.3 &   6.8 \\
Writing emails & AUTO &  51.3 &   7.0 \\
Creating images & AUTO &  30.3 &   8.0 \\
\midrule
\textit{All tasks} & &  57.3 &   5.5 \\
\textit{AUG tasks} & &  60.5 &   5.2 \\
\textit{AUTO tasks} & &  55.2 &   5.7 \\

%% file: results/class_forwith_aiusers_table.tex
Explaining concepts &  61.6 &  35.8 &  64.7 &  35.3 & AUG & AUG \\
Summarizing texts &  22.8 &  74.6 &  23.6 &  76.4 & AUTO & AUTO \\
\textbf{Finding information} &  48.5 &  48.8 &  50.5 &  49.5 & AUTO & \textbf{AUG} \\
Generating ideas &  56.8 &  40.5 &  62.5 &  37.5 & AUG & AUG \\
Proofreading &  37.9 &  59.4 &  42.5 &  57.5 & AUTO & AUTO \\
Editing essays &  50.1 &  47.3 &  55.7 &  44.3 & AUG & AUG \\
Writing emails &  36.4 &  60.9 &  37.3 &  62.7 & AUTO & AUTO \\
Coding help &  61.8 &  35.5 &  64.9 &  35.1 & AUG & AUG \\
Writing essays &  34.4 &  62.9 &  35.1 &  64.9 & AUTO & AUTO \\
Creating images &  32.9 &  64.4 &  33.2 &  66.8 & AUTO & AUTO \\